\RequirePackage[l2tabu,orthodox]{nag}
\documentclass[copyright,creativecommons,noderivs,10pt]{eptcs}
\providecommand{\event}{QPL 2013}

\pdfoutput=1

\usepackage[T1]{fontenc}
\usepackage{fixltx2e}
\usepackage{microtype} 
\usepackage{xspace}

\makeatletter
\newcommand\etc{etc\@ifnextchar.{}{.\@}\xspace}
\makeatother

\usepackage{graphicx}

\newcommand{\inlinegraphic}[2]{
  \dimendef\grafheight=255\dimendef\grafvshift=254
  \grafheight=#1
  \grafvshift=-0.5\grafheight
  \advance\grafvshift by 0.5ex
  \raisebox{\grafvshift}{\includegraphics[height=\grafheight]{images/#2}\xspace}
}

\newcommand{\ninlinegraphic}[2][1.0]{
  \dimendef\grafheight=255\dimendef\grafvshift=254
  \setbox0 = \hbox{\scalebox{#1}{\includegraphics{images/#2}}}
  \grafheight=\the\ht0
  \grafvshift=-0.5\grafheight
  \advance\grafvshift by 0.5ex
  \raisebox{\grafvshift}{\includegraphics[height=\grafheight]{images/#2}\xspace}
}

\newcommand{\inline}[1]{
  \raisebox{0.5ex}{\;#1\;}
}

\usepackage{latexsym}
\usepackage{amssymb}
\usepackage{amsmath} % can collide with some journals
\usepackage{stmaryrd}
\usepackage{mathtools}
\usepackage{amsthm}
\newtheorem{theorem}{Theorem}[section]

\newtheorem{lemma}[theorem]{Lemma}

\theoremstyle{definition}\newtheorem{example}[theorem]{Example}
\theoremstyle{definition}
\theoremstyle{definition}\newtheorem{definition}[theorem]{Definition}
\theoremstyle{definition}
\theoremstyle{definition}
\theoremstyle{definition}

\newcommand{\denote}[1]{% --``semantic'' brakets
\llbracket #1 \rrbracket} 
\newcommand{\ldenote}[1]{\left\llbracket #1 \right\rrbracket}

\newcommand{\bra}[1]{%  dirac 'bra'
    \ensuremath{\left\langle #1 \right|}\xspace}
\newcommand{\ket}[1]{%  dirac 'ket'
    \ensuremath{\left|  #1 \right\rangle}\xspace}

\newcommand{\CZ}{\ensuremath{\wedge Z}\xspace}
\newcommand{\CX}{\ensuremath{\wedge X}\xspace}

\usepackage{diagrams} % remember to give shout-out to Paul Taylor
\newarrow{Equals}{}{=}{}{=}{}

\ifx\numericids\undefined

\else

\fi

\newcommand{\fdhilb}{% the category of finite dimensional hilbert space
\ensuremath{\mathbf{fdHilb}}\xspace}
\newcommand{\fdHilb}{\fdhilb}

\newcommand{\zxcalculus}{\textsc{zx}-calculus\xspace}

\usepackage{dsfont}
\newcommand{\ID}{\mathds{1}}

\usepackage{hyperref}

\usepackage{tikz}
\usepackage{pgfplots}
\usetikzlibrary{trees}
\usetikzlibrary{topaths}
\usetikzlibrary{decorations.pathmorphing}
\usetikzlibrary{decorations.markings}
\usetikzlibrary{matrix,backgrounds,folding}
\usetikzlibrary{chains,scopes,positioning,fit}
\usetikzlibrary{arrows,shadows}
\usetikzlibrary{calc} 
\usetikzlibrary{chains}
\usetikzlibrary{shapes,shapes.geometric,shapes.misc}

\usepackage{tikzquanto}
\usepackage{tikzcircuits}

\newcommand{\ctikzfig}[1]{%
\begin{center}\rm
  \inline{%
\beginpgfgraphicnamed{#1}
\InputIfFileExists{#1.tikz}{}{\input{./tikz/#1.tikz}}
\endpgfgraphicnamed}
\end{center}}

\pgfdeclarelayer{edgelayer}
\pgfdeclarelayer{nodelayer}
\pgfsetlayers{background,edgelayer,nodelayer,main}

\tikzstyle{every picture}=[baseline=(current bounding box).east,scale=0.5,node distance=5mm]
\tikzstyle{none}=[inner sep=0mm]
\tikzstyle{every loop}=[]

\tikzstyle{(null)}=[]
\tikzstyle{plain}=[]

\newcommand{\greenspider}[1]{\inline{%
\begin{tikzpicture}[quanto]
  \spider{green vertex}{spideri}{0,0}
  \node [green angle] at (spideri) {#1};
\end{tikzpicture}
}}

\newcommand{\redspider}[1]{\inline{%
\begin{tikzpicture}[quanto]
  \spider{red vertex}{spideri}{0,0}
  \node [red angle] at (spideri) {#1};
\end{tikzpicture}
}}

\newcommand{\hgate}{\inline{%
\begin{tikzpicture}[quanto]
  \node [boundary vertex] (a) at (1.05,-1.0) {};
  \node [hadamard vertex] (c) at (1.05,-2.25) {};
  \node [boundary vertex] (d) at (1.05,-3.525) {};
  \draw [] (a) to (c);
  \draw [] (c) to (d);
\end{tikzpicture}
}}

\newcommand{\greenphase}[1]{\inline{%
  \begin{tikzpicture}[quanto]
    \node [green vertex] (c) at (1.175,-2.0) {};
    \node [boundary vertex] (b) at (1.175,-3.0) {};
    \node [boundary vertex] (a) at (1.175,-1.0) {};
    \draw [] (a) to (c);
    \draw [] (c) to (b);
    \node [green angle] at (c) {$#1$};
  \end{tikzpicture}
}}

\newcommand{\redphase}[1]{\inline{%
  \begin{tikzpicture}[quanto]
    \node [red vertex] (c) at (1.175,-2.0) {};
    \node [boundary vertex] (b) at (1.175,-3.0) {};
    \node [boundary vertex] (a) at (1.175,-1.0) {};
    \draw [] (a) to (c);
    \draw [] (c) to (b);
    \node [red angle] at (c) {$#1$};
  \end{tikzpicture}
}}

\newcommand{\czed}{\inline{%
  \begin{tikzpicture}[quanto]
    \node [boundary vertex] (a) at (1.0,-1.0) {};
    \node [boundary vertex] (d) at (3.0,-3.0) {};
    \node [boundary vertex] (c) at (1.0,-3.0) {};
    \node [boundary vertex] (b) at (3.0,-1) {};
    \node [hadamard vertex] (g) at (2.0,-2.0) {};
    \node [green vertex] (f) at (1.0,-2.0) {};
    \node [green vertex] (e) at (3.0,-2.0) {};
    \draw [] (f) to (g);
    \draw [] (e) to (d);
    \draw [] (f) to (c);
    \draw [] (b) to (e);
    \draw [] (a) to (f);
    \draw [] (g) to (e);
  \end{tikzpicture}
}}

\newcommand{\cex}{\inline{%
  \begin{tikzpicture}[quanto]
    \node [boundary vertex] (a) at (1.0,-1.0) {};
    \node [boundary vertex] (d) at (3.0,-3.0) {};
    \node [boundary vertex] (c) at (1.0,-3.0) {};
    \node [boundary vertex] (b) at (3.0,-1) {};
    \node [green vertex] (f) at (1.0,-2.0) {};
    \node [red vertex] (e) at (3.0,-2.0) {};
    \draw [] (f) to (e);
    \draw [] (e) to (d);
    \draw [] (f) to (c);
    \draw [] (b) to (e);
    \draw [] (a) to (f);
  \end{tikzpicture}
}}

\newcommand{\meas}{\inline{%
  \begin{tikzpicture}[quanto]
    \node [green vertex] (c) at (1.3,-2.25) {};
    \node [boundary vertex] (d) at (1.3,-1.0) {};
    \node [green vertex] (b) at (1.3,-3.5) {};
    \draw [] (c) to (b);
    \draw [] (d) to (c);
    \node [green angle] at (c) {$\pi,\{i\}$};
    \node [green angle] at (b) {$-\alpha$};
  \end{tikzpicture}
}}

\usetikzlibrary{fit}

\newcommand{\redKzero}{%
  \inline{
    \begin{tikzpicture}[quanto]
      \node [boundary vertex] (a) at (1.05,-2.25) {};
      \node [red vertex] (b) at (1.05,-1.0) {};
      \draw [] (b) to (a);
    \end{tikzpicture}
  }
}

\newcommand{\redKalpha}[1]{%
  \inline{
    \begin{tikzpicture}[quanto]
      \node [boundary vertex] (a) at (1.05,-2.25) {};
      \node [red vertex] (b) at (1.05,-1.0) {};
      \draw [] (b) to (a);
      \node [red angle] at (b) {$#1$};
    \end{tikzpicture}
  }
}

\newcommand{\redKone}{\redKalpha{\pi}}

\newcommand{\CZviaCX}{\inline{%
\begin{tikzpicture}[quanto]
  \node [style=boundary vertex] (0) at (-1, 4) {};
  \node [style=boundary vertex] (1) at (1, 4) {};
  \node [style=hadamard vertex] (2) at (1, 3) {};
  \node [style=green vertex] (3) at (-1, 2) {};
  \node [style=red vertex] (4) at (1, 2) {};
  \node [style=hadamard vertex] (5) at (1, 1) {};
  \node [style=boundary vertex] (6) at (-1, 0) {};
  \node [style=boundary vertex] (7) at (1, 0) {};
  \draw  (3) to (4);
  \draw  (2) to (1);
  \draw  (7) to (5);
  \draw  (0) to (3);
  \draw  (3) to (6);
  \draw  (4) to (2);
  \draw  (5) to (4);
\end{tikzpicture}
}}

\newcommand{\circbox}[1]{\inline{%
\begin{tikzpicture}[circuit]
  \node  (0) at (0, 1.5) {};
  \node [box] (1) at (0, 0) {$#1$};
  \node  (2) at (0, -1.5) {};
  \draw  (0) to (1);
  \draw  (1) to (2);
\end{tikzpicture}
}}

\newcommand{\circX}[1]{\circbox{X_#1}}
\newcommand{\circZ}[1]{\circbox{Z_#1}}
\newcommand{\circH}{\circbox{H}}

\newcommand{\circket}[1]{\inline{%
\begin{tikzpicture}[circuit]
  \node [box] (1) at (0, 0) {$\ket{#1}$};
  \node  (2) at (0, -1.5) {};
  \draw  (1) to (2);
\end{tikzpicture}
}}

\newcommand{\circCX}{\inline{%
\begin{tikzpicture}[circuit]
  \node  (0) at (-1, 1) {};
  \node  (1) at (1, 1) {};
  \circcnot{2}{-1, 0}{3}{1, 0}
  \node  (4) at (-1, -1) {};
  \node  (5) at (1, -1) {};
  \draw  (1) to (3);
  \draw  (0) to (2);
  \draw  (3) to (5);
  \draw  (2) to (4);
\end{tikzpicture}
}}

\newcommand{\circCZcirc}{\inline{%
\begin{tikzpicture}[circuit]
  \node  (0) at (-1, 2.5) {};
  \node  (1) at (1, 2.5) {};
  \node [style=H box] (2) at (1, 1.25) {};
  \circcnot{3}{-1, 0}{4}{1, 0}
  \node [style=H box] (5) at (1, -1.25) {};
  \node  (6) at (-1, -2.5) {};
  \node  (7) at (1, -2.5) {};
  \draw  (5) to (7);
  \draw  (3) to (6);
  \draw  (4) to (5);
  \draw  (1) to (2);
  \draw  (2) to (4);
  \draw  (0) to (3);
\end{tikzpicture}
}}

\newcommand{\greenSpiderLHS}{%
  \begin{tikzpicture}[quanto]
    \upspider{green vertex}{spideri}{0,0}
    \downspider{green vertex}{spiderii}{0,-1.5}
    \draw [] (spideri) to (spiderii) ;
    \node [green angle] at (spideri) {$\alpha,\Phi$};
    \node [green angle] at (spiderii) {$\beta, \Phi$};
  \end{tikzpicture}
}

\newcommand{\greenSpiderRHS}{%
  \begin{tikzpicture}[quanto]
    \spider{green vertex}{a}{0,0}
    \node [green angle] at (a) {$\alpha + \beta,\Phi$};
  \end{tikzpicture}
}

\newcommand{\greenTrueLHS}{%
  \begin{tikzpicture}[quanto]
    \spider{green vertex}{a}{0,0}
    \node [green angle] at (a) {$\alpha,\mathbf{1}$};
  \end{tikzpicture}
}

\newcommand{\greenTrueRHS}{%
  \begin{tikzpicture}[quanto]
    \spider{green vertex}{a}{0,0}
    \node [green angle] at (a) {$\alpha$};
  \end{tikzpicture}
}
\newcommand{\greenFalseLHS}{%
  \begin{tikzpicture}[quanto]
    \spider{green vertex}{a}{0,0}
    \node [green angle] at (a) {$\alpha,\mathbf{0}$};
  \end{tikzpicture}
}

\newcommand{\greenFalseRHS}{%
  \begin{tikzpicture}[quanto]
    \spider{green vertex}{a}{0,0}
  \end{tikzpicture}
}

\newcommand{\greenNewSpiderLHS}{%
  \begin{tikzpicture}[quanto]
    \upspider{green vertex}{spideri}{0,0}
    \downspider{green vertex}{spiderii}{0,-1.5}
    \draw [] (spideri) to (spiderii) ;
    \node [green angle] at (spideri) {$\pi,\Phi$};
    \node [green angle] at (spiderii) {$\pi, \Phi'$};
  \end{tikzpicture}
}

\newcommand{\greenNewSpiderRHS}{%
  \begin{tikzpicture}[quanto]
    \spider{green vertex}{a}{0,0}
    \node [green angle] at (a) {$\pi,\Phi\oplus\Phi'$};
  \end{tikzpicture}
}

\newcommand{\greenAntiLoopLHS}{%
  \begin{tikzpicture}[quanto]
    \downspider{green vertex}{c}{1.75,-1.0}
    \draw (c) to [in=40,out=140,loop] ();
    \node [green angle] at (c) {$\alpha,\Phi$};
  \end{tikzpicture}
}

\newcommand{\greenAntiLoopRHS}{%
  \begin{tikzpicture}[quanto]
    \downspider{green vertex}{c}{1.75,-1.0}
    \node [green angle] at (c) {$\alpha,\Phi$};
  \end{tikzpicture}
}

\newcommand{\greenIdentityLHS}{%
\begin{tikzpicture}[quanto]
\node [boundary vertex] (c) at (1.05,-3.525) {};
\node [boundary vertex] (a) at (1.05,-1.0) {};
\node [green vertex] (b) at (1.05,-2.25) {};
\draw [] (a) to (b);
\draw [] (b) to (c);
\node [green angle] at (b) {$0,\Phi$};
\end{tikzpicture}
}

\newcommand{\greenIdentityRHS}{%
\begin{tikzpicture}[quanto]
\node [boundary vertex] (c) at (1.05,-3.525) {};
\node [boundary vertex] (a) at (1.05,-1.0) {};
\draw [] (a) to (c);
\end{tikzpicture}
}

\newcommand{\piCommutesLHS}{%
\begin{tikzpicture}[quanto]
\node [boundary vertex] (d) at (1.05,-4.8) {};
\node [boundary vertex] (e) at (2.5,-4.8) {};
\node [green vertex] (b) at (1.75,-3.525) {};
\node [boundary vertex] (c) at (1.75,-1.0) {};
\node [red vertex] (a) at (1.75,-2.25) {};
\node [ellipses] (f) at ($ 0.5*(d) + 0.5*(e) + (0,0.4)$) {} ;
\draw [] (b) to (d);
\draw [] (a) to (b);
\draw [] (c) to (a);
\draw [] (b) to (e);
\node [green angle] at (b) {$\alpha,\Phi$};
\node [red angle] at (a) {$\pi,\Phi$};
\end{tikzpicture}
}

\newcommand{\piCommutesRHS}{%
\begin{tikzpicture}[quanto]
\node [red vertex] (c) at (1.175,-3.525) {};
\node [boundary vertex] (f) at (2.85,-4.8) {};
\node [green vertex] (a) at (2.0,-2.25) {};
\node [red vertex] (b) at (2.85,-3.525) {};
\node [boundary vertex] (e) at (1.175,-4.8) {};
\node [boundary vertex] (d) at (2.0,-1.0) {};
\node [ellipses] (g) at ($ 0.5*(f) + 0.5*(e) + (0,0.33)$) {} ;
\draw [] (c) to (e);
\draw [] (d) to (a);
\draw [] (b) to (f);
\draw [] (a) to (c);
\draw [] (a) to (b);
\node [red angle,left] at ($(c)-(0.8,0)$) {$\pi,\Phi$};
\node [green angle] at (a) {$-\alpha,\Phi$};
\node [red angle] at (b) {$\pi,\Phi$};
\end{tikzpicture}
}

\newcommand{\bialgebraLHS}{%
\begin{tikzpicture}[quanto]
\node [boundary vertex] (f) at (1.05,-1.0) {};
\node [boundary vertex] (h) at (2.5,-4.8) {};
\node [green vertex] (d) at (1.05,-2.25) {};
\node [boundary vertex] (g) at (1.05,-4.8) {};
\node [boundary vertex] (e) at (2.5,-1.0) {};
\node [red vertex] (c) at (1.05,-3.525) {};
\node [green vertex] (b) at (2.5,-2.25) {};
\node [red vertex] (a) at (2.5,-3.525) {};
\draw [] (e) to (b);
\draw [] (a) to (h);
\draw [] (c) to (g);
\draw [] (d) to (c);
\draw [] (b) to (a);
\draw [] (f) to (d);
\draw [] (b) to (c);
\draw [] (d) to (a);
\end{tikzpicture}
}

\newcommand{\bialgebraRHS}{%
\begin{tikzpicture}[quanto]
\node [red vertex] (e) at (1.75,-2.25) {};
\node [boundary vertex] (a) at (1.05,-1.0) {};
\node [green vertex] (f) at (1.75,-3.525) {};
\node [boundary vertex] (c) at (1.05,-4.8) {};
\node [boundary vertex] (d) at (2.5,-4.8) {};
\node [boundary vertex] (b) at (2.5,-1.0) {};
\draw [] (a) to (e);
\draw [] (f) to (c);
\draw [] (e) to (f);
\draw [] (f) to (d);
\draw [] (b) to (e);
\end{tikzpicture}
}

\newcommand{\copyingLHS}{%
\begin{tikzpicture}[quanto]
\node [boundary vertex] (d) at (1.05,-3.525) {};
\node [red vertex] (b) at (1.75,-1.0) {};
\node [green vertex] (a) at (1.75,-2.25) {};
\node [boundary vertex] (c) at (2.5,-3.525) {};
\node [ellipses] (e) at ($ 0.5*(d) + 0.5*(c) + (0,0.4)$) {} ;
\draw [] (a) to (c);
\draw [] (a) to (d);
\draw [] (b) to (a);
\node [green angle] at (a) {$\alpha , \Phi$};
\end{tikzpicture}
}

\newcommand{\copyingRHS}{%
\begin{tikzpicture}[quanto]
\node [red vertex] (b) at (1.05,-1.0) {};
\node [boundary vertex] (d) at (1.05,-2.25) {};
\node [boundary vertex] (c) at (2.5,-2.25) {};
\node [red vertex] (a) at (2.5,-1.0) {};
\node [ellipses] (e) at ($ 0.5*(d) + 0.5*(c) + (0,0.5)$) {} ;
\draw [] (a) to (c);
\draw [] (b) to (d);
\end{tikzpicture}
}

\newcommand{\hopfLawLHS}{%
\begin{tikzpicture}[quanto]
\downspider{red vertex}{a}{1.05,-3.525}
\upspider{green vertex}{b}{1.05,-2}
\draw [,bend left=30] (b) to (a);
\draw [,bend left=-30] (b) to (a);
\node [green angle] at (b) {$\alpha , \Phi$};
\node [red angle] at (a) {$\beta , \Phi'$};
\end{tikzpicture}
}

\newcommand{\hopfLawRHS}{%
\begin{tikzpicture}[quanto]
\downspider{red vertex}{a}{1.05,-3.525}
\upspider{green vertex}{b}{1.05,-2}
\node [green angle] at (b) {$\alpha , \Phi$};
\node [red angle] at (a) {$\beta , \Phi'$};
\end{tikzpicture}
}

\newcommand{\removeGreenLHS}{%
\begin{tikzpicture}[quanto]
\node [boundary vertex] (b) at (1.05,-4.8) {};
\node [boundary vertex] (c) at (2.5,-4.8) {};
\node [green vertex] (e) at (1.75,-3.525) {};
\node [hadamard vertex] (d) at (1.75,-2.25) {};
\node [boundary vertex] (a) at (1.75,-1.0) {};
\node [ellipses] (f) at ($ 0.5*(b) + 0.5*(c) + (0,0.4)$) {} ;
\draw [] (e) to (c);
\draw [] (a) to (d);
\draw [] (e) to (b);
\draw [] (d) to (e);
\node [green angle] at (e) {$\alpha,\Phi$};
\end{tikzpicture}
}

\newcommand{\removeGreenRHS}{%
\begin{tikzpicture}[quanto]
\node [boundary vertex] (a) at (1.75,-1.0) {};
\node [hadamard vertex] (f) at (1.05,-3.525) {};
\node [boundary vertex] (b) at (1.05,-4.8) {};
\node [boundary vertex] (c) at (2.5,-4.8) {};
\node [red vertex] (d) at (1.75,-2.25) {};
\node [hadamard vertex] (e) at (2.5,-3.525) {};
\node [ellipses] (g) at ($ 0.5*(b) + 0.5*(c) + (0,0.33)$) {} ;
\draw [] (f) to (b);
\draw [] (a) to (d);
\draw [] (d) to (f);
\draw [] (d) to (e);
\draw [] (e) to (c);
\node [red angle] at (d) {$\alpha,\Phi$};
\end{tikzpicture}
}

\newcommand{\HsquaredLHS}{%
\begin{tikzpicture}[quanto]
\node [boundary vertex] (a) at (1.05,-1.0) {};
\node [hadamard vertex] (c) at (1.05,-2.25) {};
\node [hadamard vertex] (d) at (1.05,-3.525) {};
\node [boundary vertex] (b) at (1.05,-4.8) {};
\draw [] (d) to (b);
\draw [] (a) to (c);
\draw [] (c) to (d);
\end{tikzpicture}
}

\newcommand{\HsquaredRHS}{%
\begin{tikzpicture}[quanto]
\node [boundary vertex] (c) at (1.05,-3.525) {};
\node [boundary vertex] (a) at (1.05,-1.0) {};
\draw [] (a) to (c);
\end{tikzpicture}
}

\newcommand{\ilemmai}{
\inline{
\begin{tikzpicture}
  \begin{pgfonlayer}{nodelayer}
    \node [style=red vertex] (0) at (-3, 1) {};
    \node [style=green vertex] (1) at (-1, 1) {};
    \node [style=red vertex] (2) at (1, 1) {};
    \node [style=newstyle] (3) at (1.5, 1.5) {$\pi, v$};
    \node [style=green vertex] (4) at (-2, -2) {};
    \node [style=green vertex] (5) at (0, -2) {};
    \node [style=none] (6) at (-3, -2) {};
    \node [style=hadamard vertex] (7) at (0, 1) {};
    \node [style=hadamard vertex] (8) at (-2, 1) {};
    \node [style=hadamard vertex] (9) at (-1.5, -0.75) {};
    \node [style=hadamard vertex] (10) at (-0.5, -0.75) {};
    \node [style=red vertex] (11) at (2, 1) {};
    \node [style=none] (12) at (-2, -3) {};
    \node [style=none] (13) at (0, -3) {};
    \node [style=none] (14) at (1, -2) {};
    \node [style=none] (15) at (-1, -2) {. . .};
  \end{pgfonlayer}
  \begin{pgfonlayer}{edgelayer}
    \draw (1) to (4);
    \draw (1) to (5);
    \draw (0) to (11);
    \draw (13) to (5);
    \draw (5) to (14);
    \draw (6) to (4);
    \draw (4) to (12);
  \end{pgfonlayer}
\end{tikzpicture}
}
}

\newcommand{\ilemmaii}{
\inline{
\begin{tikzpicture}
  \begin{pgfonlayer}{nodelayer}
    \node [style=red vertex] (0) at (-2, 1) {};
    \node [style=red vertex] (1) at (0, 1) {};
    \node [style=newstyle] (2) at (0.5, 1.5) {$\pi, v$};
    \node [style=green vertex] (3) at (-1.75, -0.5) {};
    \node [style=green vertex] (4) at (-0.25, -0.5) {};
    \node [style=none] (5) at (-2.75, -0.5) {};
    \node [style=none] (6) at (-1.75, -1.5) {};
    \node [style=none] (7) at (-0.25, -1.5) {};
    \node [style=none] (8) at (0.75, -0.5) {};
    \node [style=green vertex] (9) at (-1, 1) {};
    \node [style=green vertex] (10) at (-1, 1) {};
    \node [style=red vertex] (11) at (-1, 1) {};
    \node [style=none] (12) at (-1, -0.5) {. . .};
  \end{pgfonlayer}
  \begin{pgfonlayer}{edgelayer}
    \draw (7) to (4);
    \draw (4) to (8);
    \draw (5) to (3);
    \draw (3) to (6);
    \draw (0) to (1);
    \draw (10) to (3);
    \draw (10) to (4);
  \end{pgfonlayer}
\end{tikzpicture}
}
}

\newcommand{\ilemmaiii}{
\inline{
\begin{tikzpicture}
  \begin{pgfonlayer}{nodelayer}
    \node [style=newstyle] (0) at (-0.5, 1.5) {$\pi, v$};
    \node [style=green vertex] (1) at (-1.75, -0.5) {};
    \node [style=green vertex] (2) at (-0.25, -0.5) {};
    \node [style=none] (3) at (-2.75, -0.5) {};
    \node [style=none] (4) at (-1.75, -1.5) {};
    \node [style=none] (5) at (-0.25, -1.5) {};
    \node [style=none] (6) at (0.75, -0.5) {};
    \node [style=green vertex] (7) at (-1, 1) {};
    \node [style=green vertex] (8) at (-1, 1) {};
    \node [style=red vertex] (9) at (-1, 1) {};
    \node [style=none] (10) at (-1, -0.5) {. . .};
  \end{pgfonlayer}
  \begin{pgfonlayer}{edgelayer}
    \draw (5) to (2);
    \draw (2) to (6);
    \draw (3) to (1);
    \draw (1) to (4);
    \draw (8) to (1);
    \draw (8) to (2);
  \end{pgfonlayer}
\end{tikzpicture}
}
}

\newcommand{ \iilemmai}{
\inline{
\begin{tikzpicture}
  \begin{pgfonlayer}{nodelayer}
    \node [style=newstyle] (0) at (1.5, 1.5) {$\pi, v$};
    \node [style=none] (1) at (-3, -1) {};
    \node [style=none] (2) at (-2, -2) {};
    \node [style=none] (3) at (0, -2) {};
    \node [style=none] (4) at (1, -1) {};
    \node [style=green vertex] (5) at (-1, 1) {};
    \node [style=red vertex] (6) at (1, 1) {};
    \node [style=red vertex] (7) at (-3, 1) {};
    \node [style=red vertex] (8) at (2, 1) {};
    \node [style=hadamard vertex] (9) at (-2, 1) {};
    \node [style=hadamard vertex] (10) at (0, 1) {};
    \node [style=red vertex] (11) at (-2, -1) {};
    \node [style=red vertex] (12) at (0, -1) {};
    \node [style=none] (13) at (-1, -1) {. . .};
  \end{pgfonlayer}
  \begin{pgfonlayer}{edgelayer}
    \draw (7) to (8);
    \draw (1) to (11);
    \draw (11) to (2);
    \draw (12) to (3);
    \draw (12) to (4);
    \draw (12) to (5);
    \draw (5) to (11);
  \end{pgfonlayer}
\end{tikzpicture}
}
}

\newcommand{\iilemmaii}{
\inline{
\begin{tikzpicture}
  \begin{pgfonlayer}{nodelayer}
    \node [style=none] (0) at (-3, -1) {};
    \node [style=none] (1) at (-2, -2) {};
    \node [style=none] (2) at (0, -2) {};
    \node [style=none] (3) at (1, -1) {};
    \node [style=green vertex] (4) at (-1, 1) {};
    \node [style=red vertex] (5) at (-2, 1) {};
    \node [style=red vertex] (6) at (-2, -1) {};
    \node [style=red vertex] (7) at (0, -1) {};
    \node [style=none] (8) at (-1, -1) {. . .};
    \node [style=green vertex] (9) at (-2, 1) {};
    \node [style=green vertex] (10) at (0, 1) {};
    \node [style=newgreen] (11) at (0.5, 1.5) {$\pi, v$};
  \end{pgfonlayer}
  \begin{pgfonlayer}{edgelayer}
    \draw (0) to (6);
    \draw (6) to (1);
    \draw (7) to (2);
    \draw (7) to (3);
    \draw (7) to (4);
    \draw (4) to (6);
    \draw (5) to (4);
    \draw (4) to (10);
  \end{pgfonlayer}
\end{tikzpicture}
}
}

\newcommand{\iilemmaiii}{
\inline{
\begin{tikzpicture}
  \begin{pgfonlayer}{nodelayer}
    \node [style=none] (0) at (-3, -1) {};
    \node [style=none] (1) at (-2, -2) {};
    \node [style=none] (2) at (0, -2) {};
    \node [style=none] (3) at (1, -1) {};
    \node [style=green vertex] (4) at (-1, 1) {};
    \node [style=red vertex] (5) at (-2, -1) {};
    \node [style=red vertex] (6) at (0, -1) {};
    \node [style=none] (7) at (-1, -1) {. . .};
    \node [style=newgreen] (8) at (-0.5, 1.5) {$\pi, v$};
  \end{pgfonlayer}
  \begin{pgfonlayer}{edgelayer}
    \draw (0) to (5);
    \draw (5) to (1);
    \draw (6) to (2);
    \draw (6) to (3);
    \draw (6) to (4);
    \draw (4) to (5);
  \end{pgfonlayer}
\end{tikzpicture}
}
}

\newcommand{\viienci}{
\inline{
\begin{tikzpicture}
  \begin{pgfonlayer}{nodelayer}
    \node [style=green vertex] (0) at (-2, -0) {};
    \node [style=green vertex] (1) at (-0.75, -1) {};
    \node [style=green vertex] (2) at (0.25, -2) {};
    \node [style=green vertex] (3) at (1.25, -3) {};
    \node [style=red vertex] (4) at (-0.5, -0) {};
    \node [style=red vertex] (5) at (0.5, -0) {};
    \node [style=red vertex] (6) at (-1.75, 1) {};
    \node [style=red vertex] (7) at (-0.75, 1) {};
    \node [style=red vertex] (8) at (-2.5, 2) {};
    \node [style=red vertex] (9) at (0.25, 2) {};
    \node [style=red vertex] (10) at (1.25, 2) {};
    \node [style=red vertex] (11) at (0.75, 3) {};
    \node [style=red vertex] (12) at (-0.25, 3) {};
    \node [style=red vertex] (13) at (-1.5, 3) {};
    \node [style=red vertex] (14) at (1.75, 1) {};
    \node [style=hadamard vertex] (15) at (-3, -1) {};
    \node [style=hadamard vertex] (16) at (-3, -2) {};
    \node [style=hadamard vertex] (17) at (-3, -3) {};
    \node [style=red vertex] (18) at (-4, 3) {};
    \node [style=red vertex] (19) at (-4, 2) {};
    \node [style=red vertex] (20) at (-4, 1) {};
    \node [style=red vertex] (21) at (-4, -1) {};
    \node [style=red vertex] (22) at (-4, -2) {};
    \node [style=red vertex] (23) at (-4, -3) {};
    \node [style=none] (24) at (-5, -0) {};
    \node [style=none] (25) at (2.5, 3) {6};
    \node [style=none] (26) at (2.5, 2) {5};
    \node [style=none] (27) at (2.5, 1) {4};
    \node [style=none] (28) at (2.5, -0) {3};
    \node [style=none] (29) at (2.5, -1) {2};
    \node [style=none] (30) at (2.5, -2) {1};
    \node [style=none] (31) at (2.5, -3) {0};
  \end{pgfonlayer}
  \begin{pgfonlayer}{edgelayer}
    \draw (21) to (1);
    \draw (22) to (2);
    \draw (23) to (3);
    \draw (18) to (25);
    \draw (19) to (26);
    \draw (20) to (27);
    \draw (24) to (28);
    \draw (1) to (29);
    \draw (2) to (30);
    \draw (3) to (31);
    \draw (8) to (0);
    \draw (12) to (2);
    \draw (11) to (3);
    \draw (4) to (1);
    \draw (0) to (6);
    \draw (1) to (13);
    \draw (7) to (1);
    \draw (9) to (2);
    \draw (5) to (2);
    \draw (10) to (3);
    \draw (14) to (3);
  \end{pgfonlayer}
\end{tikzpicture}
}
}

\newcommand{\corleftiv}{
\inline{
\begin{tikzpicture}
  \node [style=red] (Vds) at (19.45,-12.55) {};
  \node [style=red] (j) at (30.075,-11.225) {};
  \node [style=wh] (3) at (39.775,-13.65) {};
  \node [style=wh] (5) at (38.775,-10.775) {};
  \node [style=green] (Vdw) at (17.0,-10.425) {};
  \node [style=red] (t) at (5.6,-15.0) {};
  \node [style=red] (Vao) at (36.075,-15.325) {};
  \node [style=wh] (6) at (38.85,-9.175) {};
  \node [style=green] (h) at (24.4,-12.875) {};
  \node [style=wh] (4) at (40.05,-12.25) {};
  \node [style=wh] (Vj) at (2.775,-12.75) {};
  \node [style=red] (k) at (28.675,-9.35) {};
  \node [style=wh] (0) at (39.475,-16.85) {};
  \node [style=red] (m) at (34.125,-14.4) {};
  \node [style=red] (q) at (11.325,-13.325) {};
  \node [style=green] (e) at (33.9,-1.1) {};
  \node [style=green] (Vm) at (30.35,-1.775) {};
  \node [style=green] (g) at (21.575,-9.2) {};
  \node [style=red] (Vba) at (30.1,-16.925) {};
  \node [style=red] (Vdn) at (10.8,-13.7) {};
  \node [style=red] (Vdg) at (19.35,-14.775) {};
  \node [style=green] (Vdv) at (18.55,-11.275) {};
  \node [style=green] (Vab) at (28.0,-2.65) {};
  \node [style=green] (Vdj) at (16.55,-10.85) {};
  \node [style=red] (Vdt) at (16.225,-9.65) {};
  \node [style=red] (l) at (31.875,-13.2) {};
  \node [style=red] (Vde) at (17.6,-11.9) {};
  \node [style=green] (Vdi) at (17.05,-12.725) {};
  \node [style=wh] (2) at (38.75,-14.5) {};
  \node [style=red] (Vau) at (32.875,-15.95) {};
  \node [style=green] (Vdp) at (9.8,-12.7) {};
  \node [style=green] (i) at (25.875,-14.275) {};
  \node [style=red] (Va) at (21.05,-4.35) {};
  \node [style=wh] (1) at (39.625,-15.575) {};
  \node [style=green] (Vdq) at (10.3,-13.2) {};
  \draw [] (Vde) to (i);
  \draw [] (k) to (6);
  \draw [] (q) to (Vdw);
  \draw [] (Vdg) to (Vdj);
  \draw [] (Vdt) to (Vdv);
  \draw [] (Vds) to (i);
  \draw [] (j) to (Vm);
  \draw [] (Vba) to (Vab);
  \draw [] (Vab) to (k);
  \draw [] (g) to (k);
  \draw [] (Vdv) to (j);
  \draw [] (q) to (g);
  \draw [] (e) to (l);
  \draw [] (Vdn) to (Vdq);
  \draw [] (e) to (k);
  \draw [] (Vde) to (g);
  \draw [] (Vm) to (m);
  \draw [] (Vdn) to (Vdp);
  \draw [] (Vba) to (0);
  \draw [] (i) to (m);
  \draw [] (Vau) to (1);
  \draw [] (Vds) to (g);
  \draw [] (Va) to (Vdv);
  \draw [] (Va) to (Vdj);
  \draw [] (Vba) to (Vdp);
  \draw [] (q) to (Vdp);
  \draw [] (e) to (Vao);
  \draw [] (j) to (Vab);
  \draw [] (Vau) to (Vm);
  \draw [] (Vdt) to (Vdw);
  \draw [] (Va) to (g);
  \draw [] (h) to (Vdn);
  \draw [] (h) to (Vdt);
  \draw [] (5) to (j);
  \draw [] (Vdg) to (Vdi);
  \draw [] (Vdw) to (Vau);
  \draw [] (3) to (m);
  \draw [] (Vdq) to (t);
  \draw [] (Vm) to (k);
  \draw [] (4) to (l);
  \draw [] (Vdq) to (Vj);
  \draw [] (Vdp) to (Va);
  \draw [] (e) to (m);
  \draw [] (Vao) to (Vdi);
  \draw [] (Vdw) to (t);
  \draw [] (t) to (i);
  \draw [] (Vao) to (2);
  \draw [] (Vdj) to (l);
  \draw [] (Vdi) to (t);
  \draw [] (Vds) to (h);
  \draw [] (h) to (Vdg);
  \draw [] (h) to (Vde);
  \draw [] (Vab) to (l);
  \draw [] (q) to (Vdi);
  \node [style=wh] at (3) {$$};
  \node [style=wh] at (5) {$$};
  \node [style=wh] at (6) {$$};
  \node [style=wh] at (4) {$$}; 
  \node [style=wh] at (Vj) {$$};
  \node [style=wh] at (0) {$$};
  \node [style=wh] at (2) {$$};
  \node [style=wh] at (1) {$$};
\end{tikzpicture}
}}

\newcommand{\corleftv}{
\inline{
\begin{tikzpicture}
\node [style=green] (Vcz) at (23.725,-15.075) {};
\node [style=red] (Vda) at (23.225,-14.575) {};
\node [style=green] (Vdq) at (10.3,-13.2) {};
\node [style=wh] (2) at (38.475,-15.0) {};
\node [style=red] (Vdb) at (22.225,-13.575) {};
\node [style=green] (Vdd) at (22.725,-14.075) {};
\node [style=green] (Vdf) at (14.95,-10.425) {};
\node [style=red] (Vdm) at (11.8,-14.7) {};
\node [style=wh] (6) at (37.875,-9.15) {};
\node [style=green] (Vdr) at (17.225,-10.95) {};
\node [style=red] (Vdl) at (12.3,-15.2) {};
\node [style=wh] (0) at (38.475,-18.7) {};
\node [style=wh] (Vj) at (4.3,-13.35) {};
\node [style=red] (Vcv) at (16.55,-9.0) {};
\node [style=red] (Vdn) at (10.8,-13.7) {};
\node [style=wh] (5) at (38.225,-10.85) {};
\node [style=red] (Vdt) at (16.225,-9.65) {};
\node [style=red] (j) at (30.075,-11.225) {};
\node [style=red] (Vao) at (36.075,-15.325) {};
\node [style=red] (Vau) at (32.875,-15.95) {};
\node [style=wh] (4) at (38.425,-12.3) {};
\node [style=green] (Vds) at (19.45,-12.55) {};
\node [style=green] (Vdh) at (16.55,-11.55) {};
\node [style=green] (Vdj) at (16.55,-10.85) {};
\node [style=red] (Vcx) at (18.625,-11.0) {};
\node [style=red] (l) at (31.875,-13.2) {};
\node [style=green] (Vab) at (28.0,-2.65) {};
\node [style=red] (Vdc) at (24.225,-15.575) {};
\node [style=green] (c) at (23.725,-15.075) {};
\node [style=green] (e) at (33.9,-1.1) {};
\node [style=wh] (3) at (38.275,-13.65) {};
\node [style=green] (Vdv) at (18.55,-11.275) {};
\node [style=red] (Vdo) at (11.3,-14.2) {};
\node [style=red] (Vba) at (30.1,-16.925) {};
\node [style=green] (Vdp) at (9.8,-12.7) {};
\node [style=red] (Vdg) at (19.35,-14.775) {};
\node [style=green] (Vdk) at (12.375,-13.325) {};
\node [style=wh] (1) at (38.425,-16.75) {};
\draw [] (Vcx) to (Vcz);
\draw [] (Vdn) to (Vdq);
\draw [] (Vda) to (Vdf);
\draw [] (Vdb) to (Vdd);
\draw [] (Vdc) to (Vdk);
\draw [] (Vcv) to (Vdv);
\draw [] (Vdt) to (Vdv);
\draw [] (Vdg) to (Vdk);
\draw [] (Vdd) to (Vdg);
\draw [] (Vcv) to (Vdj);
\draw [] (Vdf) to (Vdl);
\draw [] (Vdp) to (Vdl);
\draw [] (Vdt) to (Vdh);
\draw [] (Vcx) to (c);
\draw [] (Vba) to (Vcz);
\draw [] (3) to (Vdl);
\draw [] (j) to (Vab);
\draw [] (Vab) to (Vdm);
\draw [] (Vao) to (Vdk);
\draw [] (e) to (Vcx);
\draw [] (Vdq) to (Vj);
\draw [] (Vdd) to (Vdt);
\draw [] (Vdr) to (l);
\draw [] (5) to (j);
\draw [] (Vba) to (Vab);
\draw [] (c) to (Vdb);
\draw [] (j) to (Vds);
\draw [] (Vdv) to (j);
\draw [] (e) to (Vcv);
\draw [] (Vdb) to (Vdf);
\draw [] (Vda) to (e);
\draw [] (Vau) to (1);
\draw [] (Vdn) to (Vcz);
\draw [] (Vdm) to (Vdp);
\draw [] (c) to (Vdm);
\draw [] (Vdd) to (Vdn);
\draw [] (Vba) to (0);
\draw [] (Vdq) to (Vda);
\draw [] (Vdh) to (Vau);
\draw [] (Vau) to (Vds);
\draw [] (Vdp) to (Vdo);
\draw [] (6) to (Vdm);
\draw [] (e) to (Vdc);
\draw [] (Vdj) to (l);
\draw [] (Vdo) to (Vdr);
\draw [] (Vdo) to (Vds);
\draw [] (Vdc) to (Vdh);
\draw [] (Vao) to (2);
\draw [] (Vdg) to (Vdj);
\draw [] (Vab) to (l);
\draw [] (Vdr) to (Vao);
\draw [] (4) to (l);
\node [style=wh] at (2) {$$};
\node [style=wh] at (6) {$$};
\node [style=wh] at (0) {$$};
\node [style=wh] at (Vj) {$$};
\node [style=wh] at (5) {$$};
\node [style=wh] at (4) {$$};
\node [style=wh] at (3) {$$};
\node [style=wh] at (1) {$$};
\end{tikzpicture}
}
}

\newcommand{\corleftvi}{
\inline{
\begin{tikzpicture}
\node [style=red] (Vcx) at (18.825,-4.875) {};
\node [style=wh] (4) at (35.375,-11.475) {};
\node [style=green] (Vcp) at (26.95,-12.525) {};
\node [style=green] (Vdq) at (10.3,-13.2) {};
\node [style=red] (i) at (25.875,-14.275) {};
\node [style=wh] (2) at (35.5,-14.175) {};
\node [style=green] (Vcz) at (22.825,-19.85) {};
\node [style=green] (Vcv) at (16.55,-9.0) {};
\node [style=red] (Vda) at (19.325,-18.425) {};
\node [style=green] (Vcu) at (11.875,-12.825) {};
\node [style=wh] (1) at (35.35,-15.45) {};
\node [style=wh] (0) at (35.225,-16.975) {};
\node [style=red] (Vba) at (30.1,-16.925) {};
\node [style=red] (Vdm) at (10.025,-19.3) {};
\node [style=wh] (3) at (35.45,-12.825) {};
\node [style=wh] (5) at (35.875,-10.45) {};
\node [style=green] (Vab) at (28.0,-2.65) {};
\node [style=green] (Vde) at (17.6,-11.9) {};
\node [style=green] (Vdp) at (9.8,-12.7) {};
\node [style=red] (Vdo) at (11.325,-7.55) {};
\node [style=green] (Vco) at (28.1,-13.7) {};
\node [style=red] (j) at (30.075,-11.225) {};
\node [style=green] (d) at (33.575,-2.175) {};
\node [style=green] (Vds) at (19.45,-12.55) {};
\node [style=green] (Vdf) at (14.95,-10.425) {};
\node [style=red] (Vcl) at (29.0,-15.325) {};
\node [style=red] (Vdb) at (22.225,-13.575) {};
\node [style=red] (Vcs) at (25.0,-7.55) {};
\node [style=red] (Vcm) at (28.575,-14.2) {};
\node [style=red] (Veb) at (20.5,-13.95) {};
\node [style=wh] (6) at (35.35,-9.275) {};
\node [style=wh] (Vj) at (3.425,-12.925) {};
\node [style=red] (Vdl) at (12.3,-15.2) {};
\node [style=green] (c) at (27.0,-18.95) {};
\node [style=green] (Vcn) at (27.625,-13.225) {};
\node [style=red] (Vdn) at (10.525,-17.05) {};
\draw [] (j) to (Vab);
\draw [] (Vde) to (j);
\draw [] (Vcs) to (Vcu);
\draw [] (Vba) to (0);
\draw [] (Vcx) to (Vcz);
\draw [] (Veb) to (Vcp);
\draw [] (Vdo) to (Vco);
\draw [] (Vcm) to (Vco);
\draw [] (Vda) to (Vdf);
\draw [] (Vba) to (Vcz);
\draw [] (d) to (Veb);
\draw [] (Veb) to (Vde);
\draw [] (Vab) to (Vdm);
\draw [] (Vdb) to (Vcu);
\draw [] (6) to (Vdm);
\draw [] (Vcn) to (2);
\draw [] (Vcs) to (Vcv);
\draw [] (Vcv) to (Vcx);
\draw [] (Vdm) to (Vdp);
\draw [] (3) to (Vdl);
\draw [] (4) to (Vcl);
\draw [] (i) to (Vds);
\draw [] (5) to (j);
\draw [] (Vcu) to (Vdn);
\draw [] (Vdq) to (Vda);
\draw [] (i) to (Vde);
\draw [] (Vdb) to (Vdf);
\draw [] (j) to (Vds);
\draw [] (Vdp) to (Vdo);
\draw [] (Vdp) to (Vdl);
\draw [] (c) to (Vdm);
\draw [] (c) to (Vdb);
\draw [] (Vdf) to (Vdl);
\draw [] (Vba) to (Vab);
\draw [] (Vdo) to (Vds);
\draw [] (d) to (Vcs);
\draw [] (Vdn) to (Vdq);
\draw [] (Vab) to (Vcl);
\draw [] (Vda) to (Vcv);
\draw [] (Vdq) to (Vj);
\draw [] (1) to (i);
\draw [] (Vcu) to (i);
\draw [] (Vcl) to (Vcn);
\draw [] (Vcx) to (c);
\draw [] (Vcl) to (Vcu);
\draw [] (Vdn) to (Vcz);
\draw [] (Vcm) to (Vcp);
\draw [] (Vcn) to (Vcm);
\node [style=wh] at (4) {$$};
\node [style=wh] at (2) {$$};
\node [style=wh] at (1) {$$};
\node [style=wh] at (0) {$$};
\node [style=wh] at (3) {$$};
\node [style=wh] at (5) {$$};
\node [style=wh] at (6) {$$};
\node [style=wh] at (Vj) {$$};
\end{tikzpicture}
}
}

\newcommand{\corleftvii}{
\inline{
\begin{tikzpicture}
\node [style=red] (Vdb) at (22.225,-13.575) {};
\node [style=green] (Vcp) at (26.95,-12.525) {};
\node [style=red] (Vcs) at (25.0,-7.55) {};
\node [style=green] (Vdf) at (14.95,-10.425) {};
\node [style=green] (Vab) at (28.0,-2.65) {};
\node [style=wh] (2) at (37.525,-15.725) {};
\node [style=red] (Vck) at (29.25,-14.775) {};
\node [style=wh] (1) at (37.65,-18.175) {};
\node [style=red] (Vdn) at (10.525,-17.05) {};
\node [style=red] (Vdl) at (12.3,-15.2) {};
\node [style=green] (c) at (27.0,-18.95) {};
\node [style=red] (Vcx) at (18.825,-4.875) {};
\node [style=green] (Vdp) at (9.8,-12.7) {};
\node [style=red] (Vdm) at (10.025,-19.3) {};
\node [style=red] (Vcq) at (26.825,-8.725) {};
\node [style=green] (Vcv) at (16.55,-9.0) {};
\node [style=wh] (Vj) at (5.9,-13.275) {};
\node [style=red] (Vdo) at (11.325,-7.55) {};
\node [style=green] (Vco) at (28.1,-13.7) {};
\node [style=red] (Veb) at (20.5,-13.95) {};
\node [style=red] (Vcj) at (24.4,-12.85) {};
\node [style=green] (Vcr) at (26.525,-7.825) {};
\node [style=wh] (6) at (36.825,-9.325) {};
\node [style=wh] (0) at (37.6,-19.525) {};
\node [style=wh] (4) at (37.3,-12.55) {};
\node [style=green] (Vcu) at (11.875,-12.825) {};
\node [style=red] (Vcl) at (29.0,-15.325) {};
\node [style=green] (Vct) at (27.55,-8.925) {};
\node [style=green] (Vcw) at (17.15,-9.675) {};
\node [style=red] (Vcm) at (28.575,-14.2) {};
\node [style=red] (Vba) at (30.1,-16.925) {};
\node [style=red] (Vda) at (19.325,-18.425) {};
\node [style=green] (Vcn) at (27.625,-13.225) {};
\node [style=wh] (3) at (37.45,-14.125) {};
\node [style=green] (Vdq) at (10.3,-13.2) {};
\node [style=green] (Vcz) at (22.825,-19.85) {};
\node [style=wh] (5) at (36.825,-10.95) {};
\node [style=green] (d) at (33.575,-2.175) {};
\draw [] (Vba) to (Vab);
\draw [] (Vdf) to (Vdl);
\draw [] (Vcl) to (Vcu);
\draw [] (Vcs) to (Vcv);
\draw [] (Vcu) to (Vdn);
\draw [] (Vba) to (Vcz);
\draw [] (Vdn) to (Vdq);
\draw [] (d) to (Vcs);
\draw [] (Vcm) to (Vco);
\draw [] (4) to (Vcl);
\draw [] (Vcu) to (Vck);
\draw [] (Vcr) to (Vcq);
\draw [] (d) to (Veb);
\draw [] (Vdq) to (Vda);
\draw [] (6) to (Vdm);
\draw [] (Vdp) to (Vdo);
\draw [] (Vcr) to (Vcj);
\draw [] (Vcn) to (Vcm);
\draw [] (Veb) to (Vcp);
\draw [] (Vcx) to (c);
\draw [] (Vcq) to (Vct);
\draw [] (Vcn) to (2);
\draw [] (3) to (Vdl);
\draw [] (Vdm) to (Vdp);
\draw [] (Vab) to (Vdm);
\draw [] (Vdq) to (Vj);
\draw [] (c) to (Vdm);
\draw [] (Vdn) to (Vcz);
\draw [] (5) to (Vcj);
\draw [] (1) to (Vck);
\draw [] (Vcm) to (Vcp);
\draw [] (Vdp) to (Vdl);
\draw [] (Vda) to (Vcv);
\draw [] (Vcl) to (Vcn);
\draw [] (Vdo) to (Vco);
\draw [] (Vab) to (Vcl);
\draw [] (Vck) to (Vcr);
\draw [] (Veb) to (Vct);
\draw [] (Vdb) to (Vcu);
\draw [] (Vdb) to (Vdf);
\draw [] (Vcx) to (Vcz);
\draw [] (Vcj) to (Vab);
\draw [] (Vcq) to (Vcw);
\draw [] (Vdo) to (Vcw);
\draw [] (Vcs) to (Vcu);
\draw [] (c) to (Vdb);
\draw [] (Vda) to (Vdf);
\draw [] (Vcv) to (Vcx);
\draw [] (Vba) to (0);
\node [style=wh] at (2) {$$};
\node [style=wh] at (1) {$$};
\node [style=wh] at (Vj) {$$};
\node [style=wh] at (6) {$$};
\node [style=wh] at (0) {$$};
\node [style=wh] at (4) {$$};
\node [style=wh] at (3) {$$};
\node [style=wh] at (5) {$$};
\end{tikzpicture}
}
}

\newcommand{\totii}{
\inline{
\begin{tikzpicture}
\node [style=green] (Vdq) at (4.225,-13.7) {};
\node [style=red] (k) at (31.225,-19.275) {};
\node [style=green] (Vcn) at (21.55,-13.725) {};
\node [style=red] (Vda) at (13.25,-18.925) {};
\node [style=red] (Vdo) at (5.25,-8.05) {};
\node [style=red] (Vba) at (24.025,-17.425) {};
\node [style=red] (Vdm) at (3.95,-19.8) {};
\node [style=red] (Vcs) at (18.925,-8.05) {};
\node [style=green] (f) at (33.55,-19.45) {};
\node [style=red] (h) at (35.225,-16.575) {};
\node [style=red] (Vcx) at (12.75,-5.375) {};
\node [style=red] (Vdn) at (4.45,-17.55) {};
\node [style=red] (Vdb) at (16.15,-14.075) {};
\node [style=red] (Vcl) at (22.925,-15.825) {};
\node [style=red] (Veb) at (14.425,-14.45) {};
\node [style=green] (Vo) at (30.325,-16.575) {};
\node [style=green] (Vcv) at (10.475,-9.5) {};
\node [style=green] (Vcu) at (5.8,-13.325) {};
\node [style=red] (Vdl) at (6.225,-15.7) {};
\node [style=wh] (Vk) at (38.85,-16.45) {};
\node [style=green] (g) at (33.725,-17.05) {};
\node [style=red] (Vcj) at (18.35,-13.375) {};
\node [style=green] (Vcw) at (11.075,-10.175) {};
\node [style=green] (c) at (20.925,-19.45) {};
\node [style=red] (i) at (31.55,-16.725) {};
\node [style=green] (Vcz) at (16.75,-20.35) {};
\node [style=green] (Vdp) at (3.725,-13.2) {};
\node [style=green] (Vco) at (22.025,-14.2) {};
\node [style=green] (e) at (33.225,-13.5) {};
\node [style=green] (Vab) at (21.925,-3.15) {};
\node [style=red] (Vcm) at (22.5,-14.7) {};
\node [style=red] (j) at (31.625,-13.725) {};
\node [style=green] (Vdf) at (8.875,-10.925) {};
\node [style=wh] (Vj) at (0.5,-13.75) {};
\node [style=red] (Vck) at (23.175,-15.275) {};
\node [style=green] (Vcp) at (20.875,-13.025) {};
\node [style=green] (Vct) at (21.475,-9.425) {};
\node [style=green] (Vcr) at (20.45,-8.325) {};
\node [style=green] (d) at (27.5,-2.675) {};
\node [style=red] (Vcq) at (20.75,-9.225) {};
\draw [] (Vck) to (Vcr);
\draw [] (Vba) to (Vab);
\draw [] (i) to (Vo);
\draw [] (Vdp) to (Vdo);
\draw [] (Vcr) to (Vcj);
\draw [] (e) to (j);
\draw [] (Veb) to (Vcp);
\draw [] (Vdn) to (Vcz);
\draw [] (Vdf) to (Vdl);
\draw [] (h) to (f);
\draw [] (j) to (g);
\draw [] (Vk) to (Vo);
\draw [] (Vba) to (Vcz);
\draw [] (Vcu) to (Vck);
\draw [] (Vdm) to (h);
\draw [] (Vcm) to (Vcp);
\draw [] (Vcn) to (g);
\draw [] (Vcv) to (Vcx);
\draw [] (Vcl) to (Vcn);
\draw [] (g) to (k);
\draw [] (Vda) to (Vcv);
\draw [] (Vda) to (Vdf);
\draw [] (h) to (g);
\draw [] (h) to (e);
\draw [] (Vba) to (e);
\draw [] (Vcu) to (Vdn);
\draw [] (k) to (Vdl);
\draw [] (Vdq) to (Vj);
\draw [] (Vcx) to (c);
\draw [] (Vdb) to (Vdf);
\draw [] (Veb) to (Vct);
\draw [] (Vcs) to (Vcu);
\draw [] (Vcr) to (Vcq);
\draw [] (f) to (k);
\draw [] (Vcl) to (j);
\draw [] (Vcq) to (Vct);
\draw [] (Vdo) to (Vcw);
\draw [] (Vcn) to (Vcm);
\draw [] (Vcq) to (Vcw);
\draw [] (Vdm) to (Vdp);
\draw [] (Vab) to (Vdm);
\draw [] (i) to (f);
\draw [] (Vcx) to (Vcz);
\draw [] (Vdo) to (Vco);
\draw [] (Vcm) to (Vco);
\draw [] (Vdq) to (Vda);
\draw [] (c) to (Vdm);
\draw [] (Vck) to (f);
\draw [] (Vdn) to (Vdq);
\draw [] (Vdp) to (Vdl);
\draw [] (Vo) to (k);
\draw [] (c) to (Vdb);
\draw [] (i) to (e);
\draw [] (Vcj) to (Vab);
\draw [] (Vcj) to (i);
\draw [] (j) to (Vo);
\draw [] (d) to (Veb);
\draw [] (Vcl) to (Vcu);
\draw [] (d) to (Vcs);
\draw [] (Vcs) to (Vcv);
\draw [] (Vdb) to (Vcu);
\draw [] (Vab) to (Vcl);
\node [style=wh] at (Vk) {$$};
\node [style=wh] at (Vj) {$$};
\end{tikzpicture}
}
}

\newcommand{\totiii}{
\inline{
\begin{tikzpicture}
\node [style=green] (e) at (33.225,-13.5) {};
\node [style=green] (o) at (29.525,-18.075) {};
\node [style=green] (d) at (27.5,-2.675) {};
\node [style=red] (l) at (19.225,-17.975) {};
\node [style=green] (Vcz) at (16.75,-20.35) {};
\node [style=green] (c) at (20.925,-19.45) {};
\node [style=red] (Vcx) at (12.75,-5.375) {};
\node [style=green] (Vdf) at (8.875,-10.925) {};
\node [style=red] (Veh) at (28.65,-14.825) {};
\node [style=green] (h) at (35.225,-16.575) {};
\node [style=wh] (Vj) at (1.4,-14.2) {};
\node [style=green] (Vab) at (21.925,-3.15) {};
\node [style=green] (Vo) at (30.325,-16.575) {};
\node [style=red] (Vda) at (13.25,-18.925) {};
\node [style=red] (Vdb) at (16.15,-14.075) {};
\node [style=green] (Vco) at (22.025,-14.2) {};
\node [style=red] (Vcq) at (20.75,-9.225) {};
\node [style=red] (Vdn) at (4.45,-17.55) {};
\node [style=green] (Vct) at (21.475,-9.425) {};
\node [style=green] (Vcv) at (10.475,-9.5) {};
\node [style=red] (Vcs) at (18.925,-8.05) {};
\node [style=green] (Vek) at (28.75,-15.65) {};
\node [style=wh] (Vk) at (38.575,-15.3) {};
\node [style=green] (Vdq) at (4.225,-13.7) {};
\node [style=red] (Vdo) at (5.25,-8.05) {};
\node [style=green] (Vcp) at (20.875,-13.025) {};
\node [style=red] (i) at (31.55,-16.725) {};
\node [style=red] (m) at (20.075,-18.675) {};
\node [style=green] (Vcw) at (11.075,-10.175) {};
\node [style=red] (Veb) at (14.425,-14.45) {};
\node [style=green] (Vel) at (29.7,-17.175) {};
\node [style=green] (Vdp) at (3.725,-13.2) {};
\node [style=red] (Vba) at (24.025,-17.425) {};
\node [style=red] (Vcm) at (22.5,-14.7) {};
\node [style=red] (Vei) at (28.0,-14.875) {};
\draw [] (Vek) to (Vcq);
\draw [] (h) to (Vdn);
\draw [] (Veh) to (Vo);
\draw [] (Vda) to (Vdf);
\draw [] (Vcs) to (h);
\draw [] (Vdb) to (Vdf);
\draw [] (c) to (Vdb);
\draw [] (Vdo) to (Vcw);
\draw [] (Vo) to (l);
\draw [] (o) to (l);
\draw [] (Vdf) to (l);
\draw [] (Veb) to (Vct);
\draw [] (Veh) to (Vab);
\draw [] (Vcq) to (Vct);
\draw [] (Vel) to (l);
\draw [] (m) to (Vdp);
\draw [] (d) to (Veb);
\draw [] (m) to (e);
\draw [] (Vab) to (i);
\draw [] (Vdn) to (Vcz);
\draw [] (Vdn) to (Vdq);
\draw [] (Vei) to (Vek);
\draw [] (Vba) to (Vab);
\draw [] (o) to (Vcm);
\draw [] (h) to (Vei);
\draw [] (Vdq) to (Vda);
\draw [] (Vba) to (e);
\draw [] (Veh) to (e);
\draw [] (d) to (Vcs);
\draw [] (e) to (i);
\draw [] (c) to (m);
\draw [] (Vk) to (Vo);
\draw [] (Vda) to (Vcv);
\draw [] (Veh) to (h);
\draw [] (Vcx) to (Vcz);
\draw [] (Vab) to (m);
\draw [] (Vcm) to (Vco);
\draw [] (Vdq) to (Vj);
\draw [] (Vdp) to (l);
\draw [] (Vei) to (Vel);
\draw [] (Vcs) to (Vcv);
\draw [] (m) to (Vel);
\draw [] (Vcm) to (Vcp);
\draw [] (Vcq) to (Vcw);
\draw [] (i) to (h);
\draw [] (Vba) to (Vcz);
\draw [] (Vdo) to (Vco);
\draw [] (Vcx) to (c);
\draw [] (i) to (Vo);
\draw [] (Vcv) to (Vcx);
\draw [] (Vdb) to (h);
\draw [] (Veb) to (Vcp);
\draw [] (Vdp) to (Vdo);
\draw [] (m) to (o);
\node [style=wh] at (Vj) {$$};
\node [style=wh] at (Vk) {$$};
\end{tikzpicture}
}
}

\newcommand{\totiv}{
\inline{
\begin{tikzpicture}
\node [style=green] (Vdq) at (4.225,-13.7) {};
\node [style=green] (Vcv) at (10.475,-9.5) {};
\node [style=red] (Ved) at (34.425,-17.6) {};
\node [style=red] (Vcm) at (22.5,-14.7) {};
\node [style=red] (Vcx) at (12.75,-5.375) {};
\node [style=red] (Vcs) at (18.925,-8.05) {};
\node [style=green] (e) at (33.225,-13.5) {};
\node [style=green] (Vcp) at (22.2,-12.0) {};
\node [style=red] (Vdn) at (4.45,-17.55) {};
\node [style=green] (Vef) at (33.425,-16.6) {};
\node [style=green] (Vdy) at (25.175,-15.75) {};
\node [style=wh] (Vk) at (40.65,-19.225) {};
\node [style=green] (Vcw) at (11.075,-10.175) {};
\node [style=red] (Veo) at (29.475,-11.275) {};
\node [style=green] (Vel) at (29.7,-17.175) {};
\node [style=green] (f) at (33.55,-19.45) {};
\node [style=red] (Vdb) at (16.15,-14.075) {};
\node [style=green] (Vco) at (22.025,-14.2) {};
\node [style=green] (d) at (27.5,-2.675) {};
\node [style=red] (Vdo) at (5.25,-8.05) {};
\node [style=wh] (Vj) at (1.225,-13.6) {};
\node [style=red] (Vdw) at (27.8,-18.55) {};
\node [style=red] (l) at (19.225,-17.975) {};
\node [style=green] (Vdz) at (26.85,-17.45) {};
\node [style=green] (Vdp) at (3.725,-13.2) {};
\node [style=red] (Vcq) at (20.75,-9.225) {};
\node [style=red] (Vda) at (13.25,-18.925) {};
\node [style=red] (Veb) at (14.425,-14.45) {};
\node [style=green] (Vdf) at (8.875,-10.925) {};
\node [style=red] (Vdv) at (26.6,-17.95) {};
\node [style=red] (Vei) at (28.0,-14.875) {};
\node [style=green] (Veq) at (30.65,-12.5) {};
\node [style=green] (o) at (29.525,-18.075) {};
\node [style=green] (Vct) at (21.475,-9.425) {};
\node [style=green] (Vek) at (28.75,-15.65) {};
\node [style=green] (c) at (20.925,-19.45) {};
\draw [] (Vda) to (Vcv);
\draw [] (Vei) to (Vek);
\draw [] (f) to (Vk);
\draw [] (Vcs) to (Vef);
\draw [] (Vef) to (Vei);
\draw [] (Veb) to (Vcp);
\draw [] (Vdv) to (e);
\draw [] (Vcm) to (Vcp);
\draw [] (Vcq) to (Vcw);
\draw [] (Veo) to (Veq);
\draw [] (Vdo) to (Vcw);
\draw [] (Vcq) to (Vct);
\draw [] (Veo) to (Vdz);
\draw [] (Vdw) to (Vdz);
\draw [] (f) to (Veo);
\draw [] (o) to (l);
\draw [] (Vei) to (Vel);
\draw [] (Vdb) to (Vdf);
\draw [] (Vdb) to (Vef);
\draw [] (c) to (Vdb);
\draw [] (c) to (Vdv);
\draw [] (Veb) to (Vct);
\draw [] (Vdp) to (l);
\draw [] (Vdp) to (Vdo);
\draw [] (d) to (Vcs);
\draw [] (Vdv) to (o);
\draw [] (o) to (Vcm);
\draw [] (Vcv) to (Vcx);
\draw [] (Vdw) to (Vdy);
\draw [] (Vda) to (Vdf);
\draw [] (Vcm) to (Vco);
\draw [] (Vdv) to (Vel);
\draw [] (Vdf) to (l);
\draw [] (Vdq) to (Vj);
\draw [] (Vcx) to (e);
\draw [] (d) to (Veb);
\draw [] (Vdn) to (e);
\draw [] (Vdn) to (Vdq);
\draw [] (Ved) to (Vdy);
\draw [] (Veq) to (Ved);
\draw [] (Vdq) to (Vda);
\draw [] (Vcx) to (c);
\draw [] (Vdv) to (Vdp);
\draw [] (f) to (l);
\draw [] (Vel) to (l);
\draw [] (Vdo) to (Vco);
\draw [] (Ved) to (Vef);
\draw [] (Vcs) to (Vcv);
\draw [] (Vek) to (Vcq);
\draw [] (e) to (Vdw);
\draw [] (Vef) to (Vdn);
\node [style=wh] at (Vk) {$$};
\node [style=wh] at (Vj) {$$};
\end{tikzpicture}

}
}

\newcommand{\totv}{
\inline{
\begin{tikzpicture}
\node [style=red] (Vdk) at (21.475,-16.225) {};
\node [style=green] (Vef) at (33.425,-16.6) {};
\node [style=green] (Vct) at (21.475,-9.425) {};
\node [style=green] (Vdp) at (3.725,-13.2) {};
\node [style=green] (Vet) at (19.075,-16.1) {};
\node [style=red] (Vdn) at (4.45,-17.55) {};
\node [style=red] (Ved) at (34.425,-17.6) {};
\node [style=red] (Vds) at (23.475,-12.9) {};
\node [style=red] (Vcs) at (18.925,-8.05) {};
\node [style=green] (h) at (35.225,-16.575) {};
\node [style=green] (d) at (27.5,-2.675) {};
\node [style=green] (Ves) at (19.625,-15.625) {};
\node [style=red] (Veb) at (14.425,-14.45) {};
\node [style=red] (Vcq) at (20.75,-9.225) {};
\node [style=red] (Veo) at (29.475,-11.275) {};
\node [style=green] (Vej) at (29.45,-16.275) {};
\node [style=green] (Veh) at (28.65,-14.825) {};
\node [style=green] (Vco) at (22.025,-14.2) {};
\node [style=wh] (Vj) at (2.85,-15.425) {};
\node [style=green] (Vdz) at (26.85,-17.45) {};
\node [style=green] (Vek) at (28.75,-15.65) {};
\node [style=green] (Vdx) at (26.275,-16.95) {};
\node [style=red] (Vee) at (32.425,-15.6) {};
\node [style=green] (Vdy) at (25.175,-15.75) {};
\node [style=red] (Vdw) at (27.8,-18.55) {};
\node [style=green] (Vdm) at (3.95,-19.8) {};
\node [style=red] (Vdo) at (5.25,-8.05) {};
\node [style=red] (Vei) at (28.0,-14.875) {};
\node [style=red] (Vdb) at (16.15,-14.075) {};
\node [style=green] (Vdu) at (25.35,-16.575) {};
\node [style=red] (Vcm) at (22.5,-14.7) {};
\node [style=green] (Veq) at (30.65,-12.5) {};
\node [style=red] (Vep) at (28.35,-10.225) {};
\node [style=wh] (Vk) at (37.375,-16.75) {};
\node [style=green] (Vcp) at (22.2,-12.0) {};
\node [style=green] (Vcw) at (11.075,-10.175) {};
\draw [] (Vee) to (Vej);
\draw [] (Vdb) to (Vef);
\draw [] (Vk) to (h);
\draw [] (h) to (Vj);
\draw [] (Vds) to (Vdx);
\draw [] (Vdn) to (h);
\draw [] (Vef) to (Vdn);
\draw [] (Vep) to (Vet);
\draw [] (Vdw) to (Vdz);
\draw [] (Ved) to (Vef);
\draw [] (Vcm) to (Vcp);
\draw [] (Vdm) to (Vep);
\draw [] (Veo) to (Veq);
\draw [] (h) to (Veo);
\draw [] (Vcs) to (Vdm);
\draw [] (Vdx) to (Vdw);
\draw [] (Vdm) to (Vds);
\draw [] (Vdo) to (Vco);
\draw [] (h) to (Vdk);
\draw [] (Vdo) to (Vcw);
\draw [] (Vet) to (Vdo);
\draw [] (Veq) to (Ved);
\draw [] (d) to (Veb);
\draw [] (Vek) to (Vcq);
\draw [] (Vei) to (Vek);
\draw [] (Ved) to (Vdy);
\draw [] (Vej) to (Vcm);
\draw [] (Vcq) to (Vcw);
\draw [] (Vdn) to (Vdx);
\draw [] (Vdk) to (Vdm);
\draw [] (Ves) to (Vee);
\draw [] (Veo) to (Vdz);
\draw [] (Vef) to (Vei);
\draw [] (Vee) to (Veh);
\draw [] (Vcm) to (Vco);
\draw [] (Vcs) to (Vef);
\draw [] (Vds) to (Vdu);
\draw [] (Veb) to (Vct);
\draw [] (Vdk) to (Vdp);
\draw [] (Veb) to (Vcp);
\draw [] (Vdb) to (Vdp);
\draw [] (Vcq) to (Vct);
\draw [] (d) to (Vcs);
\draw [] (Vep) to (Ves);
\draw [] (Vdu) to (Vdb);
\draw [] (Vdw) to (Vdy);
\draw [] (Vei) to (Veh);
\node [style=wh] at (Vj) {$$};
\node [style=wh] at (Vk) {$$};
\end{tikzpicture}

}
}

\newcommand{\totvi}{
\inline{
\begin{tikzpicture}
\node [style=red] (Ved) at (34.425,-17.6) {};
\node [style=red] (Veb) at (14.425,-14.45) {};
\node [style=red] (Vep) at (28.35,-10.225) {};
\node [style=green] (Vdx) at (26.275,-16.95) {};
\node [style=red] (Vds) at (23.475,-12.9) {};
\node [style=red] (Vcm) at (22.5,-14.7) {};
\node [style=wh] (Vj) at (3.525,-14.75) {};
\node [style=red] (Veo) at (29.475,-11.275) {};
\node [style=red] (Vei) at (28.0,-14.875) {};
\node [style=red] (Vdb) at (16.15,-14.075) {};
\node [style=red] (Vcq) at (20.75,-9.225) {};
\node [style=red] (Vee) at (32.425,-15.6) {};
\node [style=red] (Vdn) at (4.45,-17.55) {};
\node [style=red] (Vdk) at (21.475,-16.225) {};
\node [style=green] (Vef) at (33.425,-16.6) {};
\node [style=green] (h) at (35.225,-16.575) {};
\node [style=red] (Vcs) at (18.925,-8.05) {};
\node [style=wh] (Vk) at (39.025,-16.65) {};
\node [style=red] (Vdo) at (5.25,-8.05) {};
\node [style=green] (Vdm) at (3.95,-19.8) {};
\node [style=red] (Vdw) at (27.8,-18.55) {};
\draw [] (Vdn) to (Vdx);
\draw [] (Vep) to (Vee);
\draw [] (Vcs) to (Vdm);
\draw [] (Vdm) to (Vep);
\draw [] (Veb) to (Vcq);
\draw [] (Ved) to (Vef);
\draw [] (Vds) to (Vdb);
\draw [] (h) to (Vdk);
\draw [] (Vep) to (Vdo);
\draw [] (Veo) to (Vdw);
\draw [] (Vei) to (Vcq);
\draw [] (Vee) to (Vcm);
\draw [] (Veo) to (Ved);
\draw [] (Vds) to (Vdx);
\draw [] (Veb) to (Vcm);
\draw [] (Vdb) to (Vdk);
\draw [] (h) to (Veo);
\draw [] (Vdx) to (Vdw);
\draw [] (Vdn) to (h);
\draw [] (Vef) to (Vei);
\draw [] (Vk) to (h);
\draw [] (Vdo) to (Vcq);
\draw [] (Vdm) to (Vds);
\draw [] (h) to (Vj);
\draw [] (Vcs) to (Vef);
\draw [] (Vei) to (Vee);
\draw [] (Vdb) to (Vef);
\draw [] (Vdk) to (Vdm);
\draw [] (Ved) to (Vdw);
\draw [] (Vef) to (Vdn);
\draw [] (Vdo) to (Vcm);
\draw [] (Vcs) to (Veb);
\node [style=wh] at (Vj) {$$};
\node [style=wh] at (Vk) {$$};
\end{tikzpicture}

}
}

\newcommand{\totvii}{
\inline{
\begin{tikzpicture}
\node [style=wh] (Vj) at (3.05,-16.175) {};
\node [style=red] (i) at (32.275,-17.0) {};
\node [style=green] (h) at (35.225,-16.575) {};
\node [style=red] (k) at (21.725,-15.725) {};
\node [style=green] (Vdx) at (26.275,-16.95) {};
\node [style=green] (Vef) at (33.425,-16.6) {};
\node [style=red] (j) at (20.3,-13.975) {};
\node [style=red] (e) at (17.3,-9.625) {};
\node [style=red] (Vcm) at (22.5,-14.7) {};
\node [style=green] (Vdm) at (3.95,-19.8) {};
\node [style=red] (g) at (30.7,-15.725) {};
\node [style=red] (Vdn) at (4.45,-17.55) {};
\node [style=wh] (Vk) at (39.275,-16.725) {};
\draw [] (i) to (Vef);
\draw [,bend left=40] (i) to (i);
\draw [] (g) to (Vcm);
\draw [,bend left=-20] (k) to (Vdm);
\draw [] (h) to (i);
\draw [] (e) to (g);
\draw [] (e) to (j);
\draw [] (k) to (Vef);
\draw [] (Vef) to (g);
\draw [] (h) to (Vj);
\draw [] (j) to (Vdm);
\draw [] (Vdn) to (Vdx);
\draw [] (j) to (Vef);
\draw [] (Vdm) to (e);
\draw [] (h) to (k);
\draw [] (e) to (Vcm);
\draw [] (k) to (Vdx);
\draw [] (Vdn) to (h);
\draw [] (g) to (j);
\draw [,bend left=-20] (Vdm) to (k);
\draw [] (Vk) to (h);
\draw [] (Vef) to (Vdn);
\draw [] (Vdx) to (i);
\draw [] (j) to (Vcm);
\node [style=wh] at (Vj) {$$};
\node [style=wh] at (Vk) {$$};
\end{tikzpicture}

}
}

\newcommand{\totviii}{
\inline{
\begin{tikzpicture}
\node [style=wh] (Vj) at (16.225,-21.125) {};
\node [style=red] (j) at (26.425,-15.725) {};
\node [style=green] (Vbj) at (31.25,-19.6) {};
\node [style=red] (e) at (23.425,-11.375) {};
\node [style=green] (Vdm) at (20.575,-22.4) {};
\node [style=red] (Vbf) at (29.6,-21.925) {};
\node [style=wh] (Vk) at (39.5,-22.025) {};
\node [style=red] (g) at (36.825,-17.475) {};
\node [style=red] (Vcm) at (29.775,-15.025) {};
\node [style=green] (Vdx) at (32.4,-25.8) {};
\node [style=green] (Vbh) at (24.3,-18.975) {};
\node [style=green] (Vaz) at (27.2,-23.65) {};
\node [style=red] (Vbg) at (26.275,-18.425) {};
\node [style=green] (Vbi) at (33.1,-21.75) {};
\draw [] (e) to (g);
\draw [] (Vbf) to (Vbh);
\draw [,bend left=-20] (Vaz) to (Vdx);
\draw [] (j) to (Vdm);
\draw [] (Vbf) to (Vdx);
\draw [] (Vbg) to (Vbj);
\draw [] (Vbg) to (Vbi);
\draw [] (Vbj) to (g);
\draw [] (g) to (j);
\draw [] (Vbi) to (Vk);
\draw [] (Vdm) to (e);
\draw [] (Vbi) to (Vj);
\draw [,bend left=-20] (Vbf) to (Vdm);
\draw [,bend left=-20] (Vdx) to (Vaz);
\draw [] (e) to (j);
\draw [] (j) to (Vbj);
\draw [] (Vaz) to (Vbh);
\draw [] (j) to (Vcm);
\draw [] (Vbh) to (Vbg);
\draw [] (g) to (Vcm);
\draw [,bend left=-20] (Vdm) to (Vbf);
\draw [] (e) to (Vcm);
\node [style=wh] at (Vj) {$$};
\node [style=wh] at (Vk) {$$};
\end{tikzpicture}

}
}

\newcommand{\totix}{
\inline{
\begin{tikzpicture}
  \begin{pgfonlayer}{nodelayer}
    \node [style=wh] (0) at (-2, -2) {};
    \node [style=wh] (1) at (2, -2) {};
    \node [style=green] (2) at (0, -2) {};
    \node [style=red] (3) at (0, -1) {};
    \node [style=green] (4) at (1.5, -1) {};
    \node [style=green] (5) at (-1.5, -1) {};
    \node [style=red] (6) at (-1.5, 0.5) {};
    \node [style=red] (7) at (1.5, 0.5) {};
    \node [style=green] (8) at (0, 1.25) {};
  \end{pgfonlayer}
  \begin{pgfonlayer}{edgelayer}
    \draw (0) to (2);
    \draw (2) to (1);
    \draw (2) to (3);
    \draw (3) to (4);
    \draw (3) to (5);
    \draw [bend left=15, looseness=1.25] (5) to (6);
    \draw [bend left, looseness=1.00] (7) to (4);
    \draw [bend left, looseness=1.00] (6) to (8);
    \draw [bend left, looseness=1.00] (8) to (7);
    \draw [bend left, looseness=1.00] (6) to (5);
    \draw [bend right=15, looseness=1.00] (6) to (8);
    \draw [bend right=15, looseness=1.00] (8) to (7);
    \draw [bend right=15, looseness=1.00] (7) to (4);
  \end{pgfonlayer}
\end{tikzpicture}

}
}

\newcommand{\totx}{
\inline{
\begin{tikzpicture}
  \begin{pgfonlayer}{nodelayer}
    \node [style=wh] (0) at (-2, -2) {};
    \node [style=wh] (1) at (2, -2) {};
    \node [style=green] (2) at (0, -2) {};
    \node [style=red] (3) at (0, -1) {};
    \node [style=green] (4) at (1.5, -1) {};
    \node [style=green] (5) at (-1.5, -1) {};
    \node [style=red] (6) at (-1.5, 0.5) {};
    \node [style=red] (7) at (1.5, 0.5) {};
    \node [style=green] (8) at (0, 1.25) {};
  \end{pgfonlayer}
  \begin{pgfonlayer}{edgelayer}
    \draw (0) to (2);
    \draw (2) to (1);
    \draw (2) to (3);
    \draw (3) to (4);
    \draw (3) to (5);
  \end{pgfonlayer}
\end{tikzpicture}

}
}

\newcommand{\totxi}{
\inline{
\begin{tikzpicture}
  \begin{pgfonlayer}{nodelayer}
    \node [style=wh] (0) at (-2, -2) {};
    \node [style=wh] (1) at (2, -2) {};
    \node [style=green] (2) at (0, -2) {};
    \node [style=red] (3) at (-1, -1) {};
    \node [style=red] (4) at (1, -1) {};
    \node [style=green] (5) at (0, -0.25) {};
  \end{pgfonlayer}
  \begin{pgfonlayer}{edgelayer}
    \draw (0) to (2);
    \draw (2) to (1);
  \end{pgfonlayer}
\end{tikzpicture}

}
}

\newcommand{\totxii}{
\inline{
\begin{tikzpicture}
\node [style=wh] (Vi) at (8.15,-5.95) {};
\node [style=wh] (Vk) at (10.5,-5.95) {};
\draw [] (Vk) to (Vi);
\node [style=wh] at (Vi) {$$};
\node [style=wh] at (Vk) {$$};
\end{tikzpicture}

}
}

\title{Verifying the Steane code with Quantomatic}
\author{Ross Duncan and Maxime Lucas}
\date{\today}

\author{Ross Duncan
\institute{University of Strathclyde\\ Glasgow, UK}
\email{ross.duncan@strath.ac.uk}
\and
Maxime Lucas
\institute{Universit\'{e} Libre de Bruxelles\\ Brussels, Belgium}
\email{\quad mlucas@ulb.ac.be}
}
\def\titlerunning{Verifying the Steane code with Quantomatic}
\def\authorrunning{R. Duncan \& M. Lucas}

\begin{document}
% For article.cls abstract goes here 
% other classes may have it in front-matter
\maketitle
\begin{abstract}
  In this paper we give a partially mechanized proof of the correctness of
  Steane's 7-qubit error correcting code, using the tool Quantomatic.
  To the best of our knowledge, this represents the largest and most
  complicated verification task yet carried out using Quantomatic.
\end{abstract}

\section{Introduction}
\label{sec:introduction}

Even more so than their classical equivalents, quantum information
processing technologies are susceptible to noise and other errors,
which can destroy stored data and render computations meaningless.
For this reason, quantum error-correcting codes are seen as a crucial
ingredient of any practical quantum information scheme.

Conceptually, error-correcting codes are simple: the space of possible
messages is embedded in a larger code-space.  If an error occurs on the
encoded message, it will (hopefully) be mapped to an element of the
code-space that does not correspond to any message.  By careful choice
of the code-space, and if the error rate is not too high, it should be
possible to recover the original message.

For example, consider the classical repetition code.  Here a bit is
mapped to a code-space consisting of three bits:
\[
0 \mapsto 000 \qquad\qquad 1 \mapsto 111
\]
If a single bit of the codeword is flipped, then the original encoded
bit can be recovered by taking the majority of the three bits.  Notice
that this code only protects against the specific error model where
only one bit can be flipped:  if two (or more) bits are flipped then
the code cannot correct the error and the encoded bit will be flipped.

The classical case is simpler than the quantum case\footnote{In fact,
  we consider the simplest possible case for both: errors affect
  only a single bit/qubit.  This is not terribly realistic.}  for two
reasons: the only possible errors are bit flips; and, it is possible
to measure a classical bit without changing its value.  However
neither of these problems is fatal.  Although the errors which may
afflict a qubit might be any unitary map, it suffices to coherently
correct Pauli $X$ and $Z$ errors, since these generate all the
others.  To detect the errors in the first place, quantum codes
typically entangle the code-word with an ancilla and measure the
ancilla.  For the full details see, for example, \cite{Mermin:2007fk},
and references therein.

While the simple repetition code described above can be adapted for
the quantum case, at a cost of 9 qubits for each encoded qubit, we
will consider a 7-qubit code due to Andrew Steane \cite{Steane:1996lr}.  Like
the repetition code, the Steane code can only correct single qubit
errors\footnote{This is not exactly true.  As we shall see later, the
  code can correct a single $Z$ error and/or a single $X$ error, but
  they do not have to affect the same qubit.}.  
Practically speaking the code consists of three circuits:  an
encoding circuit, a decoding circuit (adjoint to the encoder), and an
error correcting circuit that tests the code-space for errors and
corrects them.

In the rest of the paper, the three circuits mentioned above will be
translated into the graphical formalism of the \zxcalculus
\cite{Coecke:2009aa}.  Using the formalism, as implemented in the
automated rewriting system \emph{Quantomatic} \cite{quantomatichome},
the circuits are simplified.  We then, again by rewriting, derive the
error syndrome conditions for the correction circuit, and verify that
it does indeed correct the claimed errors.  (It is remarkable how
simple and clear this graphical proof is, compared to the usual
textbook approach.)  Finally, we show that the combined
circuit---encoder-corrector-decoder---rewrites to the identity on one
qubit, proving it does not introduce any errors itself as a
consequence of its design.

\section{The \zxcalculus}
\label{sec:zxcalculus}

The \zxcalculus is a formal language for reasoning about quantum
computational systems.  It comprises a graphical syntax and a set of
axioms presented as rewrite rules.  It has a standard semantics in
Hilbert spaces, and can represent any quantum circuit.  We briefly
review the \zxcalculus here; for full
detalis consult \cite{Coecke:2009aa}; here we make use of the
\zxcalculus extended with conditional vertices, as elaborated in
\cite{Duncan:2010aa,Duncan:2012uq}.

\begin{figure}[tb]
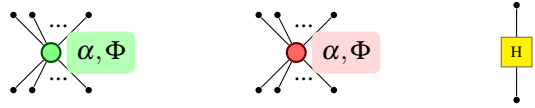

  \centering
  \greenspider{$\alpha,\Phi$}\hspace{1cm}
  \redspider{$\alpha,\Phi$}\hspace{1cm}
  \hgate
  \caption{Permitted interior vertices}
  \label{fig:components}\vspace{-6mm}
\end{figure}

\begin{definition}\label{def:diagram}
  Let $S$ be some set of variables.  
  A \emph{diagram over $S$} is an open graph  whose interior
  vertices are restricted to the following types:
  \begin{itemize}
  \item $Z$-vertices, labelled
    by an angle $\alpha \in  [0,2\pi)$ and a boolean formula $\Phi$
    with variables from $S$;  these are shown as (light) green circles,
  \item $X$ vertices, labelled by an angle $\alpha \in [0,2\pi)$ and a
    boolean formula $\Phi$ with variables from $S$; these are shown as
    (dark) red circles,
  \item $H$ (or Hadamard) vertices, restricted to degree 2; shown as squares.
  \end{itemize}
  The allowed vertices are shown in Figure~\ref{fig:components}.
\end{definition}

If an $X$ or $Z$ vertex is labelled by $\alpha = 0$ then the label is
omitted.  If the labelling formula $\Phi$ is not constant
then the vertex is called \emph{conditional}; otherwise it is
unconditional.  We omit the formulae from unconditional vertices.

For each $S$, the diagrams over $S$ form a symmetric monoidal
category (in fact compact closed) in the obvious way: the objects of
the category are natural numbers and an arrow $g:m\to n$ is a diagram
with $m$ input and $n$ outputs.  The tensor product is juxtaposition,
and composition $g\circ f$ is defined by identifying the output
vertices of $f$ with the input vertices of $g$.  For more details see
\cite{Duncan:thesis:2006,Lucas-Dixon:2009yq}.  Denote this category
$\mathbb{D}(S)$; we denote the category $\mathbb{D}(\emptyset)$ of
unconditional diagrams by $\mathbb{D}$.  Note that the components
shown in Figure~\ref{fig:components} are the generators of
$\mathbb{D}(S)$.

\begin{definition}\label{def:valuation}
  Let $v:S\to \{0,1\}$ be any boolean function; it naturally assigns a
  truth value $v(\Phi)$ to any formula $\Phi$ over $S$.  We define a
  \emph{valuation} functor $\hat{v}:\mathbb{D}(S) \to
  \mathbb{D}$ by relabelling the $Z$ and $X$ vertices as follows:
  \[
  \alpha,\Phi \mapsto \left\{
  \begin{array}{rl}
    \alpha & \text{ if } v(\Phi) = 1\\
    0 & \text{ otherwise}
  \end{array} \right.
  \]
\end{definition}

\begin{definition}\label{def:semantic-pt1}
  Let $\denote{\cdot} : \mathbb{D} \to \fdHilb$ be a traced monoidal
  functor;  define its action on objects by $\denote{n} =
  \mathbb{C}^{2^n}$;  define its action on the generators as:
  \begin{eqnarray*}
    \ldenote{\inline{\greenspider{$\alpha$}}}
    & = &
    \left\{
    \begin{array}{ccl}
      \ket{0}^{\otimes m} & \mapsto &   \ket{0}^{\otimes n}\\
      \ket{1}^{\otimes m} & \mapsto &   e^{i \alpha}\ket{1}^{\otimes n}
    \end{array}\right.
\\
    \ldenote{\inline{\redspider{$\alpha$}}}
    & = &
    \left\{
    \begin{array}{ccl}
      \ket{+}^{\otimes m} & \mapsto &   \ket{+}^{\otimes n}\\
      \ket{-}^{\otimes m} & \mapsto &   e^{i \alpha}\ket{-}^{\otimes n}
    \end{array}\right.
    \\
    \ldenote{\inline{\hgate}}
    &=& 
    \frac{1}{\sqrt{2}}
    \left(
      \begin{array}{cc}
        1&1\\1&-1
      \end{array}
    \right).
  \end{eqnarray*}  
\end{definition}

\begin{definition}\label{def:semantic-pt2}
  The denotation of a diagram $D$ over variables $S$ is a
  superoperator constructed by summing over all the valuations of
  $S$:
  \[
  \rho \mapsto \sum_{v\in 2^S}
  \denote{\hat{v}(D)}\rho\denote{\hat{v}(D)}^\dag\;.
  \]
\end{definition}

\begin{example}[Unitary gates]\label{ex:gates-to-zx}
  Any quantum circuit can be written in the \zxcalculus.  This is most
  easily seen via the  translation of a universal gate set:
  \[
  \begin{array}{ccccc}
    \circX{\alpha}  
    & = &    \begin{pmatrix*}[r]
      \cos \frac{\alpha}{2} & -i\sin\frac{\alpha}{2} \\
      -i\sin\frac{\alpha}{2} & \cos \frac{\alpha}{2}  
    \end{pmatrix*}
    &=& \ldenote{\redphase{\alpha}} 
    \\
    \circZ{\beta} 
    &= &     \begin{pmatrix*}[r]
      1 & 0 \\ 0 & e^{i\beta}
    \end{pmatrix*}
    &= & \ldenote{\greenphase{\beta}}
    \\
    \circH  
    &= & \frac{1}{\sqrt{2}}
    \begin{pmatrix*}[r]
      1 & 1 \\ 1 & -1
    \end{pmatrix*}
    & = & \ldenote{\hgate}
    \\
    \circCX 
    &= &     \begin{pmatrix*}
      1 & 0 & 0 & 0\\
      0 & 1 & 0 & 0\\
      0 & 0 & 0 & 1\\
      0 & 0 & 1 & 0
    \end{pmatrix*}\,.
    &= & \ldenote{\cex}
  \end{array}
  \]
  Of course, one easily incorporates state preparations as well.  For
  example:
  \[
\circket{0} =
\begin{pmatrix*}
  1 \\ 0
\end{pmatrix*} = 
\ldenote{\redKzero}
\qquad\qquad
\circket{1} =
\begin{pmatrix*}
  0 \\ 1
\end{pmatrix*} = 
\ldenote{\redKone}
  \]
  Notice that these components only require \emph{uncondtional} diagrams.
\end{example}

\begin{example}[Measurements and classical  control]
  \label{ex:measurements-in-zx}
  A classically controlled operation is encoded using a conditional
  operator.  For example, the following diagram represents a Pauli $Z$
  which is turned on by the classical data stored in the variable $v$.
  \[
  \ldenote{\greenphase{\pi,\{v\}} } = 
  \left\{ \;\rho \mapsto \ID\rho\ID + Z\rho Z \;\right\}
  \]
  We represent measurements by combining a conditional Pauli and a
  projection.  Let $\ket{\alpha_\pm} = (\ket{0} \pm
  e^{i\alpha}\ket{1})/\sqrt 2$, then 
  \[
  \ldenote{\meas} 
  = \left\{\rule{0pt}{2em} \; \rho \mapsto 
      \bra{\alpha_+}\left(\ID\rho\ID + Z\rho Z\right) \ket{\alpha_+} 
      \;=\; \bra{\alpha_+}\rho\ket{\alpha_+} +  \bra{\alpha_-}\rho\ket{\alpha_-}
      \;\right\}
  \]
  is the measurement in the basis $\ket{\alpha_\pm}$.  Notice that
  here the variable $i$ represents the \emph{outcome} of the
  measurement;  hence to properly encode a measurement it should
  always be a \emph{fresh variable}.
\end{example}

\paragraph{Axioms.}
\label{sec:axioms}

\begin{figure}[p]
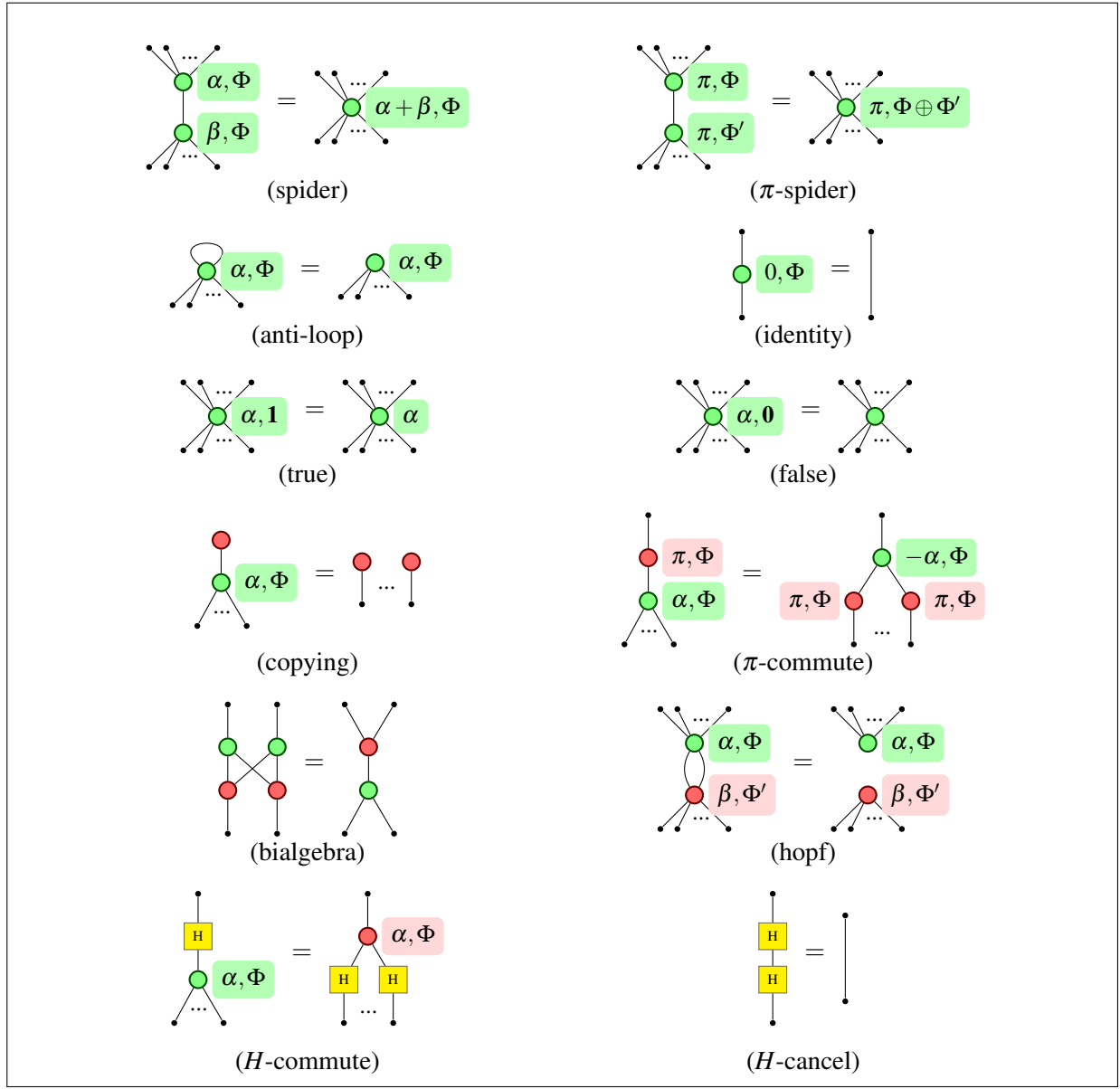

  \centering
  \fbox{
  \begin{minipage}[c]{1.0\linewidth}
  \centering
\[
\begin{array}{ccc}
\greenSpiderLHS
\;\;=\;\;
\greenSpiderRHS
& 
\qquad \quad 
&
\greenNewSpiderLHS
\;\;=\;\;
\greenNewSpiderRHS
\\
\text{(spider)}&&\text{($\pi$-spider)}
\\ \\[-1ex]
\greenAntiLoopLHS
\;\;=\;\;
\greenAntiLoopRHS
&
\qquad \quad 
&
\greenIdentityLHS
\;\;=\;\;
\greenIdentityRHS
\\
\text{(anti-loop)} && \text{(identity)} 
\\ \\[-1ex]
\greenTrueLHS
\;\;=\;\;
\greenTrueRHS
&
\qquad \quad 
&
\greenFalseLHS
\;\;=\;\;
\greenFalseRHS
\\
\text{(true)} && \text{(false)} 
\\ \\[-1ex]
\copyingLHS
\;\;=\;\;
\copyingRHS
&
\qquad \qquad
&
\piCommutesLHS
\;\;=\;\;
\piCommutesRHS
\\ 
\text{(copying)}  &&  \text{($\pi$-commute)}
\\ \\[-1ex]
\bialgebraLHS
\;\;=\;\;
\bialgebraRHS
&
\qquad \qquad
&
\hopfLawLHS
\;\;=\;\;
\hopfLawRHS
\\
\text{(bialgebra)} &&\text{(hopf)} 
\\ \\[-1ex]
\removeGreenLHS
\;\;=\;\;
\removeGreenRHS
& 
\qquad \qquad
&
\HsquaredLHS
\;\;=\;\;
\HsquaredRHS
\\ \\[-1ex]
\text{($H$-commute)}  &&\text{($H$-cancel)} 
\end{array}
\]
\end{minipage}}
\caption{Equational rules for the \zxcalculus.  We present the rules for the
  $Z$ subsystem;  to obtain the complete set of rules exchange the
  colours in the rules shown above.
}
\label{fig:rewrite-rules}
\end{figure}

The \zxcalculus calculus is also an equational theory; its axioms are
shown in Figure~\ref{fig:rewrite-rules}.  Two key properties are:
\begin{itemize}
\item \emph{Soundness:} If $D = D'$ via the equational rules, then
  $\denote{D} = \denote{D'}$.
\item \emph{Colour Duality}: If $D_1 = D_2$ is a derivable equation in the
  \zxcalculus then $D_1' = D_2'$ is also derivable, where $D_i'$ is
  obtained from $D_i$ by exchanging the colours red and green.
\end{itemize}

\begin{example}[The \CZ-gate]\label{ex:CZ}
  The \CZ-gate can be obtained by using a Hadamard gate to transform
  the target bit of a \CX gate.
  \[
  \circCZcirc  \quad \mapsto \quad \CZviaCX \quad = \quad \czed
  \]
\end{example}

%%%%%%%%%%%%%%%%%%%%%%%%%%%%%%%%%%%%%%%%%%%%%%%%%%%%%%%%%%%%%%%%%%
%%%%%%%%%%%%%%%%%%%%%%%%%%%%%%%%%%%%%%%%%%%%%%%%%%%%%%%%%%%%%%%%%%
%%%%% START OF REAL SHIT HERE
%%%%%%%%%%%%%%%%%%%%%%%%%%%%%%%%%%%%%%%%%%%%%%%%%%%%%%%%%%%%%%%%%%
%%%%%%%%%%%%%%%%%%%%%%%%%%%%%%%%%%%%%%%%%%%%%%%%%%%%%%%%%%%%%%%%%%

\section{Translation and Simplification}
\label{sec:transl-simpl}

In this section we describe how the encoding and correcting circuits
of the Steane code can be reprsented in \zxcalculus, and perform some
simplifications.  We will not at any point give a traditional
\emph{definition} of the code other than these circuits; the
discussion of Section~\ref{sec:errors-their-corr} will demonstrate quite
clearly how it works.

\noindent
\textbf{NB.} From this point onwards, we adopt the convention that
diagrams are oriented with the inputs to the \emph{left} side and
outputs to the \emph{right}.

\paragraph{The encoder}
\label{sec:encoder}

\begin{figure}[htbp]
  \centering
  \[
  \begin{array}{ccc}
    \inline{%
\beginpgfgraphicnamed{enc-circuit}
\begin{tikzpicture}
	\begin{pgfonlayer}{nodelayer}
		\node [style=keto box] (0) at (-6, -2.25) {};
		\node [style=keto box] (1) at (-6, -3.25) {};
		\node [style=H box] (2) at (-4.5, -2.25) {};
		\node [style=H box] (3) at (-4.5, -3.25) {};
		\node [style=ctrl vertex] (4) at (0.75, -3.25) {};
		\node [style=ctrl vertex] (5) at (-0.5, -2.25) {};
		\node [style=none] (6) at (2.5, -2.25) { };
		\node [style=none] (7) at (2.5, -3.25) { };
		\node [style=none] (8) at (-7, 0) {};
		\node [style=none] (9) at (2.5, 0) { };
		\node [style=none] (10) at (2.5, 1.25) { };
		\node [style=keto box] (11) at (-6, 1.25) {};
		\node [style=H box] (12) at (-4.5, -1.25) {};
		\node [style=none] (13) at (2.5, -1.25) { };
		\node [style=keto box] (14) at (-6, -1.25) {};
		\node [style=ctrl vertex] (15) at (-1.75, -1.25) {};
		\node [style=none] (16) at (2.5, 3.25) { };
		\node [style=keto box] (17) at (-6, 3.25) {};
		\node [style=none] (18) at (2.5, 2.25) { };
		\node [style=keto box] (19) at (-6, 2.25) {};
		\node [style=ctrl vertex] (20) at (-3, 0) {};
		\node [style=target vertex] (21) at (-3, 1.25) {};
		\node [style=target vertex] (22) at (-3, 2.25) {};
		\node [style=target vertex] (23) at (-1.75, 0) {};
		\node [style=target vertex] (24) at (-1.75, 1.25) {};
		\node [style=target vertex] (25) at (-1.75, 3.25) {};
		\node [style=target vertex] (26) at (-0.5, 0) {};
		\node [style=target vertex] (27) at (-0.5, 2.25) {};
		\node [style=target vertex] (28) at (-0.5, 3.25) {};
		\node [style=target vertex] (29) at (0.75, 1.25) {};
		\node [style=target vertex] (30) at (0.75, 2.25) {};
		\node [style=target vertex] (31) at (0.75, 3.25) {};
		\node [style=none] (32) at (3, 2.25) {5};
		\node [style=none] (33) at (3, -3.25) {0};
		\node [style=none] (34) at (3, -1.25) {2};
		\node [style=none] (35) at (3, 0) {3};
		\node [style=none] (36) at (3, 3.25) {6};
		\node [style=none] (37) at (3, -2.25) {1};
		\node [style=none] (38) at (3, 1.25) {4};
	\end{pgfonlayer}
	\begin{pgfonlayer}{edgelayer}
		\draw (0) to (2);
		\draw (2) to (5);
		\draw (1) to (3);
		\draw (3) to (4);
		\draw (14) to (12);
		\draw (12) to (15);
		\draw [style=(null)] (8.center) to (9.center);
		\draw [style=(null)] (15) to (13.center);
		\draw [style=(null)] (5) to (6.center);
		\draw [style=(null)] (4) to (7.center);
		\draw [style=(null)] (17) to (16.center);
		\draw [style=(null)] (19) to (18.center);
		\draw [style=(null)] (11) to (10.center);
		\draw [style=(null)] (20) to (22);
		\draw [style=(null)] (15) to (25);
		\draw [style=(null)] (5) to (28);
		\draw [style=(null)] (4) to (31);
	\end{pgfonlayer}
\end{tikzpicture}}
\endpgfgraphicnamed} 
    & \rule{2em}{0pt} &
    \viienci
    \\
    \\
    \text{(a)} && \text{(b)}
  \end{array}
  \]
  \caption{The encoder (a) as a circuit; (b) in the \zxcalculus.}
  \label{fig:encoder}
\end{figure}
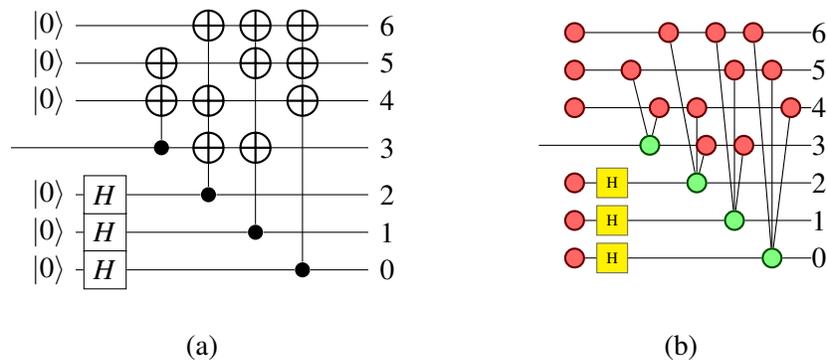

The encoder maps a single input qubit to the 7-qubit code-space via
the circuit shown in Figure~\ref{fig:encoder} (a).  Note that qubit 3
is the input.  Using the translation described in
Example~\ref{ex:gates-to-zx} we obtain the \zxcalculus term shown in
Figure~\ref{fig:encoder} (b).

The encoding circuit is simple enough that Quantomatic can reduce it
to its minimal form without human intervention.  The resulting
simplification is shown below; the final equation is merely a
rearrangment of the vertices to show the structure more clearly.
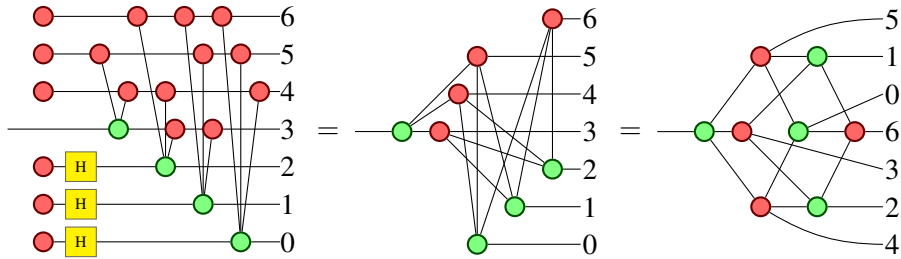
\[
\viienci
  = \inline{\begin{tikzpicture}
	\begin{pgfonlayer}{nodelayer}
		\node [style=green vertex] (0) at (0, -3) {};
		\node [style=red vertex] (1) at (-1, 0) {};
		\node [style=red vertex] (2) at (-0.5, 1) {};
		\node [style=red vertex] (3) at (0, 2) {};
		\node [style=none] (4) at (-3.25, -0) {};
		\node [style=none] (5) at (2.75, 3) {};
		\node [style=none] (6) at (2.75, 2) {};
		\node [style=none] (7) at (2.75, 1) {};
		\node [style=none] (8) at (2.75, 0) {};
		\node [style=none] (9) at (2.75, -1) {};
		\node [style=none] (10) at (2.75, -2) {};
		\node [style=none] (11) at (2.75, -3) {};
		\node [style=red vertex] (12) at (2, 3) {};
		\node [style=green vertex] (13) at (-2, 0) {};
		\node [style=green vertex] (14) at (2, -1) {};
		\node [style=green vertex] (15) at (1, -2) {};
		\node [style=none] (16) at (3, 1) {4};
		\node [style=none] (17) at (3, 0) {3};
		\node [style=none] (18) at (3, 2) {5};
		\node [style=none] (19) at (3, -3) {0};
		\node [style=none] (20) at (3, 3) {6};
		\node [style=none] (21) at (3, -2) {1};
		\node [style=none] (22) at (3, -1) {2};
	\end{pgfonlayer}
	\begin{pgfonlayer}{edgelayer}
		\draw (0) to (11.center);
		\draw (12) to (5.center);
		\draw (12) to (0);
		\draw (3) to (6.center);
		\draw (3) to (0);
		\draw (2) to (7.center);
		\draw (2) to (0);
		\draw (4.center) to (13);
		\draw (13) to (1);
		\draw (1) to (8.center);
		\draw (13) to (2);
		\draw (3) to (13);
		\draw (1) to (14);
		\draw (1) to (15);
		\draw (2) to (14);
		\draw (3) to (15);
		\draw [in=180, out=0] (14) to (9.center);
		\draw (15) to (10.center);
		\draw (14) to (12);
		\draw (15) to (12);
	\end{pgfonlayer}
\end{tikzpicture}}
  =  \inline{\begin{tikzpicture}
	\begin{pgfonlayer}{nodelayer}
		\node [style=green vertex] (0) at (0.5, 0) {};
		\node [style=red vertex] (1) at (-1, 0) {};
		\node [style=red vertex] (2) at (-0.5, -2) {};
		\node [style=red vertex] (3) at (-0.5, 2) {};
		\node [style=none] (4) at (-3.25, -0) {};
		\node [style=none] (5) at (2.75, 0) {};
		\node [style=none] (6) at (2.75, 3) {};
		\node [style=none] (7) at (2.75, -3) {};
		\node [style=none] (8) at (2.75, -1) {};
		\node [style=none] (9) at (2.75, -2) {};
		\node [style=none] (10) at (2.75, 2) {};
		\node [style=none] (11) at (2.75, 1) {};
		\node [style=red vertex] (12) at (2, 0) {};
		\node [style=green vertex] (13) at (-2, 0) {};
		\node [style=green vertex] (14) at (1, -2) {};
		\node [style=green vertex] (15) at (1, 2) {};
		\node [style=none] (16) at (3, -3) {4};
		\node [style=none] (17) at (3, -1) {3};
		\node [style=none] (18) at (3, 3) {5};
		\node [style=none] (19) at (3, 1) {0};
		\node [style=none] (20) at (3, 0) {6};
		\node [style=none] (21) at (3, 2) {1};
		\node [style=none] (22) at (3, -2) {2};
	\end{pgfonlayer}
	\begin{pgfonlayer}{edgelayer}
		\draw (0) to (11.center);
		\draw (12) to (5.center);
		\draw (12) to (0);
		\draw [bend left=15] (3) to (6.center);
		\draw (3) to (0);
		\draw [bend right=15] (2) to (7.center);
		\draw (2) to (0);
		\draw (4.center) to (13);
		\draw (13) to (1);
		\draw (1) to (8.center);
		\draw (13) to (2);
		\draw (3) to (13);
		\draw (1) to (14);
		\draw (1) to (15);
		\draw (2) to (14);
		\draw (3) to (15);
		\draw (14) to (9.center);
		\draw (15) to (10.center);
		\draw (14) to (12);
		\draw (15) to (12);
	\end{pgfonlayer}
\end{tikzpicture}}
\]
The decoding circuit is simply the  encoder in reverse.

\paragraph{The error corrector}
\label{sec:error-corrector}

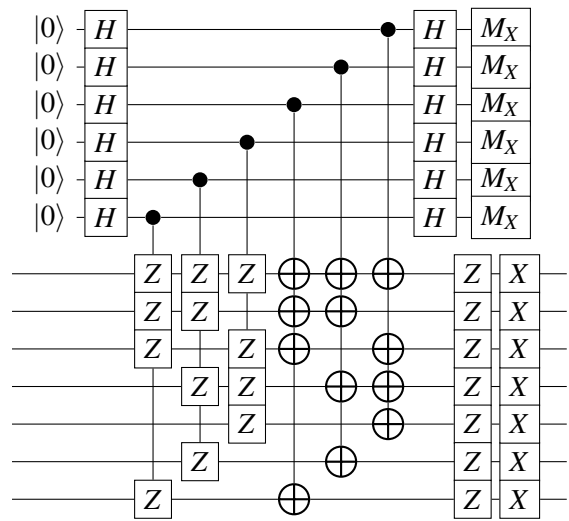
\begin{figure}[htbp]
  \centering
  \inline{\begin{tikzpicture}
	\begin{pgfonlayer}{nodelayer}
		\node [style=keto box] (0) at (-5.5, 2) {};
		\node [style=keto box] (1) at (-5.5, 1) {};
		\node [style=H box] (2) at (-4, 2) {};
		\node [style=H box] (3) at (-4, 1) {};
		\node [style=Z box] (4) at (-2.75, -0.5) {};
		\node [style=Z box] (5) at (-1.5, -0.5) {};
		\node [style=ctrl vertex] (6) at (-2.75, 1) {};
		\node [style=ctrl vertex] (7) at (-1.5, 2) {};
		\node [style=H box] (8) at (4.75, 2) {};
		\node [style=H box] (9) at (4.75, 1) {};
		\node [style=Mx box] (10) at (6.5, 2) {};
		\node [style=Mx box] (11) at (6.5, 1) {};
		\node [style=none] (12) at (-6.5, -0.5) {};
		\node [style=none] (13) at (8.25, -0.5) {};
		\node [style=Z box] (14) at (-0.25, -0.5) {};
		\node [style=Mx box] (15) at (6.5, 4) {};
		\node [style=H box] (16) at (4.75, 4) {};
		\node [style=H box] (17) at (-4, 4) {};
		\node [style=keto box] (18) at (-5.5, 4) {};
		\node [style=H box] (19) at (-4, 3) {};
		\node [style=ctrl vertex] (20) at (1, 4) {};
		\node [style=Mx box] (21) at (6.5, 3) {};
		\node [style=keto box] (22) at (-5.5, 3) {};
		\node [style=H box] (23) at (4.75, 3) {};
		\node [style=ctrl vertex] (24) at (-0.25, 3) {};
		\node [style=Mx box] (25) at (6.5, 6) {};
		\node [style=H box] (26) at (4.75, 6) {};
		\node [style=H box] (27) at (-4, 6) {};
		\node [style=keto box] (28) at (-5.5, 6) {};
		\node [style=H box] (29) at (-4, 5) {};
		\node [style=ctrl vertex] (30) at (3.5, 6) {};
		\node [style=Mx box] (31) at (6.5, 5) {};
		\node [style=keto box] (32) at (-5.5, 5) {};
		\node [style=H box] (33) at (4.75, 5) {};
		\node [style=ctrl vertex] (34) at (2.25, 5) {};
		\node [style=target vertex] (35) at (3.5, -0.5) {};
		\node [style=target vertex] (36) at (2.25, -0.5) {};
		\node [style=target vertex] (37) at (1, -0.5) {};
		\node [style=Z box] (38) at (5.75, -0.5) {};
		\node [style=X box] (39) at (7, -0.5) {};
		\node [style=Z box] (40) at (-2.75, -1.5) {};
		\node [style=target vertex] (41) at (1, -1.5) {};
		\node [style=Z box] (42) at (-1.5, -1.5) {};
		\node [style=none] (43) at (-6.5, -1.5) {};
		\node [style=target vertex] (44) at (2.25, -1.5) {};
		\node [style=none] (45) at (8.25, -1.5) {};
		\node [style=Z box] (46) at (5.75, -1.5) {};
		\node [style=X box] (47) at (7, -1.5) {};
		\node [style=target vertex] (48) at (3.5, -3.5) {};
		\node [style=target vertex] (49) at (1, -2.5) {};
		\node [style=none] (50) at (-6.5, -3.5) {};
		\node [style=none] (51) at (-6.5, -2.5) {};
		\node [style=Z box] (52) at (5.75, -3.5) {};
		\node [style=none] (53) at (8.25, -2.5) {};
		\node [style=Z box] (54) at (5.75, -2.5) {};
		\node [style=Z box] (55) at (-0.25, -3.5) {};
		\node [style=Z box] (56) at (-2.75, -2.5) {};
		\node [style=none] (57) at (8.25, -3.5) {};
		\node [style=target vertex] (58) at (2.25, -3.5) {};
		\node [style=X box] (59) at (7, -3.5) {};
		\node [style=target vertex] (60) at (3.5, -2.5) {};
		\node [style=Z box] (61) at (-1.5, -3.5) {};
		\node [style=X box] (62) at (7, -2.5) {};
		\node [style=Z box] (63) at (-0.25, -2.5) {};
		\node [style=target vertex] (64) at (3.5, -4.5) {};
		\node [style=none] (65) at (8.25, -5.5) {};
		\node [style=none] (66) at (-6.5, -4.5) {};
		\node [style=X box] (67) at (7, -5.5) {};
		\node [style=Z box] (68) at (5.75, -4.5) {};
		\node [style=X box] (69) at (7, -6.5) {};
		\node [style=Z box] (70) at (-0.25, -4.5) {};
		\node [style=none] (71) at (8.25, -6.5) {};
		\node [style=target vertex] (72) at (2.25, -5.5) {};
		\node [style=Z box] (73) at (-1.5, -5.5) {};
		\node [style=none] (74) at (8.25, -4.5) {};
		\node [style=X box] (75) at (7, -4.5) {};
		\node [style=Z box] (76) at (-2.75, -6.5) {};
		\node [style=Z box] (77) at (5.75, -5.5) {};
		\node [style=target vertex] (78) at (1, -6.5) {};
		\node [style=none] (79) at (-6.5, -5.5) {};
		\node [style=none] (80) at (-6.5, -6.5) {};
		\node [style=Z box] (81) at (5.75, -6.5) {};
	\end{pgfonlayer}
	\begin{pgfonlayer}{edgelayer}
		\draw (0) to (2);
		\draw (2) to (7);
		\draw (7) to (8);
		\draw (10) to (8);
		\draw (1) to (3);
		\draw (3) to (6);
		\draw (6) to (9);
		\draw (9) to (11);
		\draw (6) to (4);
		\draw (7) to (5);
		\draw (12.center) to (4);
		\draw (4) to (5);
		\draw (5) to (13.center);
		\draw (18) to (17);
		\draw (17) to (20);
		\draw (20) to (16);
		\draw (15) to (16);
		\draw (22) to (19);
		\draw (19) to (24);
		\draw (24) to (23);
		\draw (23) to (21);
		\draw (28) to (27);
		\draw (27) to (30);
		\draw (30) to (26);
		\draw (25) to (26);
		\draw (32) to (29);
		\draw (29) to (34);
		\draw (34) to (33);
		\draw (33) to (31);
		\draw (37) to (36);
		\draw (43.center) to (40);
		\draw (40) to (42);
		\draw (42) to (45.center);
		\draw (41) to (44);
		\draw (51.center) to (56);
		\draw (61) to (57.center);
		\draw (73) to (65.center);
		\draw (80.center) to (76);
		\draw (24) to (14);
		\draw (20) to (37);
		\draw (34) to (36);
		\draw (30) to (35);
		\draw (76) to (78);
		\draw (78) to (81);
		\draw (81) to (69);
		\draw (69) to (71.center);
		\draw (56) to (63);
		\draw (63) to (49);
		\draw (49) to (60);
		\draw (60) to (54);
		\draw (54) to (62);
		\draw (50.center) to (61);
		\draw (66.center) to (70);
		\draw (70) to (64);
		\draw (64) to (68);
		\draw (68) to (75);
		\draw (79.center) to (73);
		\draw (4) to (40);
		\draw (40) to (56);
		\draw (56) to (76);
		\draw (5) to (42);
		\draw (42) to (61);
		\draw (61) to (73);
		\draw (14) to (63);
		\draw (63) to (55);
		\draw (55) to (70);
		\draw (37) to (41);
		\draw (41) to (49);
		\draw (49) to (78);
		\draw (36) to (44);
		\draw (44) to (58);
		\draw (58) to (72);
		\draw (35) to (60);
		\draw (60) to (48);
		\draw (48) to (64);
		\draw (62) to (53.center);
		\draw (75) to (74.center);
	\end{pgfonlayer}
\end{tikzpicture}}
  \caption{The circuit to correct errors.  Note that the conditions to
  activate the Pauli $X$ and $Z$ corrections are not shown.}
  \label{fig:correct-circuit}
\end{figure}

An encoder is not very useful if we cannot detect and correct errors
on the encoded data.  The circuit for carrying out this function is shown
in Figure~\ref{fig:correct-circuit}.  Notice that the circuit
naturally splits into two parts.  The \emph{error-detecting} part introduces a 6-qubit ancilla,
entangles it with the input, and then measures the ancillae.  The
resulting measurements are called the \emph{error syndrome}.  The
\emph{error-correcting} part comprises the Pauli operations at the
end, which are applied \emph{conditionally}, depending on the
value of the error syndrome.  One crucial detail has been omitted from
this picture: we do not show the conditions which control the
error-correcting part.  We will derive these conditions in the next
section.

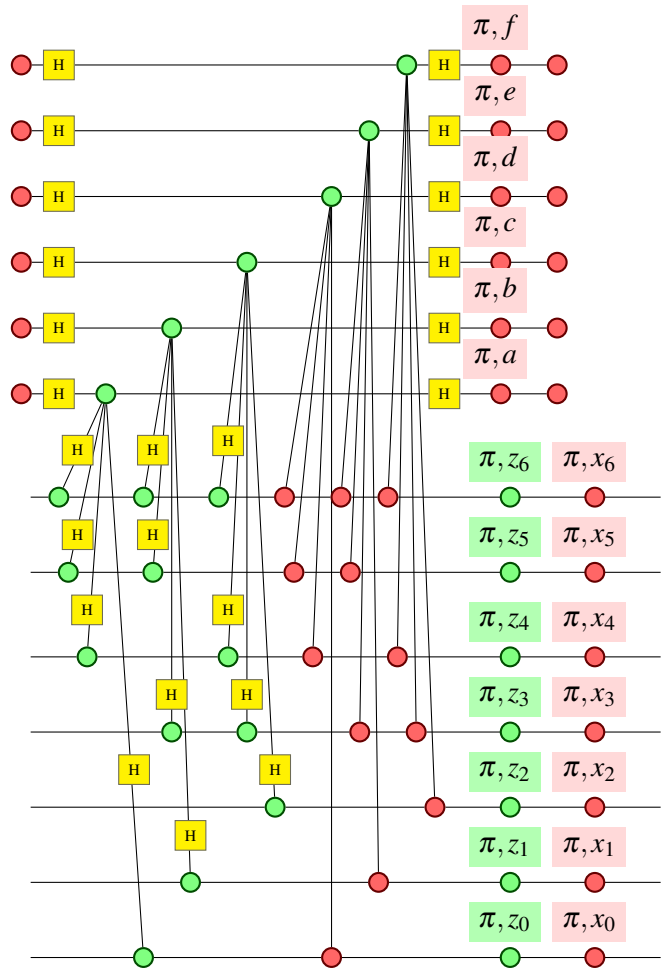
\begin{figure}[htbp]
  \centering
  \inline{%
\beginpgfgraphicnamed{cor-i}
\begin{tikzpicture}
	\begin{pgfonlayer}{nodelayer}
		\node [style=hadamard vertex] (0) at (-7, 3.5) {};
		\node [style=hadamard vertex] (1) at (-7, 5.25) {};
		\node [style=hadamard vertex] (2) at (-7, 7) {};
		\node [style=hadamard vertex] (3) at (-7, 8.75) {};
		\node [style=hadamard vertex] (4) at (-7, 10.5) {};
		\node [style=hadamard vertex] (5) at (-7, 12.25) {};
		\node [style=green vertex] (6) at (-5.75, 3.5) {};
		\node [style=green vertex] (7) at (-4, 5.25) {};
		\node [style=green vertex] (8) at (-2, 7) {};
		\node [style=green vertex] (9) at (0.25, 8.75) {};
		\node [style=green vertex] (10) at (1.25, 10.5) {};
		\node [style=green vertex] (11) at (2.25, 12.25) {};
		\node [style=hadamard vertex] (12) at (3.25, 12.25) {};
		\node [style=hadamard vertex] (13) at (3.25, 10.5) {};
		\node [style=hadamard vertex] (14) at (3.25, 8.75) {};
		\node [style=hadamard vertex] (15) at (3.25, 7) {};
		\node [style=hadamard vertex] (16) at (3.25, 5.25) {};
		\node [style=hadamard vertex] (17) at (3.25, 3.5) {};
		\node [style=none] (18) at (-7.75, 0.75) {};
		\node [style=none] (19) at (-7.75, -1.25) {};
		\node [style=none] (20) at (-7.75, -1.25) {};
		\node [style=none] (21) at (-7.75, -3.5) {};
		\node [style=none] (22) at (-7.75, -5.5) {};
		\node [style=none] (23) at (-7.75, -7.5) {};
		\node [style=none] (24) at (-7.75, -9.5) {};
		\node [style=none] (25) at (9, 0.75) {};
		\node [style=none] (26) at (9, -1.25) {};
		\node [style=none] (27) at (9, -3.5) {};
		\node [style=none] (28) at (9, -5.5) {};
		\node [style=none] (29) at (9, -7.5) {};
		\node [style=none] (30) at (9, -9.5) {};
		\node [style=green vertex] (31) at (-7, 0.75) {};
		\node [style=green vertex] (32) at (-6.75, -1.25) {};
		\node [style=green vertex] (33) at (-6.25, -3.5) {};
		\node [style=green vertex] (34) at (-4.75, -11.5) {};
		\node [style=green vertex] (35) at (-4.75, 0.75) {};
		\node [style=green vertex] (36) at (-2.75, 0.75) {};
		\node [style=green vertex] (37) at (-4.5, -1.25) {};
		\node [style=green vertex] (38) at (-4, -5.5) {};
		\node [style=green vertex] (39) at (-2.5, -3.5) {};
		\node [style=green vertex] (40) at (-2, -5.5) {};
		\node [style=green vertex] (41) at (-1.25, -7.5) {};
		\node [style=green vertex] (42) at (-3.5, -9.5) {};
		\node [style=none] (43) at (9, -11.5) {};
		\node [style=red vertex] (44) at (-1, 0.75) {};
		\node [style=red vertex] (45) at (-0.75, -1.25) {};
		\node [style=red vertex] (46) at (-0.25, -3.5) {};
		\node [style=red vertex] (47) at (0.25, -11.5) {};
		\node [style=red vertex] (48) at (0.5, 0.75) {};
		\node [style=red vertex] (49) at (0.75, -1.25) {};
		\node [style=red vertex] (50) at (1, -5.5) {};
		\node [style=red vertex] (51) at (1.5, -9.5) {};
		\node [style=red vertex] (52) at (1.75, 0.75) {};
		\node [style=red vertex] (53) at (2, -3.5) {};
		\node [style=red vertex] (54) at (2.5, -5.5) {};
		\node [style=red vertex] (55) at (3, -7.5) {};
		\node [style=none] (56) at (-7.75, -11.5) {};
		\node [style=hadamard vertex] (57) at (-6.5, 2) {};
		\node [style=hadamard vertex] (58) at (-6.5, -0.25) {};
		\node [style=hadamard vertex] (59) at (-6.25, -2.25) {};
		\node [style=hadamard vertex] (60) at (-5, -6.5) {};
		\node [style=hadamard vertex] (61) at (-4.5, 2) {};
		\node [style=hadamard vertex] (62) at (-4.5, -0.25) {};
		\node [style=hadamard vertex] (63) at (-4, -4.5) {};
		\node [style=hadamard vertex] (64) at (-3.5, -8.25) {};
		\node [style=hadamard vertex] (65) at (-1.25, -6.5) {};
		\node [style=hadamard vertex] (66) at (-2, -4.5) {};
		\node [style=hadamard vertex] (67) at (-2.5, -2.25) {};
		\node [style=hadamard vertex] (68) at (-2.5, 2.25) {};
		\node [style=red vertex] (69) at (-8, 12.25) {};
		\node [style=red vertex] (70) at (-8, 10.5) {};
		\node [style=red vertex] (71) at (-8, 8.75) {};
		\node [style=red vertex] (72) at (-8, 7) {};
		\node [style=red vertex] (73) at (-8, 5.25) {};
		\node [style=red vertex] (74) at (-8, 3.5) {};
		\node [style=red vertex] (75) at (6.25, 12.25) {};
		\node [style=red vertex] (76) at (6.25, 10.5) {};
		\node [style=red vertex] (77) at (6.25, 8.75) {};
		\node [style=red vertex] (78) at (6.25, 7) {};
		\node [style=red vertex] (79) at (6.25, 5.25) {};
		\node [style=red vertex] (80) at (6.25, 3.5) {};
		\node [style=red vertex] (81) at (4.75, 12.25) {};
		\node [style=red vertex] (82) at (4.75, 10.5) {};
		\node [style=red vertex] (83) at (4.75, 8.75) {};
		\node [style=red vertex] (84) at (4.75, 7) {};
		\node [style=red vertex] (85) at (4.75, 5.25) {};
		\node [style=red vertex] (86) at (4.75, 3.5) {};
		\node [style=red vertex] (87) at (7.25, 0.75) {};
		\node [style=red vertex] (88) at (7.25, -1.25) {};
		\node [style=red vertex] (89) at (7.25, -3.5) {};
		\node [style=red vertex] (90) at (7.25, -5.5) {};
		\node [style=red vertex] (91) at (7.25, -7.5) {};
		\node [style=red vertex] (92) at (7.25, -9.5) {};
		\node [style=red vertex] (93) at (7.25, -11.5) {};
		\node [style=newstyle] (94) at (7.5, 0.75) {$\pi,x_6$};
		\node [style=newstyle] (95) at (7.5, -1.25) {$\pi,x_5$};
		\node [style=newstyle] (96) at (7.5, -3.5) {$\pi,x_4$};
		\node [style=newstyle] (97) at (7.5, -5.5) {$\pi,x_3$};
		\node [style=newstyle] (98) at (7.5, -7.5) {$\pi,x_2$};
		\node [style=newstyle] (99) at (7.5, -9.5) {$\pi,x_1$};
		\node [style=newstyle] (100) at (5, 12.25) {$\pi,f$};
		\node [style=newstyle] (101) at (5, 10.5) {$\pi,e$};
		\node [style=newstyle] (102) at (5, 8.75) {$\pi,d$};
		\node [style=newstyle] (103) at (5, 7) {$\pi,c$};
		\node [style=newstyle] (104) at (5, 5.25) {$\pi,b$};
		\node [style=newstyle] (105) at (5, 3.5) {$\pi,a$};
		\node [style=newstyle] (106) at (7.5, -11.5) {$\pi,x_0$};
		\node [style=green vertex] (107) at (5, -11.5) {};
		\node [style=green vertex] (108) at (5, -9.5) {};
		\node [style=green vertex] (109) at (5, -7.5) {};
		\node [style=green vertex] (110) at (5, -5.5) {};
		\node [style=green vertex] (111) at (5, -3.5) {};
		\node [style=green vertex] (112) at (5, -1.25) {};
		\node [style=green vertex] (113) at (5, 0.75) {};
		\node [style=newgreen] (114) at (5.25, 0.75) {$\pi, z_6$};
		\node [style=newgreen] (115) at (5.25, -1.25) {$\pi,z_5$};
		\node [style=newgreen] (116) at (5.25, -3.5) {$\pi,z_4$};
		\node [style=newgreen] (117) at (5.25, -5.5) {$\pi,z_3$};
		\node [style=newgreen] (118) at (5.25, -7.5) {$\pi,z_2$};
		\node [style=newgreen] (119) at (5.25, -9.5) {$\pi,z_1$};
		\node [style=newgreen] (120) at (5.25, -11.5) {$\pi,z_0$};
	\end{pgfonlayer}
	\begin{pgfonlayer}{edgelayer}
		\draw (5) to (11);
		\draw (11) to (12);
		\draw (13) to (10);
		\draw (10) to (4);
		\draw (3) to (9);
		\draw (9) to (14);
		\draw (15) to (8);
		\draw (8) to (2);
		\draw (1) to (7);
		\draw (7) to (16);
		\draw (17) to (6);
		\draw (6) to (0);
		\draw (18.center) to (31);
		\draw (31) to (35);
		\draw (35) to (36);
		\draw (36) to (44);
		\draw (44) to (48);
		\draw (48) to (52);
		\draw (52) to (25.center);
		\draw (26.center) to (49);
		\draw (49) to (45);
		\draw (45) to (37);
		\draw (37) to (32);
		\draw (32) to (19.center);
		\draw (21.center) to (33);
		\draw (33) to (39);
		\draw (39) to (46);
		\draw (46) to (53);
		\draw (53) to (27.center);
		\draw (28.center) to (54);
		\draw (50) to (54);
		\draw (50) to (40);
		\draw (40) to (38);
		\draw (38) to (22.center);
		\draw (23.center) to (41);
		\draw (29.center) to (55);
		\draw (41) to (55);
		\draw (51) to (30.center);
		\draw (43.center) to (47);
		\draw (47) to (34);
		\draw (34) to (56.center);
		\draw (24.center) to (42);
		\draw (42) to (51);
		\draw (6) to (31);
		\draw (6) to (32);
		\draw (33) to (6);
		\draw (34) to (6);
		\draw (7) to (35);
		\draw (37) to (7);
		\draw (7) to (38);
		\draw (42) to (7);
		\draw (8) to (36);
		\draw (8) to (39);
		\draw (8) to (40);
		\draw (8) to (41);
		\draw (9) to (44);
		\draw (9) to (45);
		\draw (46) to (9);
		\draw (10) to (48);
		\draw (49) to (10);
		\draw (50) to (10);
		\draw (52) to (11);
		\draw (11) to (53);
		\draw (54) to (11);
		\draw (11) to (55);
		\draw (51) to (10);
		\draw (9) to (47);
		\draw (69) to (5);
		\draw (70) to (4);
		\draw (71) to (3);
		\draw (72) to (2);
		\draw (73) to (1);
		\draw (74) to (0);
		\draw (12) to (75);
		\draw (76) to (13);
		\draw (14) to (77);
		\draw (78) to (15);
		\draw (79) to (16);
		\draw (17) to (80);
	\end{pgfonlayer}
\end{tikzpicture}}
\endpgfgraphicnamed}
  \caption{The correcting circuit in the \zxcalculus}
  \label{fig:zx-corrector}
\end{figure}

We translate the circuit into the \zxcalculus as shown in
Figure~\ref{fig:zx-corrector}.  The variables $a,b,\ldots,f$ indicate
the outcomes of the six syndrome measurments, while the variables
$x_i$ and $z_i$ indicate whether a Pauli $X$ or $Z$ correction need be
applied on qubit $i$ of the codeword.

This diagram can be substantially simplified; to do so we rely on some
easy lemmas:

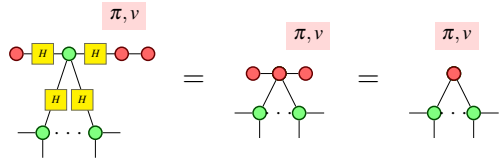
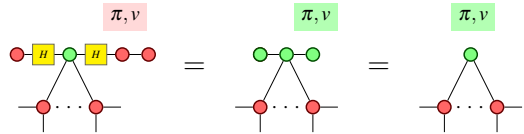
\begin{lemma}\label{lem:1}
  \begin{equation*}
    \scalebox{0.7}{\ilemmai} = \scalebox{0.7}{\ilemmaii} 
    = \scalebox{0.7}{\ilemmaiii}
  \end{equation*}
\end{lemma}

\begin{lemma}\label{lem:2}
  \begin{equation*}
    \scalebox{0.7}{\iilemmai} = \scalebox{0.7}{\iilemmaii} 
    = \scalebox{0.7}{\iilemmaiii}
  \end{equation*}
\end{lemma}

\begin{figure}[htbp]
  \centering
  \inline{%
\beginpgfgraphicnamed{cor-ii}
\begin{tikzpicture}
	\begin{pgfonlayer}{nodelayer}
		\node [style=red vertex] (0) at (-7.75, 1.5) {};
		\node [style=red vertex] (1) at (-5.75, 1.5) {};
		\node [style=red vertex] (2) at (-4, 1.5) {};
		\node [style=green vertex] (3) at (-2, 1.5) {};
		\node [style=green vertex] (4) at (0, 1.5) {};
		\node [style=green vertex] (5) at (2, 1.5) {};
		\node [style=none] (6) at (-9.25, 0) {};
		\node [style=none] (7) at (-9.25, -2) {};
		\node [style=none] (8) at (-9.25, -2) {};
		\node [style=none] (9) at (-9.25, -4.25) {};
		\node [style=none] (10) at (-9.25, -6.25) {};
		\node [style=none] (11) at (-9.25, -8.25) {};
		\node [style=none] (12) at (-9.25, -10.25) {};
		\node [style=none] (13) at (7.5, 0) {};
		\node [style=none] (14) at (7.5, -2) {};
		\node [style=none] (15) at (7.5, -4.25) {};
		\node [style=none] (16) at (7.5, -6.25) {};
		\node [style=none] (17) at (7.5, -8.25) {};
		\node [style=none] (18) at (7.5, -10.25) {};
		\node [style=green vertex] (19) at (-8.5, 0) {};
		\node [style=green vertex] (20) at (-8.25, -2) {};
		\node [style=green vertex] (21) at (-7.75, -4.25) {};
		\node [style=green vertex] (22) at (-6.25, -12.25) {};
		\node [style=green vertex] (23) at (-6.25, 0) {};
		\node [style=green vertex] (24) at (-4.25, 0) {};
		\node [style=green vertex] (25) at (-6, -2) {};
		\node [style=green vertex] (26) at (-5.5, -6.25) {};
		\node [style=green vertex] (27) at (-4, -4.25) {};
		\node [style=green vertex] (28) at (-3.5, -6.25) {};
		\node [style=green vertex] (29) at (-2.75, -8.25) {};
		\node [style=green vertex] (30) at (-4.5, -10.25) {};
		\node [style=none] (31) at (7.5, -12.25) {};
		\node [style=red vertex] (32) at (-3, 0) {};
		\node [style=red vertex] (33) at (-2.5, -2) {};
		\node [style=red vertex] (34) at (-2, -4.25) {};
		\node [style=red vertex] (35) at (-0.5, -12.25) {};
		\node [style=red vertex] (36) at (-1, 0) {};
		\node [style=red vertex] (37) at (-0.75, -2) {};
		\node [style=red vertex] (38) at (-0.25, -6.25) {};
		\node [style=red vertex] (39) at (0.75, -10.25) {};
		\node [style=red vertex] (40) at (1.25, 0) {};
		\node [style=red vertex] (41) at (1, -4.25) {};
		\node [style=red vertex] (42) at (1.75, -6.25) {};
		\node [style=red vertex] (43) at (2.25, -8.25) {};
		\node [style=none] (44) at (-9.25, -12.25) {};
		\node [style=red vertex] (45) at (5.75, 0) {};
		\node [style=red vertex] (46) at (5.75, -2) {};
		\node [style=red vertex] (47) at (5.75, -4.25) {};
		\node [style=red vertex] (48) at (5.75, -6.25) {};
		\node [style=red vertex] (49) at (5.75, -8.25) {};
		\node [style=red vertex] (50) at (5.75, -10.25) {};
		\node [style=red vertex] (51) at (5.75, -12.25) {};
		\node [style=newstyle] (52) at (6, 0) {$\pi,x_6$};
		\node [style=newstyle] (53) at (6, -2) {$\pi,x_5$};
		\node [style=newstyle] (54) at (6, -4.25) {$\pi,x_4$};
		\node [style=newstyle] (55) at (6, -6.25) {$\pi,x_3$};
		\node [style=newstyle] (56) at (6, -8.25) {$\pi,x_2$};
		\node [style=newstyle] (57) at (6, -10.25) {$\pi,x_1$};
		\node [style=newgreen] (58) at (2.25, 1.5) {$\pi,f$};
		\node [style=newgreen] (59) at (0.25, 1.5) {$\pi,e$};
		\node [style=newgreen] (60) at (-1.75, 1.5) {$\pi,d$};
		\node [style=newstyle] (61) at (-3.75, 1.5) {$\pi,c$};
		\node [style=newstyle] (62) at (-5.5, 1.5) {$\pi,b$};
		\node [style=newstyle] (63) at (-7.5, 1.5) {$\pi,a$};
		\node [style=newstyle] (64) at (6, -12.25) {$\pi,x_0$};
		\node [style=green vertex] (65) at (3.5, -12.25) {};
		\node [style=green vertex] (66) at (3.5, -10.25) {};
		\node [style=green vertex] (67) at (3.5, -8.25) {};
		\node [style=green vertex] (68) at (3.5, -6.25) {};
		\node [style=green vertex] (69) at (3.5, -4.25) {};
		\node [style=green vertex] (70) at (3.5, -2) {};
		\node [style=green vertex] (71) at (3.5, 0) {};
		\node [style=newgreen] (72) at (3.75, 0) {$\pi, z_6$};
		\node [style=newgreen] (73) at (3.75, -2) {$\pi,z_5$};
		\node [style=newgreen] (74) at (3.75, -4.25) {$\pi,z_4$};
		\node [style=newgreen] (75) at (3.75, -6.25) {$\pi,z_3$};
		\node [style=newgreen] (76) at (3.75, -8.25) {$\pi,z_2$};
		\node [style=newgreen] (77) at (3.75, -10.25) {$\pi,z_1$};
		\node [style=newgreen] (78) at (3.75, -12.25) {$\pi,z_0$};
	\end{pgfonlayer}
	\begin{pgfonlayer}{edgelayer}
		\draw (6.center) to (19);
		\draw (19) to (23);
		\draw (23) to (24);
		\draw (24) to (32);
		\draw (32) to (36);
		\draw (36) to (40);
		\draw (40) to (13.center);
		\draw (14.center) to (37);
		\draw (37) to (33);
		\draw (33) to (25);
		\draw (25) to (20);
		\draw (20) to (7.center);
		\draw (9.center) to (21);
		\draw (21) to (27);
		\draw (27) to (34);
		\draw (34) to (41);
		\draw (41) to (15.center);
		\draw (16.center) to (42);
		\draw (38) to (42);
		\draw (38) to (28);
		\draw (28) to (26);
		\draw (26) to (10.center);
		\draw (11.center) to (29);
		\draw (17.center) to (43);
		\draw (29) to (43);
		\draw (39) to (18.center);
		\draw (31.center) to (35);
		\draw (35) to (22);
		\draw (22) to (44.center);
		\draw (12.center) to (30);
		\draw (30) to (39);
		\draw (0) to (19);
		\draw (0) to (20);
		\draw (21) to (0);
		\draw (22) to (0);
		\draw (1) to (23);
		\draw (25) to (1);
		\draw (1) to (26);
		\draw (30) to (1);
		\draw (2) to (24);
		\draw (2) to (27);
		\draw (2) to (28);
		\draw (2) to (29);
		\draw (3) to (32);
		\draw (3) to (33);
		\draw (34) to (3);
		\draw (4) to (36);
		\draw (37) to (4);
		\draw (38) to (4);
		\draw (40) to (5);
		\draw (5) to (41);
		\draw (42) to (5);
		\draw (5) to (43);
		\draw (39) to (4);
		\draw (3) to (35);
	\end{pgfonlayer}
\end{tikzpicture}}
\endpgfgraphicnamed}  
  \caption{The simplified correcting circuit}
  \label{fig:-zx-simp-corrector}
\end{figure}
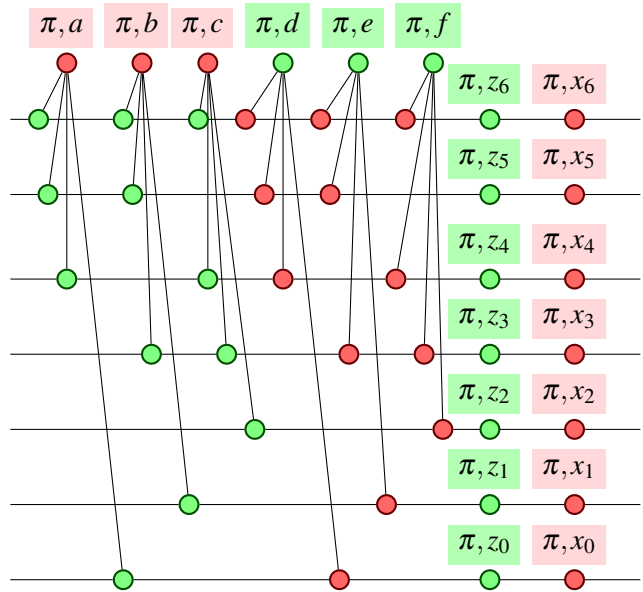

By  three applications of each lemma, the correcting circuit rewrites
to the diagram in Figure~\ref{fig:-zx-simp-corrector}.  Notice that
this is not a minimal form, however it is well adapted to the analysis of
the next section.

%%%%%%%%%%%%%%%%%%%%%%%%correction

\section{Verification}
\label{sec:errors-their-corr}

In this section we will demonstrate the correctness of the Steane
error-correcting code.  There are two parts to this argument.  Firstly
we show that the correcting circuit can remove single qubit Pauli
errors; and secondly, we show that the composition of encoder,
corrector, and decoder will reduce to the identity for one qubit.

Consider again the error-correcting circuit in
Figure~\ref{fig:-zx-simp-corrector}.  The Pauli $Z$ on qubit $i$ must
be activated exactly when there is a $Z$-error on qubit $i$:
therefore the variable $z_i$ must be true exactly in this case.
However the only  available information are the syndrome
measurements $a,b,\ldots,f$.  Therefore the first step is to derive
the value of the $z_i$ and $x_i$ in terms of these variables.

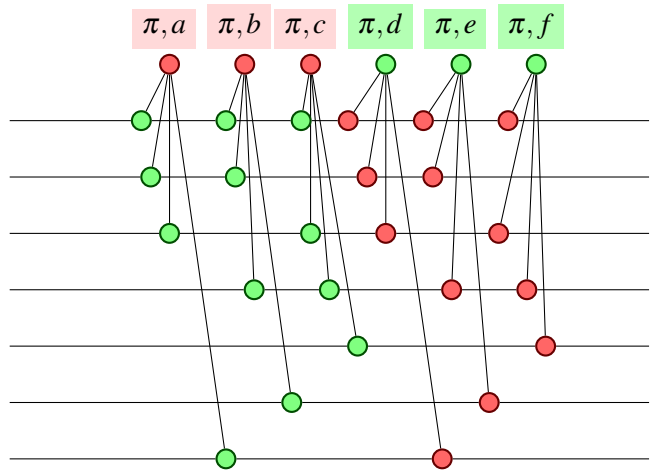
\begin{figure}[htbp]
  \centering
  \inline{%
\beginpgfgraphicnamed{cor-iii}
\begin{tikzpicture}
	\begin{pgfonlayer}{nodelayer}
		\node [style=red vertex] (0) at (-3.75, 1.5) {};
		\node [style=red vertex] (1) at (-1.75, 1.5) {};
		\node [style=red vertex] (2) at (0, 1.5) {};
		\node [style=green vertex] (3) at (2, 1.5) {};
		\node [style=green vertex] (4) at (4, 1.5) {};
		\node [style=green vertex] (5) at (6, 1.5) {};
		\node [style=none] (6) at (-8, 0) {};
		\node [style=none] (7) at (-8, -1.5) {};
		\node [style=none] (8) at (-8, -1.5) {};
		\node [style=none] (9) at (-8, -3) {};
		\node [style=none] (10) at (-8, -4.5) {};
		\node [style=none] (11) at (-8, -6) {};
		\node [style=none] (12) at (-8, -7.5) {};
		\node [style=none] (13) at (9, 0) {};
		\node [style=none] (14) at (9, -1.5) {};
		\node [style=none] (15) at (9, -3) {};
		\node [style=none] (16) at (9, -4.5) {};
		\node [style=none] (17) at (9, -6) {};
		\node [style=none] (18) at (9, -7.5) {};
		\node [style=green vertex] (19) at (-4.5, 0) {};
		\node [style=green vertex] (20) at (-4.25, -1.5) {};
		\node [style=green vertex] (21) at (-3.75, -3) {};
		\node [style=green vertex] (22) at (-2.25, -9) {};
		\node [style=green vertex] (23) at (-2.25, 0) {};
		\node [style=green vertex] (24) at (-0.25, 0) {};
		\node [style=green vertex] (25) at (-2, -1.5) {};
		\node [style=green vertex] (26) at (-1.5, -4.5) {};
		\node [style=green vertex] (27) at (0, -3) {};
		\node [style=green vertex] (28) at (0.5, -4.5) {};
		\node [style=green vertex] (29) at (1.25, -6) {};
		\node [style=green vertex] (30) at (-0.5, -7.5) {};
		\node [style=none] (31) at (9, -9) {};
		\node [style=red vertex] (32) at (1, 0) {};
		\node [style=red vertex] (33) at (1.5, -1.5) {};
		\node [style=red vertex] (34) at (2, -3) {};
		\node [style=red vertex] (35) at (3.5, -9) {};
		\node [style=red vertex] (36) at (3, 0) {};
		\node [style=red vertex] (37) at (3.25, -1.5) {};
		\node [style=red vertex] (38) at (3.75, -4.5) {};
		\node [style=red vertex] (39) at (4.75, -7.5) {};
		\node [style=red vertex] (40) at (5.25, 0) {};
		\node [style=red vertex] (41) at (5, -3) {};
		\node [style=red vertex] (42) at (5.75, -4.5) {};
		\node [style=red vertex] (43) at (6.25, -6) {};
		\node [style=none] (44) at (-8, -9) {};
		\node [style=newgreen] (45) at (6.25, 1.5) {$\pi,f$};
		\node [style=newgreen] (46) at (4.25, 1.5) {$\pi,e$};
		\node [style=newgreen] (47) at (2.25, 1.5) {$\pi,d$};
		\node [style=newstyle] (48) at (0.25, 1.5) {$\pi,c$};
		\node [style=newstyle] (49) at (-1.5, 1.5) {$\pi,b$};
		\node [style=newstyle] (50) at (-3.5, 1.5) {$\pi,a$};
	\end{pgfonlayer}
	\begin{pgfonlayer}{edgelayer}
		\draw (6.center) to (19);
		\draw (19) to (23);
		\draw (23) to (24);
		\draw (24) to (32);
		\draw (32) to (36);
		\draw (36) to (40);
		\draw (40) to (13.center);
		\draw (14.center) to (37);
		\draw (37) to (33);
		\draw (33) to (25);
		\draw (25) to (20);
		\draw (20) to (7.center);
		\draw (9.center) to (21);
		\draw (21) to (27);
		\draw (27) to (34);
		\draw (34) to (41);
		\draw (41) to (15.center);
		\draw (16.center) to (42);
		\draw (38) to (42);
		\draw (38) to (28);
		\draw (28) to (26);
		\draw (26) to (10.center);
		\draw (11.center) to (29);
		\draw (17.center) to (43);
		\draw (29) to (43);
		\draw (39) to (18.center);
		\draw (31.center) to (35);
		\draw (35) to (22);
		\draw (22) to (44.center);
		\draw (12.center) to (30);
		\draw (30) to (39);
		\draw (0) to (19);
		\draw (0) to (20);
		\draw (21) to (0);
		\draw (22) to (0);
		\draw (1) to (23);
		\draw (25) to (1);
		\draw (1) to (26);
		\draw (30) to (1);
		\draw (2) to (24);
		\draw (2) to (27);
		\draw (2) to (28);
		\draw (2) to (29);
		\draw (3) to (32);
		\draw (3) to (33);
		\draw (34) to (3);
		\draw (4) to (36);
		\draw (37) to (4);
		\draw (38) to (4);
		\draw (40) to (5);
		\draw (5) to (41);
		\draw (42) to (5);
		\draw (5) to (43);
		\draw (39) to (4);
		\draw (3) to (35);
	\end{pgfonlayer}
\end{tikzpicture}}
\endpgfgraphicnamed}
  \caption{The error-detecting circuit}
\label{fig:error-detection}
\end{figure}

Note the colour-symmetry of the error-detection sub-circuit
(Figure~\ref{fig:error-detection}). From this we can immediately see
that the first three measurements are resonsible for detecting $X$
errors, while the remaing three are responsible for $Z$ errors.  The
variables $a,b,c$ then form the $X$-syndrome, and $d,e,f$ the
$Z$-syndrome.  Viewing the each sub-syndrome as a bit-string, the
eight possible values correspond to the 7 qubits of the codeword, and
the possibility ``no error''.

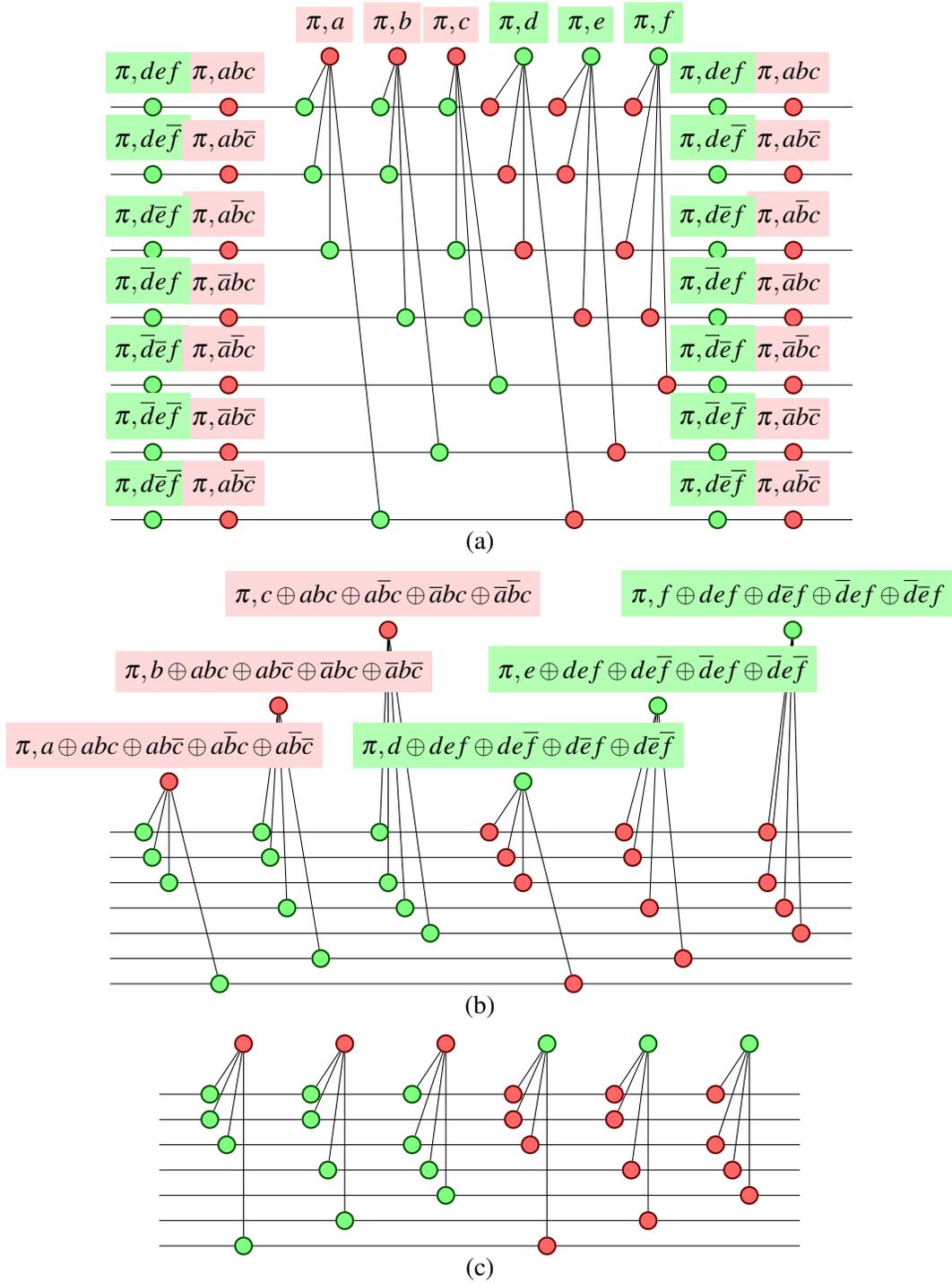
\begin{figure}[htbp]
  \centering
  \makebox[\textwidth]{\inline{%
\beginpgfgraphicnamed{cor-iv}
\begin{tikzpicture}
	\begin{pgfonlayer}{nodelayer}
		\node [style=red vertex] (0) at (9.5, 13.5) {};
		\node [style=red vertex] (1) at (11.5, 13.5) {};
		\node [style=red vertex] (2) at (13.25, 13.5) {};
		\node [style=green vertex] (3) at (15.25, 13.5) {};
		\node [style=green vertex] (4) at (17.25, 13.5) {};
		\node [style=green vertex] (5) at (19.25, 13.5) {};
		\node [style=none] (6) at (3, 12) {};
		\node [style=none] (7) at (3, 10) {};
		\node [style=none] (8) at (3, 10) {};
		\node [style=none] (9) at (3, 7.75) {};
		\node [style=none] (10) at (3, 5.75) {};
		\node [style=none] (11) at (3, 3.75) {};
		\node [style=none] (12) at (3, 1.75) {};
		\node [style=none] (13) at (25, 12) {};
		\node [style=none] (14) at (25, 10) {};
		\node [style=none] (15) at (25, 7.75) {};
		\node [style=none] (16) at (25, 5.75) {};
		\node [style=none] (17) at (25, 3.75) {};
		\node [style=none] (18) at (25, 1.75) {};
		\node [style=green vertex] (19) at (8.75, 12) {};
		\node [style=green vertex] (20) at (9, 10) {};
		\node [style=green vertex] (21) at (9.5, 7.75) {};
		\node [style=green vertex] (22) at (11, -0.25) {};
		\node [style=green vertex] (23) at (11, 12) {};
		\node [style=green vertex] (24) at (13, 12) {};
		\node [style=green vertex] (25) at (11.25, 10) {};
		\node [style=green vertex] (26) at (11.75, 5.75) {};
		\node [style=green vertex] (27) at (13.25, 7.75) {};
		\node [style=green vertex] (28) at (13.75, 5.75) {};
		\node [style=green vertex] (29) at (14.5, 3.75) {};
		\node [style=green vertex] (30) at (12.75, 1.75) {};
		\node [style=none] (31) at (25, -0.25) {};
		\node [style=red vertex] (32) at (14.25, 12) {};
		\node [style=red vertex] (33) at (14.75, 10) {};
		\node [style=red vertex] (34) at (15.25, 7.75) {};
		\node [style=red vertex] (35) at (16.75, -0.25) {};
		\node [style=red vertex] (36) at (16.25, 12) {};
		\node [style=red vertex] (37) at (16.5, 10) {};
		\node [style=red vertex] (38) at (17, 5.75) {};
		\node [style=red vertex] (39) at (18, 1.75) {};
		\node [style=red vertex] (40) at (18.5, 12) {};
		\node [style=red vertex] (41) at (18.25, 7.75) {};
		\node [style=red vertex] (42) at (19, 5.75) {};
		\node [style=red vertex] (43) at (19.5, 3.75) {};
		\node [style=none] (44) at (3, -0.25) {};
		\node [style=red vertex] (45) at (23.25, 12) {};
		\node [style=red vertex] (46) at (23.25, 10) {};
		\node [style=red vertex] (47) at (23.25, 7.75) {};
		\node [style=red vertex] (48) at (23.25, 5.75) {};
		\node [style=red vertex] (49) at (23.25, 3.75) {};
		\node [style=red vertex] (50) at (23.25, 1.75) {};
		\node [style=red vertex] (51) at (23.25, -0.25) {};
		\node [style=newstyle] (52) at (23.5, 12) {$\pi,abc$};
		\node [style=newstyle] (53) at (23.5, 10) {$\pi,ab\overline c$};
		\node [style=newstyle] (54) at (23.5, 7.75) {$\pi,a\overline b c$};
		\node [style=newstyle] (55) at (23.5, 5.75) {$\pi,\overline a bc$};
		\node [style=newstyle] (56) at (23.5, 3.75) {$\pi,\overline a \overline b c$};
		\node [style=newstyle] (57) at (23.5, 1.75) {$\pi,\overline a b \overline c$};
		\node [style=newgreen] (58) at (19.5, 13.5) {$\pi,f$};
		\node [style=newgreen] (59) at (17.5, 13.5) {$\pi,e$};
		\node [style=newgreen] (60) at (15.5, 13.5) {$\pi,d$};
		\node [style=newstyle] (61) at (13.5, 13.5) {$\pi,c$};
		\node [style=newstyle] (62) at (11.75, 13.5) {$\pi,b$};
		\node [style=newstyle] (63) at (9.75, 13.5) {$\pi,a$};
		\node [style=newstyle] (64) at (23.5, -0.25) {$\pi,a\overline b \overline c$};
		\node [style=green vertex] (65) at (21, -0.25) {};
		\node [style=green vertex] (66) at (21, 1.75) {};
		\node [style=green vertex] (67) at (21, 3.75) {};
		\node [style=green vertex] (68) at (21, 5.75) {};
		\node [style=green vertex] (69) at (21, 7.75) {};
		\node [style=green vertex] (70) at (21, 10) {};
		\node [style=green vertex] (71) at (21, 12) {};
		\node [style=newgreen] (72) at (21.25, 12) {$\pi, def$};
		\node [style=newgreen] (73) at (21.25, 10) {$\pi,de\overline f$};
		\node [style=newgreen] (74) at (21.25, 7.75) {$\pi,d\overline e f$};
		\node [style=newgreen] (75) at (21.25, 5.75) {$\pi,\overline d e f$};
		\node [style=newgreen] (76) at (21.25, 3.75) {$\pi,\overline d \overline e f$};
		\node [style=newgreen] (77) at (21.25, 1.75) {$\pi,\overline d e \overline f$};
		\node [style=newgreen] (78) at (21.25, -0.25) {$\pi,d \overline e \overline f$};
		\node [style=red vertex] (79) at (6.5, 1.75) {};
		\node [style=green vertex] (80) at (4.25, 10) {};
		\node [style=green vertex] (81) at (4.25, 1.75) {};
		\node [style=newgreen] (82) at (4.5, 10) {$\pi,de\overline f$};
		\node [style=red vertex] (83) at (6.5, -0.25) {};
		\node [style=green vertex] (84) at (4.25, 3.75) {};
		\node [style=newgreen] (85) at (4.5, 3.75) {$\pi,\overline d \overline e f$};
		\node [style=red vertex] (86) at (6.5, 3.75) {};
		\node [style=green vertex] (87) at (4.25, -0.25) {};
		\node [style=newstyle] (88) at (6.75, 12) {$\pi,abc$};
		\node [style=red vertex] (89) at (6.5, 10) {};
		\node [style=newgreen] (90) at (4.5, -0.25) {$\pi,d \overline e \overline f$};
		\node [style=newstyle] (91) at (6.75, 3.75) {$\pi,\overline a \overline b c$};
		\node [style=green vertex] (92) at (4.25, 12) {};
		\node [style=newstyle] (93) at (6.75, 5.75) {$\pi,\overline a bc$};
		\node [style=green vertex] (94) at (4.25, 7.75) {};
		\node [style=newstyle] (95) at (6.75, 10) {$\pi,ab\overline c$};
		\node [style=newstyle] (96) at (6.75, 1.75) {$\pi,\overline a b \overline c$};
		\node [style=red vertex] (97) at (6.5, 7.75) {};
		\node [style=green vertex] (98) at (4.25, 5.75) {};
		\node [style=newgreen] (99) at (4.5, 1.75) {$\pi,\overline d e \overline f$};
		\node [style=red vertex] (100) at (6.5, 5.75) {};
		\node [style=newstyle] (101) at (6.75, 7.75) {$\pi,a\overline b c$};
		\node [style=newgreen] (102) at (4.5, 7.75) {$\pi,d\overline e f$};
		\node [style=newstyle] (103) at (6.75, -0.25) {$\pi,a\overline b \overline c$};
		\node [style=newgreen] (104) at (4.5, 5.75) {$\pi,\overline d e f$};
		\node [style=newgreen] (105) at (4.5, 12) {$\pi, def$};
		\node [style=red vertex] (106) at (6.5, 12) {};
	\end{pgfonlayer}
	\begin{pgfonlayer}{edgelayer}
		\draw (6.center) to (19);
		\draw (19) to (23);
		\draw (23) to (24);
		\draw (24) to (32);
		\draw (32) to (36);
		\draw (36) to (40);
		\draw (40) to (13.center);
		\draw (14.center) to (37);
		\draw (37) to (33);
		\draw (33) to (25);
		\draw (25) to (20);
		\draw (20) to (7.center);
		\draw (9.center) to (21);
		\draw (21) to (27);
		\draw (27) to (34);
		\draw (34) to (41);
		\draw (41) to (15.center);
		\draw (16.center) to (42);
		\draw (38) to (42);
		\draw (38) to (28);
		\draw (28) to (26);
		\draw (26) to (10.center);
		\draw (11.center) to (29);
		\draw (17.center) to (43);
		\draw (29) to (43);
		\draw (39) to (18.center);
		\draw (31.center) to (35);
		\draw (35) to (22);
		\draw (22) to (44.center);
		\draw (12.center) to (30);
		\draw (30) to (39);
		\draw (0) to (19);
		\draw (0) to (20);
		\draw (21) to (0);
		\draw (22) to (0);
		\draw (1) to (23);
		\draw (25) to (1);
		\draw (1) to (26);
		\draw (30) to (1);
		\draw (2) to (24);
		\draw (2) to (27);
		\draw (2) to (28);
		\draw (2) to (29);
		\draw (3) to (32);
		\draw (3) to (33);
		\draw (34) to (3);
		\draw (4) to (36);
		\draw (37) to (4);
		\draw (38) to (4);
		\draw (40) to (5);
		\draw (5) to (41);
		\draw (42) to (5);
		\draw (5) to (43);
		\draw (39) to (4);
		\draw (3) to (35);
	\end{pgfonlayer}
\end{tikzpicture}}
\endpgfgraphicnamed}}\\
  (a)\\
  \medskip
  \makebox[\textwidth]{\inline{%
\beginpgfgraphicnamed{cor-v}
\begin{tikzpicture}
	\begin{pgfonlayer}{nodelayer}
		\node [style=red vertex] (0) at (3, 3.75) {};
		\node [style=red vertex] (1) at (6.25, 6) {};
		\node [style=red vertex] (2) at (9.5, 8.25) {};
		\node [style=green vertex] (3) at (13.5, 3.75) {};
		\node [style=green vertex] (4) at (17.5, 6) {};
		\node [style=green vertex] (5) at (21.5, 8.25) {};
		\node [style=none] (6) at (1.25, 2.25) {};
		\node [style=none] (7) at (1.25, 1.5) {};
		\node [style=none] (8) at (1.25, 1.5) {};
		\node [style=none] (9) at (1.25, 0.75) {};
		\node [style=none] (10) at (1.25, 0) {};
		\node [style=none] (11) at (1.25, -0.75) {};
		\node [style=none] (12) at (1.25, -1.5) {};
		\node [style=none] (13) at (23.25, 2.25) {};
		\node [style=none] (14) at (23.25, 1.5) {};
		\node [style=none] (15) at (23.25, 0.75) {};
		\node [style=none] (16) at (23.25, 0) {};
		\node [style=none] (17) at (23.25, -0.75) {};
		\node [style=none] (18) at (23.25, -1.5) {};
		\node [style=green vertex] (19) at (2.25, 2.25) {};
		\node [style=green vertex] (20) at (2.5, 1.5) {};
		\node [style=green vertex] (21) at (3, 0.75) {};
		\node [style=green vertex] (22) at (4.5, -2.25) {};
		\node [style=green vertex] (23) at (5.75, 2.25) {};
		\node [style=green vertex] (24) at (9.25, 2.25) {};
		\node [style=green vertex] (25) at (6, 1.5) {};
		\node [style=green vertex] (26) at (6.5, 0) {};
		\node [style=green vertex] (27) at (9.5, 0.75) {};
		\node [style=green vertex] (28) at (10, 0) {};
		\node [style=green vertex] (29) at (10.75, -0.75) {};
		\node [style=green vertex] (30) at (7.5, -1.5) {};
		\node [style=none] (31) at (23.25, -2.25) {};
		\node [style=red vertex] (32) at (12.5, 2.25) {};
		\node [style=red vertex] (33) at (13, 1.5) {};
		\node [style=red vertex] (34) at (13.5, 0.75) {};
		\node [style=red vertex] (35) at (15, -2.25) {};
		\node [style=red vertex] (36) at (16.5, 2.25) {};
		\node [style=red vertex] (37) at (16.75, 1.5) {};
		\node [style=red vertex] (38) at (17.25, 0) {};
		\node [style=red vertex] (39) at (18.25, -1.5) {};
		\node [style=red vertex] (40) at (20.75, 2.25) {};
		\node [style=red vertex] (41) at (20.75, 0.75) {};
		\node [style=red vertex] (42) at (21.25, 0) {};
		\node [style=red vertex] (43) at (21.75, -0.75) {};
		\node [style=none] (44) at (1.25, -2.25) {};
		\node [style=newgreen] (45) at (21.75, 8.25) {$\pi,f\oplus def \oplus d\overline e f \oplus \overline d e f \oplus \overline d \overline e f$};
		\node [style=newgreen] (46) at (17.75, 6) {$\pi,e \oplus def \oplus d e \overline f \oplus \overline d e f \oplus
\overline d  e \overline f$};
		\node [style=newgreen] (47) at (13.75, 3.75) {$\pi,d \oplus def \oplus d e \overline f \oplus  d \overline e f \oplus
 d  \overline e \overline f$};
		\node [style=newstyle] (48) at (9.75, 8.25) {$\pi,c\oplus abc \oplus a\overline b c \oplus \overline a b c \oplus \overline a \overline b c$};
		\node [style=newstyle] (49) at (6.5, 6) {$\pi,b \oplus abc \oplus a b \overline c \oplus \overline a b c \oplus
\overline a  b \overline c$};
		\node [style=newstyle] (50) at (3.25, 3.75) {$\pi,a \oplus abc \oplus a b \overline c \oplus  a \overline b c \oplus
 a  \overline b \overline c$};
	\end{pgfonlayer}
	\begin{pgfonlayer}{edgelayer}
		\draw (6.center) to (19);
		\draw (19) to (23);
		\draw (23) to (24);
		\draw (24) to (32);
		\draw (32) to (36);
		\draw (36) to (40);
		\draw (40) to (13.center);
		\draw (14.center) to (37);
		\draw (37) to (33);
		\draw (33) to (25);
		\draw (25) to (20);
		\draw (20) to (7.center);
		\draw (9.center) to (21);
		\draw (21) to (27);
		\draw (27) to (34);
		\draw (34) to (41);
		\draw (41) to (15.center);
		\draw (16.center) to (42);
		\draw (38) to (42);
		\draw (38) to (28);
		\draw (28) to (26);
		\draw (26) to (10.center);
		\draw (11.center) to (29);
		\draw (17.center) to (43);
		\draw (29) to (43);
		\draw (39) to (18.center);
		\draw (31.center) to (35);
		\draw (35) to (22);
		\draw (22) to (44.center);
		\draw (12.center) to (30);
		\draw (30) to (39);
		\draw (0) to (19);
		\draw (0) to (20);
		\draw (21) to (0);
		\draw (22) to (0);
		\draw (1) to (23);
		\draw (25) to (1);
		\draw (1) to (26);
		\draw (30) to (1);
		\draw (2) to (24);
		\draw (2) to (27);
		\draw (2) to (28);
		\draw (2) to (29);
		\draw (3) to (32);
		\draw (3) to (33);
		\draw (34) to (3);
		\draw (4) to (36);
		\draw (37) to (4);
		\draw (38) to (4);
		\draw (40) to (5);
		\draw (5) to (41);
		\draw (42) to (5);
		\draw (5) to (43);
		\draw (39) to (4);
		\draw (3) to (35);
	\end{pgfonlayer}
\end{tikzpicture}}
\endpgfgraphicnamed}}\\
  (b)\\
  \medskip
  \makebox[\textwidth]{\inline{%
\beginpgfgraphicnamed{cor-vi}
\begin{tikzpicture}
	\begin{pgfonlayer}{nodelayer}
		\node [style=red vertex] (0) at (5, 3.75) {};
		\node [style=red vertex] (1) at (8, 3.75) {};
		\node [style=red vertex] (2) at (11, 3.75) {};
		\node [style=green vertex] (3) at (14, 3.75) {};
		\node [style=green vertex] (4) at (17, 3.75) {};
		\node [style=green vertex] (5) at (20, 3.75) {};
		\node [style=none] (6) at (2.5, 2.25) {};
		\node [style=none] (7) at (2.5, 1.5) {};
		\node [style=none] (8) at (2.5, 1.5) {};
		\node [style=none] (9) at (2.5, 0.75) {};
		\node [style=none] (10) at (2.5, 0) {};
		\node [style=none] (11) at (2.5, -0.75) {};
		\node [style=none] (12) at (2.5, -1.5) {};
		\node [style=none] (13) at (21.5, 2.25) {};
		\node [style=none] (14) at (21.5, 1.5) {};
		\node [style=none] (15) at (21.5, 0.75) {};
		\node [style=none] (16) at (21.5, 0) {};
		\node [style=none] (17) at (21.5, -0.75) {};
		\node [style=none] (18) at (21.5, -1.5) {};
		\node [style=green vertex] (19) at (4, 2.25) {};
		\node [style=green vertex] (20) at (4, 1.5) {};
		\node [style=green vertex] (21) at (4.5, 0.75) {};
		\node [style=green vertex] (22) at (5, -2.25) {};
		\node [style=green vertex] (23) at (7, 2.25) {};
		\node [style=green vertex] (24) at (10, 2.25) {};
		\node [style=green vertex] (25) at (7, 1.5) {};
		\node [style=green vertex] (26) at (7.5, 0) {};
		\node [style=green vertex] (27) at (10, 0.75) {};
		\node [style=green vertex] (28) at (10.5, 0) {};
		\node [style=green vertex] (29) at (11, -0.75) {};
		\node [style=green vertex] (30) at (8, -1.5) {};
		\node [style=none] (31) at (21.5, -2.25) {};
		\node [style=red vertex] (32) at (13, 2.25) {};
		\node [style=red vertex] (33) at (13, 1.5) {};
		\node [style=red vertex] (34) at (13.5, 0.75) {};
		\node [style=red vertex] (35) at (14, -2.25) {};
		\node [style=red vertex] (36) at (16, 2.25) {};
		\node [style=red vertex] (37) at (16, 1.5) {};
		\node [style=red vertex] (38) at (16.5, 0) {};
		\node [style=red vertex] (39) at (17, -1.5) {};
		\node [style=red vertex] (40) at (19, 2.25) {};
		\node [style=red vertex] (41) at (19, 0.75) {};
		\node [style=red vertex] (42) at (19.5, 0) {};
		\node [style=red vertex] (43) at (20, -0.75) {};
		\node [style=none] (44) at (2.5, -2.25) {};
	\end{pgfonlayer}
	\begin{pgfonlayer}{edgelayer}
		\draw (6.center) to (19);
		\draw (19) to (23);
		\draw (23) to (24);
		\draw (24) to (32);
		\draw (32) to (36);
		\draw (36) to (40);
		\draw (40) to (13.center);
		\draw (14.center) to (37);
		\draw (37) to (33);
		\draw (33) to (25);
		\draw (25) to (20);
		\draw (20) to (7.center);
		\draw (9.center) to (21);
		\draw (21) to (27);
		\draw (27) to (34);
		\draw (34) to (41);
		\draw (41) to (15.center);
		\draw (16.center) to (42);
		\draw (38) to (42);
		\draw (38) to (28);
		\draw (28) to (26);
		\draw (26) to (10.center);
		\draw (11.center) to (29);
		\draw (17.center) to (43);
		\draw (29) to (43);
		\draw (39) to (18.center);
		\draw (31.center) to (35);
		\draw (35) to (22);
		\draw (22) to (44.center);
		\draw (12.center) to (30);
		\draw (30) to (39);
		\draw (0) to (19);
		\draw (0) to (20);
		\draw (21) to (0);
		\draw (22) to (0);
		\draw (1) to (23);
		\draw (25) to (1);
		\draw (1) to (26);
		\draw (30) to (1);
		\draw (2) to (24);
		\draw (2) to (27);
		\draw (2) to (28);
		\draw (2) to (29);
		\draw (3) to (32);
		\draw (3) to (33);
		\draw (34) to (3);
		\draw (4) to (36);
		\draw (37) to (4);
		\draw (38) to (4);
		\draw (40) to (5);
		\draw (5) to (41);
		\draw (42) to (5);
		\draw (5) to (43);
		\draw (39) to (4);
		\draw (3) to (35);
	\end{pgfonlayer}
\end{tikzpicture}}
\endpgfgraphicnamed}}\\
  (c)\\
  \caption{(a) The error-correcting circuit with errors. (b) The
    reduced form (c) The unconditional form. \textbf{NB}: we write $ab$
  for $a\wedge b$ and $\overline a$ for $\neg a$.}
  \label{fig:corrector-with-errors}
\end{figure}

As a typical example, we derive the value of $z_6$, indicating a
$Z$-error on qubit~6.  Ignoring the  other qubits of the codeword we
have the following situation:
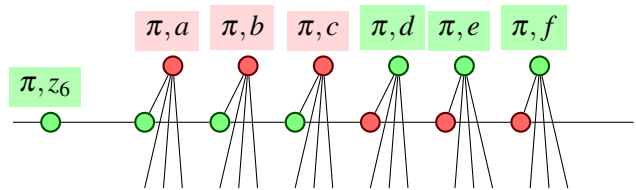
\ctikzfig{cor-syndrome-i}
By repeatedly applying the $\alpha$-commute, copying, and spider
rules, this can be rewritten as follows:
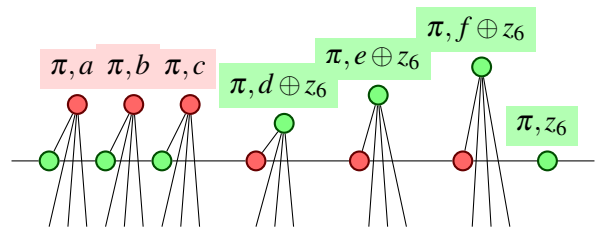
\ctikzfig{cor-syndrome-ii}
Now it is easy to see that the diagram will only be deterministic
(with respect to  $Z$ errors) on the condition that 
\[
(z_6 \leftrightarrow d) \wedge 
(z_6 \leftrightarrow e) \wedge 
(z_6 \leftrightarrow f)
\]
from which we derive $z_6 = d\wedge e \wedge f$. Similar analysis can
derive the values for the other variables, shown in the table below.
\begin{center}
\begin{tabular}{c|cc}
\hline 
Qubit $i$ & $x_i$  & $z_i$ \\
\hline
6 &$ a \wedge b \wedge c $&$ d \wedge e \wedge f $\\
5 &$ a \wedge b \wedge \neg c $&$ d \wedge e \wedge \neg f $\\
4 &$ a \wedge \neg b \wedge c $&$ d \wedge \neg e \wedge f $\\
3 &$ \neg a \wedge b \wedge c $&$ \neg d \wedge e \wedge f $\\
2 &$ \neg a \wedge \neg b \wedge c $&$ \neg d \wedge \neg e \wedge f $\\
1 &$ \neg a \wedge b \wedge \neg c $&$ \neg d \wedge e \wedge \neg f $\\
0 &$ a \wedge \neg b \wedge \neg c $&$ d \wedge \neg e \wedge \neg f $\\
\hline
\end{tabular}
\end{center}
Now we can write down the complete circuit, including the errors, as
shown in Figure~\ref{fig:corrector-with-errors} (a).  Notice that the
choice of conditional formulae for the errors enforces the desired error model.
By propagating the errors forward, as above, we arrive at the diagram
of Figure~\ref{fig:corrector-with-errors}~(b).  It is easily verified
that the boolean expressions in the remaining conditional vertices are
all uniformly false, hence they can be dropped.  Therefore we conclude
that the correcting circuit does indeed correct any of the claimed
errors.

The final thing we will check is that the composition of the encoding
circuit, the correcting circuit, and the decoding circuit combine to
yield the identity for a single qubit.  The combined graph is rather
large, so we will not show it here.  Quantomatic was not able to carry
out the proof completely automatically, and  several lemmas had to be
added as new rules.  However, with some human intervention the circuit
does indeed reduce to the desired form.  The proof can
be found in the appendix.

\section{Discussion}
\label{sec:discussion}

We have shown that Steane's 7-qubit error-correcting code performs
as advertised.  This is perhaps not surprising.  The main achievement
here was to carry out the proof in the automated rewriting system
Quantomatic.  Since any practical quantum program will have to be
implemented ``underneath'' an error-correcting code of some kind we
view this as a first step towards mechanically verifying practical
programs.  

Since (to our knowledge) this is the largest test yet carried out
using Quantomatic, a few comments are in order.  There were two
obstacles to this work.  Firstly, the \zxcalculus with conditional
vertices is not implemented in Quantomatic, so this part had to be
done either by hand or via case analysis.  In particular it was
necessary to simplify the boolean expressions.  It should be possible
to add such features to the program.  The more significant problem was
the lack of good strategies for fully automatic work.  Since many of
the rewrite rules must be used as expansions rather than reductions,
manual intervention is required.  As glance at the appendix will tell
the reader that we have already reached the stage where it is not easy
for the human operator to see what he is doing.

The Steane code is not optimal, in the sense that it uses more qubits
than the theoretical minimum of 5, which is attained by other codes.
The reason we chose to study Steane code is that quantum logic gates
can be implemented directly on the code-space, and these have a
particularly simple form.  The study of the encoded operations, and
the verification of encoded circuits is a subject for future work.

{\small
\bibliography{all}
}

\newpage
\appendix
\section{Putting it all together}
\label{sec:putting-it-all}

In this appendix, we describe the mechanised proof that the complete
Steane code is correct by composing the encoding circuit, the
unconditional diagram for the error-correcting circuit (derived in
Section~\ref{sec:errors-their-corr}), and the decoding circuit, and
then showing that this combined diagram reduces to the identity. This
calculation was done with Quantomatic.

\paragraph{Phase 1}
\label{sec:phase-1}

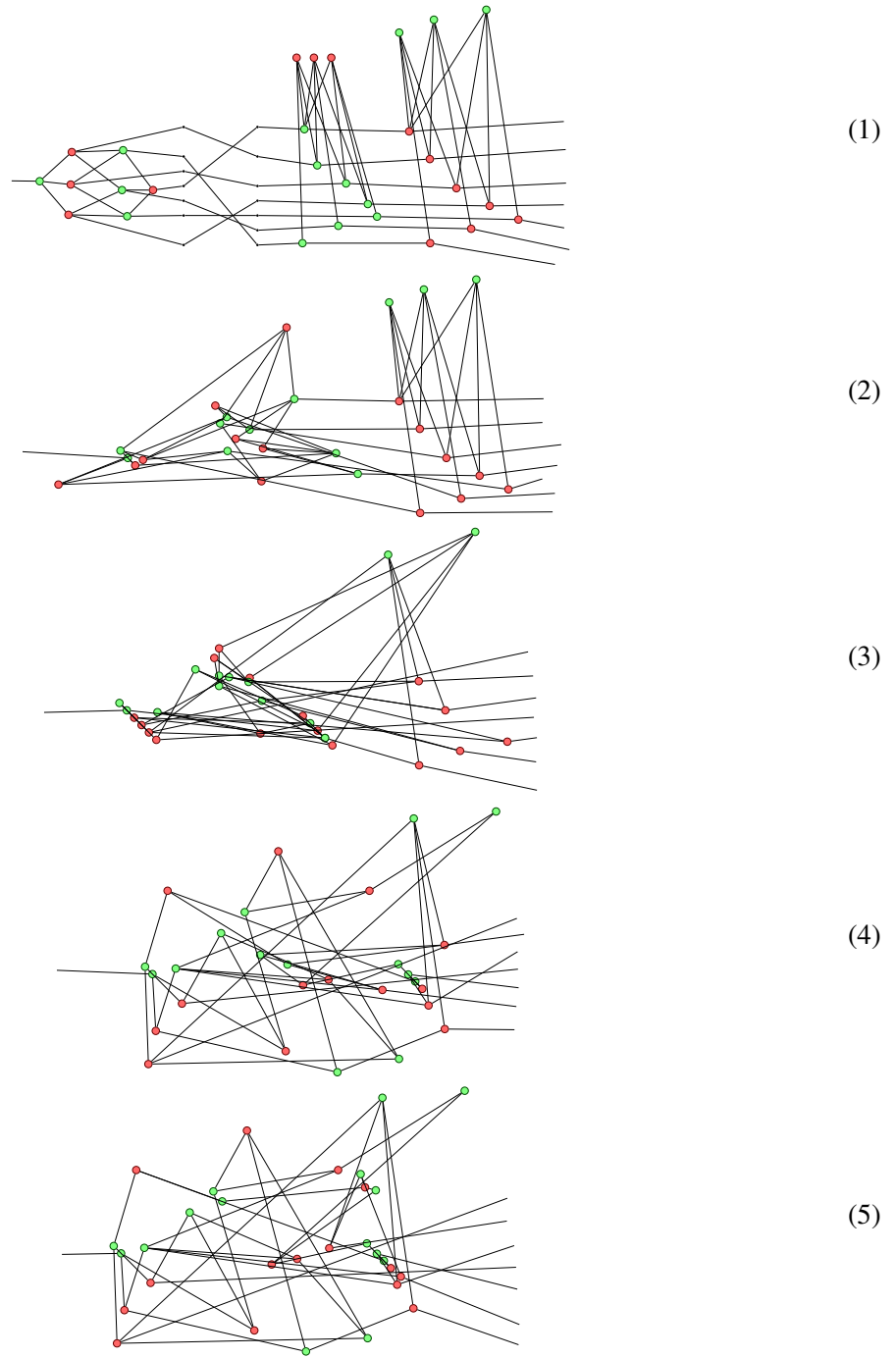
\begin{figure}[htbp]
  \centering
\begin{align}
&  \scalebox{0.4}{\makebox[\textwidth]{\inline{%
\beginpgfgraphicnamed{bigproof-i-bis}
\begin{tikzpicture} 
	\begin{pgfonlayer}{nodelayer}   
\node [style=none] (Vj) at (28.3,-29.65) {};
\node [style=none] (2) at (65.825,-32.85) {};
\node [style=red vertex] (q) at (37.925,-30.275) {};

\node [style=red vertex] (Va) at (47.65,-21.3) {};
\node [style=none] (4) at (65.95,-29.8) {};
\node [style=green] (h) at (51.0,-29.825) {};
\node [style=red vertex] (r) at (32.425,-27.7) {};
\node [style=red vertex] (Vau) at (59.475,-32.9) {};
\node [style=red vertex] (k) at (55.275,-26.3) {};
\node [style=red vertex] (s) at (32.35,-29.9) {};
\node [style=green] (n) at (35.9,-27.575) {};
\node [style=red vertex] (Vc) at (50.0,-21.3) {};
\node [style=green] (Val) at (53.1,-32.075) {};
\node [style=none] (5) at (65.925,-27.525) {};
\node [style=none] (3) at (65.8,-31.275) {};
\node [style=green] (i) at (52.475,-31.225) {};
\node [style=green] (Var) at (50.475,-32.7) {};
\node [style=red vertex] (Vba) at (56.7,-33.875) {};
\node [style=red vertex] (Vb) at (48.825,-21.3) {};
\node [style=green] (o) at (35.825,-30.275) {};
\node [style=red vertex] (j) at (56.675,-28.175) {};
\node [style=red vertex] (m) at (60.725,-31.35) {};
\node [style=green] (e) at (60.5,-18.05) {};
\node [style=none] (6) at (65.8,-25.625) {};
\node [style=green] (Vab) at (54.6,-19.6) {};
\node [style=none] (0) at (65.2,-35.325) {};
\node [style=green] (Vm) at (56.95,-18.725) {};
\node [style=green] (p) at (36.175,-32.05) {};
\node [style=green] (g) at (48.175,-26.15) {};
\node [style=green] (Vo) at (30.25,-29.675) {};
\node [style=red vertex] (l) at (58.475,-30.15) {};
\node [style=red vertex] (t) at (32.2,-31.95) {};
\node [style=red vertex] (Vao) at (62.675,-32.275) {};
\node [style=green] (Vax) at (48.05,-33.875) {};
\node [style=none] (1) at (66.175,-34.325) {};
\node [style=green] (f) at (49.05,-28.6) {};

\node [style=boundary vertex] (7) at (40,-28) {};
\node [style=boundary vertex] (8) at (40,-31) {};
\node [style=boundary vertex] (9) at (40,-32) {};
\node [style=boundary vertex] (10) at (40,-34) {};
\node [style=boundary vertex] (11) at (40,-29) {};
\node [style=boundary vertex] (12) at (40,-26) {};
\node [style=boundary vertex] (13) at (40,-30) {};

\node [style=boundary vertex] (Vaw) at (45,-34) {};
\node [style=boundary vertex] (Vaq) at (45,-33) {};
\node [style=boundary vertex] (Vak) at (45,-32) {};
\node [style=boundary vertex] (Vae) at (45,-31) {};
\node [style=boundary vertex] (Vbi) at (45,-30) {};
\node [style=boundary vertex] (Vbc) at (45,-28) {};
\node [style=boundary vertex] (Vbo) at (45,-26) {};

	\end{pgfonlayer}
	\begin{pgfonlayer}{edgelayer}

\draw [style=plain] (Vab) to (l);
\draw [style=plain] (Va) to (g);
\draw [style=plain] (Vaw) to (Vax);
\draw [style=plain] (4) to (l);
\draw [style=plain] (e) to (m);
\draw [style=plain] (r) to (n);
\draw [style=plain] (Vao) to (Val);
\draw [style=plain] (12) to (Vbc);
\draw [style=plain] (Vba) to (Vab);
\draw [style=plain] (q) to (o);
\draw [style=plain] (10) to (Vae);
\draw [style=plain] (Vau) to (1);
\draw [style=plain] (Vak) to (Val);
\draw [style=plain] (Vau) to (Vm);
\draw [style=plain] (Vb) to (g);
\draw [style=plain] (p) to (t);
\draw [style=plain] (5) to (j);
\draw [style=plain] (f) to (j);
\draw [style=plain] (f) to (Vbc);
\draw [style=plain] (o) to (t);
\draw [style=plain] (e) to (l);
\draw [style=plain] (Va) to (h);
\draw [style=plain] (h) to (Vc);
\draw [style=plain] (11) to (s);
\draw [style=plain] (h) to (l);
\draw [style=plain] (Vbi) to (11);
\draw [style=plain] (h) to (Vbi);
\draw [style=plain] (Vm) to (k);
\draw [style=plain] (Vax) to (Va);
\draw [style=plain] (9) to (Vak);
\draw [style=plain] (i) to (Vae);
\draw [style=plain] (13) to (Vbo);
\draw [style=plain] (7) to (n);
\draw [style=plain] (g) to (k);
\draw [style=plain] (Vc) to (g);
\draw [style=plain] (13) to (q);
\draw [style=plain] (Var) to (Vau);
\draw [style=plain] (7) to (Vaw);
\draw [style=plain] (3) to (m);
\draw [style=plain] (s) to (p);
\draw [style=plain] (Vaq) to (Var);
\draw [style=plain] (Va) to (f);
\draw [style=plain] (8) to (Vaq);
\draw [style=plain] (r) to (Vo);
\draw [style=plain] (Vbo) to (g);
\draw [style=plain] (s) to (Vo);
\draw [style=plain] (f) to (Vb);
\draw [style=plain] (Vm) to (m);
\draw [style=plain] (Vb) to (i);
\draw [style=plain] (Vb) to (Var);
\draw [style=plain] (j) to (Vab);
\draw [style=plain] (e) to (Vao);
\draw [style=plain] (Vao) to (2);
\draw [style=plain] (n) to (s);
\draw [style=plain] (q) to (n);
\draw [style=plain] (e) to (k);
\draw [style=plain] (j) to (Vm);
\draw [style=plain] (Vba) to (Vax);
\draw [style=plain] (r) to (o);
\draw [style=plain] (8) to (o);
\draw [style=plain] (Vo) to (t);
\draw [style=plain] (i) to (m);
\draw [style=plain] (q) to (p);
\draw [style=plain] (t) to (10);
\draw [style=plain] (Val) to (Vc);
\draw [style=plain] (Vc) to (i);
\draw [style=plain] (12) to (r);
\draw [style=plain] (Vab) to (k);
\draw [style=plain] (Vj) to (Vo);
\draw [style=plain] (9) to (p);
\draw [style=plain] (k) to (6);
\draw [style=plain] (Vba) to (0);
\end{pgfonlayer}
\end{tikzpicture}}
\endpgfgraphicnamed}}}\label{left1} \\
&  \scalebox{0.4}{\makebox[\textwidth]{\corleftiv}} \label{left4} \\
&  \scalebox{0.4}{\makebox[\textwidth]{\corleftv}} \label{left5}\\
&  \scalebox{0.4}{\makebox[\textwidth]{\corleftvi}} \label{left6}\\
&  \scalebox{0.4}{\makebox[\textwidth]{\corleftvii}} \label{left7}
\end{align}  
  \caption{Rewrite sequence 1}
  \label{fig:rwi}
\end{figure}

We begin by composing the encoder to the error-corrector:
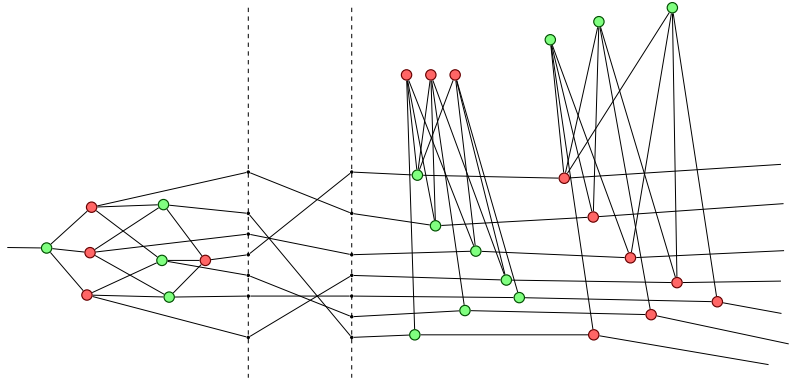
\begin{equation*}
    \scalebox{0.55}{\makebox[\textwidth]{\inline{%
\beginpgfgraphicnamed{bigproof-i}
\begin{tikzpicture} 
	\begin{pgfonlayer}{nodelayer}   
\node [style=none] (Vj) at (28.3,-29.65) {};
\node [style=none] (2) at (65.825,-32.85) {};
\node [style=red vertex] (q) at (37.925,-30.275) {};

\node [style=red vertex] (Va) at (47.65,-21.3) {};
\node [style=none] (4) at (65.95,-29.8) {};
\node [style=green] (h) at (51.0,-29.825) {};
\node [style=red vertex] (r) at (32.425,-27.7) {};
\node [style=red vertex] (Vau) at (59.475,-32.9) {};
\node [style=red vertex] (k) at (55.275,-26.3) {};
\node [style=red vertex] (s) at (32.35,-29.9) {};
\node [style=green] (n) at (35.9,-27.575) {};
\node [style=red vertex] (Vc) at (50.0,-21.3) {};
\node [style=green] (Val) at (53.1,-32.075) {};
\node [style=none] (5) at (65.925,-27.525) {};
\node [style=none] (3) at (65.8,-31.275) {};
\node [style=green] (i) at (52.475,-31.225) {};
\node [style=green] (Var) at (50.475,-32.7) {};
\node [style=red vertex] (Vba) at (56.7,-33.875) {};
\node [style=red vertex] (Vb) at (48.825,-21.3) {};
\node [style=green] (o) at (35.825,-30.275) {};
\node [style=red vertex] (j) at (56.675,-28.175) {};
\node [style=red vertex] (m) at (60.725,-31.35) {};
\node [style=green] (e) at (60.5,-18.05) {};
\node [style=none] (6) at (65.8,-25.625) {};
\node [style=green] (Vab) at (54.6,-19.6) {};
\node [style=none] (0) at (65.2,-35.325) {};
\node [style=green] (Vm) at (56.95,-18.725) {};
\node [style=green] (p) at (36.175,-32.05) {};
\node [style=green] (g) at (48.175,-26.15) {};
\node [style=green] (Vo) at (30.25,-29.675) {};
\node [style=red vertex] (l) at (58.475,-30.15) {};
\node [style=red vertex] (t) at (32.2,-31.95) {};
\node [style=red vertex] (Vao) at (62.675,-32.275) {};
\node [style=green] (Vax) at (48.05,-33.875) {};
\node [style=none] (1) at (66.175,-34.325) {};
\node [style=green] (f) at (49.05,-28.6) {};

\node [style=boundary vertex] (7) at (40,-28) {};
\node [style=boundary vertex] (8) at (40,-31) {};
\node [style=boundary vertex] (9) at (40,-32) {};
\node [style=boundary vertex] (10) at (40,-34) {};
\node [style=boundary vertex] (11) at (40,-29) {};
\node [style=boundary vertex] (12) at (40,-26) {};
\node [style=boundary vertex] (13) at (40,-30) {};

\node [style=boundary vertex] (Vaw) at (45,-34) {};
\node [style=boundary vertex] (Vaq) at (45,-33) {};
\node [style=boundary vertex] (Vak) at (45,-32) {};
\node [style=boundary vertex] (Vae) at (45,-31) {};
\node [style=boundary vertex] (Vbi) at (45,-30) {};
\node [style=boundary vertex] (Vbc) at (45,-28) {};
\node [style=boundary vertex] (Vbo) at (45,-26) {};

\node [style=none] (line1a) at (40,-18) {};
\node [style=none] (line1b) at (40,-36) {};
\node [style=none] (line2a) at (45,-18) {};
\node [style=none] (line2b) at (45,-36) {};
	\end{pgfonlayer}
	\begin{pgfonlayer}{edgelayer}

\draw[style=dashed] (line1a) to (line1b);
\draw[style=dashed] (line2a) to (line2b);
\draw [style=plain] (Vab) to (l);
\draw [style=plain] (Va) to (g);
\draw [style=plain] (Vaw) to (Vax);
\draw [style=plain] (4) to (l);
\draw [style=plain] (e) to (m);
\draw [style=plain] (r) to (n);
\draw [style=plain] (Vao) to (Val);
\draw [style=plain] (12) to (Vbc);
\draw [style=plain] (Vba) to (Vab);
\draw [style=plain] (q) to (o);
\draw [style=plain] (10) to (Vae);
\draw [style=plain] (Vau) to (1);
\draw [style=plain] (Vak) to (Val);
\draw [style=plain] (Vau) to (Vm);
\draw [style=plain] (Vb) to (g);
\draw [style=plain] (p) to (t);
\draw [style=plain] (5) to (j);
\draw [style=plain] (f) to (j);
\draw [style=plain] (f) to (Vbc);
\draw [style=plain] (o) to (t);
\draw [style=plain] (e) to (l);
\draw [style=plain] (Va) to (h);
\draw [style=plain] (h) to (Vc);
\draw [style=plain] (11) to (s);
\draw [style=plain] (h) to (l);
\draw [style=plain] (Vbi) to (11);
\draw [style=plain] (h) to (Vbi);
\draw [style=plain] (Vm) to (k);
\draw [style=plain] (Vax) to (Va);
\draw [style=plain] (9) to (Vak);
\draw [style=plain] (i) to (Vae);
\draw [style=plain] (13) to (Vbo);
\draw [style=plain] (7) to (n);
\draw [style=plain] (g) to (k);
\draw [style=plain] (Vc) to (g);
\draw [style=plain] (13) to (q);
\draw [style=plain] (Var) to (Vau);
\draw [style=plain] (7) to (Vaw);
\draw [style=plain] (3) to (m);
\draw [style=plain] (s) to (p);
\draw [style=plain] (Vaq) to (Var);
\draw [style=plain] (Va) to (f);
\draw [style=plain] (8) to (Vaq);
\draw [style=plain] (r) to (Vo);
\draw [style=plain] (Vbo) to (g);
\draw [style=plain] (s) to (Vo);
\draw [style=plain] (f) to (Vb);
\draw [style=plain] (Vm) to (m);
\draw [style=plain] (Vb) to (i);
\draw [style=plain] (Vb) to (Var);
\draw [style=plain] (j) to (Vab);
\draw [style=plain] (e) to (Vao);
\draw [style=plain] (Vao) to (2);
\draw [style=plain] (n) to (s);
\draw [style=plain] (q) to (n);
\draw [style=plain] (e) to (k);
\draw [style=plain] (j) to (Vm);
\draw [style=plain] (Vba) to (Vax);
\draw [style=plain] (r) to (o);
\draw [style=plain] (8) to (o);
\draw [style=plain] (Vo) to (t);
\draw [style=plain] (i) to (m);
\draw [style=plain] (q) to (p);
\draw [style=plain] (t) to (10);
\draw [style=plain] (Val) to (Vc);
\draw [style=plain] (Vc) to (i);
\draw [style=plain] (12) to (r);
\draw [style=plain] (Vab) to (k);
\draw [style=plain] (Vj) to (Vo);
\draw [style=plain] (9) to (p);
\draw [style=plain] (k) to (6);
\draw [style=plain] (Vba) to (0);
\end{pgfonlayer}
\end{tikzpicture}}
\endpgfgraphicnamed}}}
\end{equation*}
Note that the permutation between the two circuits is not strictly
necessary.  We rewrite this diagram as described in 
Table \ref{tab:proof1}, and illustrated in Figure~\ref{fig:rwi}.

\begin{table}[h]
\centering
\begin{tabular}{c|l}
\hline
Graph & Rewrite sequence from previous\\
\hline
\eqref{left1} & Initial graph \\
\eqref{left4} & 3 $\times$ spider, 3 $\times$ bialg, 4 $\times$ spider \\
\eqref{left5} & bialg, 3 $\times$ spider, bialg, spider, 2 $\times$ bialg\\
\eqref{left6} & altcycle4, bialg, 4 $\times$ spider, bialg, spider \\
\eqref{left7} &  bialg \\
\hline
\end{tabular}
\caption{Descripition of rewrite sequence for Figure~\ref{fig:rwi}}
\label{tab:proof1}
\end{table}

\paragraph{Phase 2}
\label{sec:phase-2}

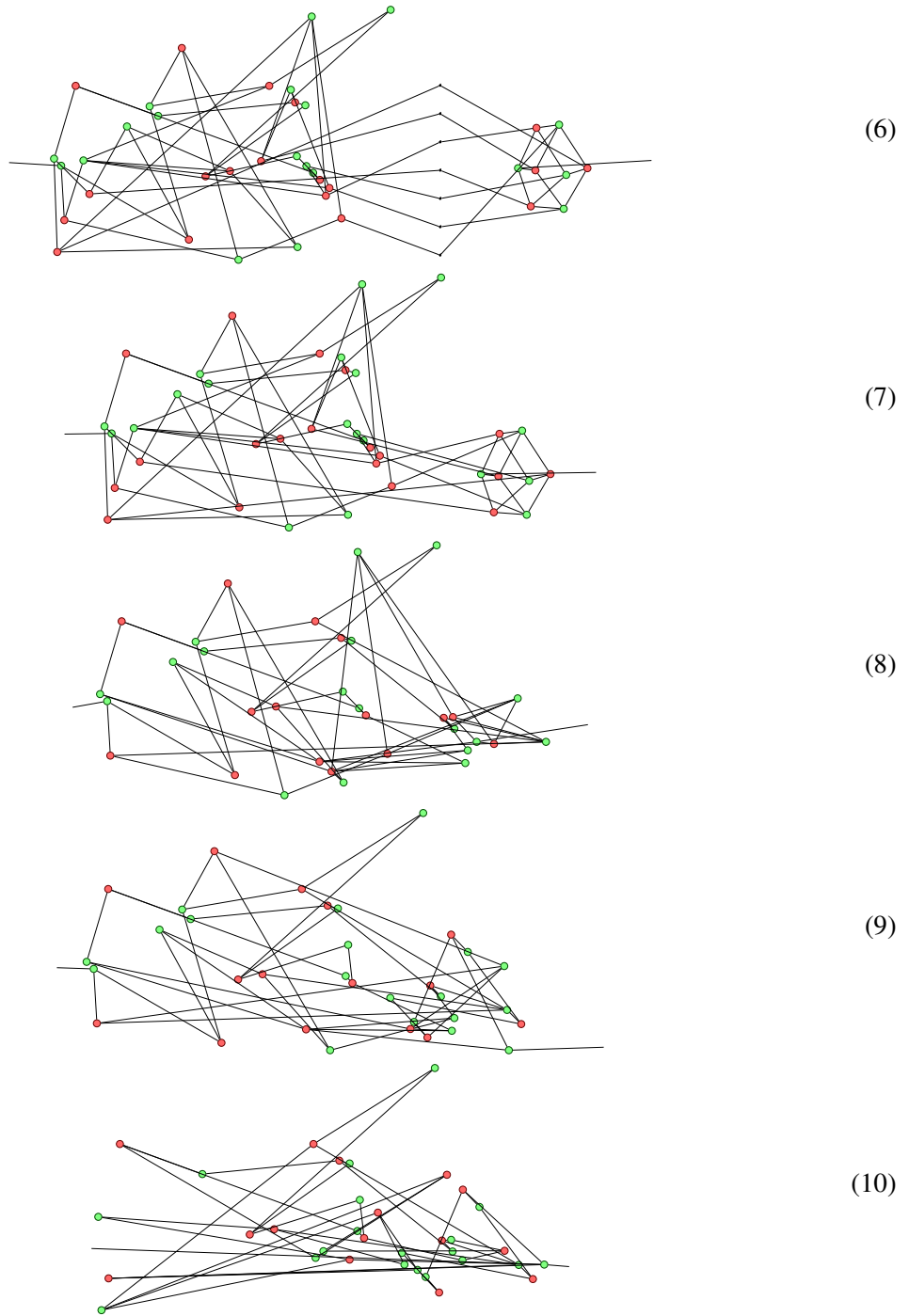
\begin{figure}[htbp]
  \centering
\begin{gather}
  \scalebox{0.4}{\inline{%
\beginpgfgraphicnamed{bigproof-ii-bis}
\begin{tikzpicture}
\node [style=red vertex] (Vdo) at (5.25,-13.025) {};
\node [style=green vertex] (d) at (27.5,-7.65) {};
\node [style=red vertex] (Vcq) at (20.75,-14.2) {};
\node [style=green vertex] (Vcu) at (5.8,-18.3) {};
\node [style=green vertex] (Vo) at (36.5,-18.85) {};
\node [style=red vertex] (k) at (37.4,-21.55) {};
\node [style=green vertex] (Vcv) at (10.475,-14.475) {};
\node [style=red vertex] (j) at (37.8,-16.0) {};
\node [style=green vertex] (Vab) at (21.925,-8.125) {};
\node [style=red vertex] (Vba) at (24.025,-22.4) {};
\node [style=red vertex] (Vdn) at (4.45,-22.525) {};
\node [style=red vertex] (Vcm) at (22.5,-19.675) {};
\node [style=red vertex] (h) at (41.4,-18.85) {};
\node [style=green vertex] (Vcz) at (16.75,-25.325) {};
\node [style=green vertex] (Vcr) at (20.45,-13.3) {};
\node [style=red vertex] (Vdb) at (16.15,-19.05) {};
\node [style=green vertex] (e) at (39.4,-15.775) {};
\node [style=red vertex] (Vdl) at (6.225,-20.675) {};
\node [style=red vertex] (Vcj) at (18.35,-18.35) {};
\node [style=red vertex] (Veb) at (14.425,-19.425) {};
\node [style=green vertex] (f) at (39.725,-21.725) {};
\node [style=none] (Vk) at (45.975,-18.3) {};
\node [style=green vertex] (Vco) at (22.025,-19.175) {};
\node [style=green vertex] (g) at (39.9,-19.325) {};
\node [style=red vertex] (Vdm) at (3.95,-24.775) {};
\node [style=red vertex] (Vcl) at (22.925,-20.8) {};
\node [style=green vertex] (Vcp) at (20.875,-18.0) {};
\node [style=red vertex] (i) at (37.725,-19.0) {};
\node [style=red vertex] (Vcs) at (18.925,-13.025) {};
\node [style=green vertex] (Vdq) at (4.225,-18.675) {};
\node [style=red vertex] (Vck) at (23.175,-20.25) {};
\node [style=none] (Vj) at (0.5,-18.45) {};
\node [style=green vertex] (Vcn) at (21.55,-18.7) {};
\node [style=green vertex] (Vdf) at (8.875,-15.9) {};
\node [style=red vertex] (Vcx) at (12.75,-10.35) {};
\node [style=green vertex] (Vct) at (21.475,-14.4) {};
\node [style=green vertex] (Vdp) at (3.725,-18.175) {};
\node [style=green vertex] (c) at (20.925,-24.425) {};
\node [style=red vertex] (Vda) at (13.25,-23.9) {};
\node [style=green vertex] (Vcw) at (11.075,-15.15) {};

\node [style=boundary vertex] (0) at (31,-25) {};
\node [style=boundary vertex] (1) at (31,-23) {};
\node [style=boundary vertex] (2) at (31,-21) {};
\node [style=boundary vertex] (3) at (31,-19) {};
\node [style=boundary vertex] (4) at (31,-17) {};
\node [style=boundary vertex] (5) at (31,-15) {};
\node [style=boundary vertex] (6) at (31,-13) {};

\node [style=boundary vertex] (7) at (0) {};
\node [style=boundary vertex] (8) at (1) {};
\node [style=boundary vertex] (9) at (2) {};
\node [style=boundary vertex] (10) at (3) {};
\node [style=boundary vertex] (11) at (4) {};
\node [style=boundary vertex] (12) at (5) {};
\node [style=boundary vertex] (13) at (6) {};

\draw [] (Vba) to (Vcz);
\draw [] (h) to (g);
\draw [] (Vcr) to (Vcq);
\draw [] (8) to (f);
\draw [] (Vcq) to (Vct);
\draw [] (Vdp) to (Vdo);
\draw [] (h) to (e);
\draw [] (Vcj) to (Vab);
\draw [] (Vdq) to (Vj);
\draw [] (Vdf) to (Vdl);
\draw [] (e) to (j);
\draw [] (11) to (j);
\draw [] (Veb) to (Vct);
\draw [] (Vcr) to (Vcj);
\draw [] (Vck) to (Vcr);
\draw [] (Vcn) to (2);
\draw [] (Vda) to (Vcv);
\draw [] (f) to (k);
\draw [] (i) to (f);
\draw [] (Vdo) to (Vcw);
\draw [] (Vdn) to (Vdq);
\draw [] (Vcl) to (Vcn);
\draw [] (Vcm) to (Vcp);
\draw [] (c) to (Vdb);
\draw [] (d) to (Vcs);
\draw [] (2) to (9);
\draw [] (1) to (Vck);
\draw [] (12) to (i);
\draw [] (Vab) to (Vcl);
\draw [] (Vk) to (Vo);
\draw [] (Vda) to (Vdf);
\draw [] (i) to (e);
\draw [] (j) to (g);
\draw [] (Vdo) to (Vco);
\draw [] (3) to (Vdl);
\draw [] (c) to (Vdm);
\draw [] (13) to (h);
\draw [] (Vab) to (Vdm);
\draw [] (k) to (10);
\draw [] (5) to (12);
\draw [] (Vcs) to (Vcv);
\draw [] (h) to (f);
\draw [] (Vcx) to (Vcz);
\draw [] (Vdp) to (Vdl);
\draw [] (Vba) to (Vab);
\draw [] (Vcm) to (Vco);
\draw [] (Vcu) to (Vck);
\draw [] (Vcv) to (Vcx);
\draw [] (Vcq) to (Vcw);
\draw [] (9) to (g);
\draw [] (Vcs) to (Vcu);
\draw [] (d) to (Veb);
\draw [] (1) to (8);
\draw [] (Vcn) to (Vcm);
\draw [] (Vdm) to (Vdp);
\draw [] (i) to (Vo);
\draw [] (j) to (Vo);
\draw [] (g) to (k);
\draw [] (Vdb) to (Vcu);
\draw [] (7) to (e);
\draw [] (Vdb) to (Vdf);
\draw [] (4) to (11);
\draw [] (6) to (13);
\draw [] (Vcx) to (c);
\draw [] (Vcu) to (Vdn);
\draw [] (3) to (10);
\draw [] (Vo) to (k);
\draw [] (4) to (Vcl);
\draw [] (Vcl) to (Vcu);
\draw [] (Veb) to (Vcp);
\draw [] (6) to (Vdm);
\draw [] (Vdn) to (Vcz);
\draw [] (0) to (7);
\draw [] (Vdq) to (Vda);
\draw [] (5) to (Vcj);
\draw [] (Vba) to (0);

\end{tikzpicture}}
\endpgfgraphicnamed}} \label{tot1}\\
  \scalebox{0.4}{\totii}\label{tot2} \\
  \scalebox{0.4}{\totiii} \label{tot3}\\
  \scalebox{0.4}{\totiv} \label{tot4}\\
  \scalebox{0.4}{\totv} \label{tot5}
\end{gather}
  \caption{Rewrite sequence 2(a)}
  \label{fig:rwiia}
\end{figure}

Next, we adjoin the decoding circuit to diagram \eqref{left7}.
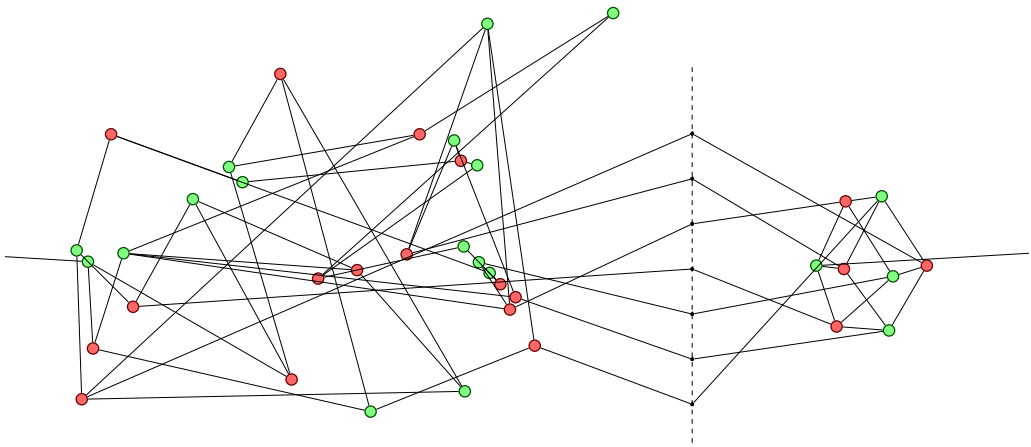
\begin{equation*}
  \scalebox{0.6}{\inline{%
\beginpgfgraphicnamed{bigproof-ii}
\begin{tikzpicture}
\node [style=red vertex] (Vdo) at (5.25,-13.025) {};
\node [style=green vertex] (d) at (27.5,-7.65) {};
\node [style=red vertex] (Vcq) at (20.75,-14.2) {};
\node [style=green vertex] (Vcu) at (5.8,-18.3) {};
\node [style=green vertex] (Vo) at (36.5,-18.85) {};
\node [style=red vertex] (k) at (37.4,-21.55) {};
\node [style=green vertex] (Vcv) at (10.475,-14.475) {};
\node [style=red vertex] (j) at (37.8,-16.0) {};
\node [style=green vertex] (Vab) at (21.925,-8.125) {};
\node [style=red vertex] (Vba) at (24.025,-22.4) {};
\node [style=red vertex] (Vdn) at (4.45,-22.525) {};
\node [style=red vertex] (Vcm) at (22.5,-19.675) {};
\node [style=red vertex] (h) at (41.4,-18.85) {};
\node [style=green vertex] (Vcz) at (16.75,-25.325) {};
\node [style=green vertex] (Vcr) at (20.45,-13.3) {};
\node [style=red vertex] (Vdb) at (16.15,-19.05) {};
\node [style=green vertex] (e) at (39.4,-15.775) {};
\node [style=red vertex] (Vdl) at (6.225,-20.675) {};
\node [style=red vertex] (Vcj) at (18.35,-18.35) {};
\node [style=red vertex] (Veb) at (14.425,-19.425) {};
\node [style=green vertex] (f) at (39.725,-21.725) {};
\node [style=none] (Vk) at (45.975,-18.3) {};
\node [style=green vertex] (Vco) at (22.025,-19.175) {};
\node [style=green vertex] (g) at (39.9,-19.325) {};
\node [style=red vertex] (Vdm) at (3.95,-24.775) {};
\node [style=red vertex] (Vcl) at (22.925,-20.8) {};
\node [style=green vertex] (Vcp) at (20.875,-18.0) {};
\node [style=red vertex] (i) at (37.725,-19.0) {};
\node [style=red vertex] (Vcs) at (18.925,-13.025) {};
\node [style=green vertex] (Vdq) at (4.225,-18.675) {};
\node [style=red vertex] (Vck) at (23.175,-20.25) {};
\node [style=none] (Vj) at (0.5,-18.45) {};
\node [style=green vertex] (Vcn) at (21.55,-18.7) {};
\node [style=green vertex] (Vdf) at (8.875,-15.9) {};
\node [style=red vertex] (Vcx) at (12.75,-10.35) {};
\node [style=green vertex] (Vct) at (21.475,-14.4) {};
\node [style=green vertex] (Vdp) at (3.725,-18.175) {};
\node [style=green vertex] (c) at (20.925,-24.425) {};
\node [style=red vertex] (Vda) at (13.25,-23.9) {};
\node [style=green vertex] (Vcw) at (11.075,-15.15) {};

\node [style=boundary vertex] (0) at (31,-25) {};
\node [style=boundary vertex] (1) at (31,-23) {};
\node [style=boundary vertex] (2) at (31,-21) {};
\node [style=boundary vertex] (3) at (31,-19) {};
\node [style=boundary vertex] (4) at (31,-17) {};
\node [style=boundary vertex] (5) at (31,-15) {};
\node [style=boundary vertex] (6) at (31,-13) {};

\node [style=boundary vertex] (7) at (0) {};
\node [style=boundary vertex] (8) at (1) {};
\node [style=boundary vertex] (9) at (2) {};
\node [style=boundary vertex] (10) at (3) {};
\node [style=boundary vertex] (11) at (4) {};
\node [style=boundary vertex] (12) at (5) {};
\node [style=boundary vertex] (13) at (6) {};

\node [style=none] (line1) at (31,-10) {};
\node [style=none] (line2) at (31,-27) {};
\draw [style=dashed] (line1) to (line2);

\draw [] (Vba) to (Vcz);
\draw [] (h) to (g);
\draw [] (Vcr) to (Vcq);
\draw [] (8) to (f);
\draw [] (Vcq) to (Vct);
\draw [] (Vdp) to (Vdo);
\draw [] (h) to (e);
\draw [] (Vcj) to (Vab);
\draw [] (Vdq) to (Vj);
\draw [] (Vdf) to (Vdl);
\draw [] (e) to (j);
\draw [] (11) to (j);
\draw [] (Veb) to (Vct);
\draw [] (Vcr) to (Vcj);
\draw [] (Vck) to (Vcr);
\draw [] (Vcn) to (2);
\draw [] (Vda) to (Vcv);
\draw [] (f) to (k);
\draw [] (i) to (f);
\draw [] (Vdo) to (Vcw);
\draw [] (Vdn) to (Vdq);
\draw [] (Vcl) to (Vcn);
\draw [] (Vcm) to (Vcp);
\draw [] (c) to (Vdb);
\draw [] (d) to (Vcs);
\draw [] (2) to (9);
\draw [] (1) to (Vck);
\draw [] (12) to (i);
\draw [] (Vab) to (Vcl);
\draw [] (Vk) to (Vo);
\draw [] (Vda) to (Vdf);
\draw [] (i) to (e);
\draw [] (j) to (g);
\draw [] (Vdo) to (Vco);
\draw [] (3) to (Vdl);
\draw [] (c) to (Vdm);
\draw [] (13) to (h);
\draw [] (Vab) to (Vdm);
\draw [] (k) to (10);
\draw [] (5) to (12);
\draw [] (Vcs) to (Vcv);
\draw [] (h) to (f);
\draw [] (Vcx) to (Vcz);
\draw [] (Vdp) to (Vdl);
\draw [] (Vba) to (Vab);
\draw [] (Vcm) to (Vco);
\draw [] (Vcu) to (Vck);
\draw [] (Vcv) to (Vcx);
\draw [] (Vcq) to (Vcw);
\draw [] (9) to (g);
\draw [] (Vcs) to (Vcu);
\draw [] (d) to (Veb);
\draw [] (1) to (8);
\draw [] (Vcn) to (Vcm);
\draw [] (Vdm) to (Vdp);
\draw [] (i) to (Vo);
\draw [] (j) to (Vo);
\draw [] (g) to (k);
\draw [] (Vdb) to (Vcu);
\draw [] (7) to (e);
\draw [] (Vdb) to (Vdf);
\draw [] (4) to (11);
\draw [] (6) to (13);
\draw [] (Vcx) to (c);
\draw [] (Vcu) to (Vdn);
\draw [] (3) to (10);
\draw [] (Vo) to (k);
\draw [] (4) to (Vcl);
\draw [] (Vcl) to (Vcu);
\draw [] (Veb) to (Vcp);
\draw [] (6) to (Vdm);
\draw [] (Vdn) to (Vcz);
\draw [] (0) to (7);
\draw [] (Vdq) to (Vda);
\draw [] (5) to (Vcj);
\draw [] (Vba) to (0);

\end{tikzpicture}}
\endpgfgraphicnamed}}
\end{equation*}
The rewriting sequence is
described in Table~\ref{tab:proof2} and illustrated in
Figures~\ref{fig:rwiia} and \ref{fig:rwiib}.  The final diagram is a
single wire, hence the cicruit rewrites to the identity, as required.

\begin{table}[h]
\centering
\begin{tabular}{c|l}
\hline
Graph & Rewrite sequence from previous\\
\hline

\eqref{tot1} & Initial graph \\
\eqref{tot2} & 6 $\times$ spider, 1 $\times$ bialg, 2 $\times$ spider\\
\eqref{tot3} & 2 $\times$ bialg, 4 $\times$ spider, 1 $\times$ bialg, 2 $\times$ spider \\
\eqref{tot4} & 3 $\times$ bialg, 4 $\times$ spider, 1 $\times$ bialg, 4 $\times$ spider\\
\eqref{tot5} & 15 $\times$ spider (only on green)\\
\eqref{tot6} & 8 $\times$  spider (only on red)  \\
\eqref{tot7} & 1 $\times$ antiloop, 2 $\times$ bialg, 3 $\times$ x$\_$abelian1 \\
\eqref{tot8} & 1 $\times$ hopf-banged, 1 $\times$ spider\\
\eqref{tot9} & 6 $\times$ spider \\
\eqref{tot10} & 4 $\times$ hopf-banged \\ 
\eqref{tot11} & 1 $\times$  bialgebra1-rev, 4 $\times$ green spiders,
3 $\times$  red spiders \\
\eqref{tot12} & drop scalars ($=1$)\\
\hline
\end{tabular}
\caption{Description of rewrite sequence for Figures ~\ref{fig:rwiia} and \ref{fig:rwiib}.}
\label{tab:proof2}
\end{table}

\begin{figure}[htbp]
  \centering
\begin{gather}
  \scalebox{0.5}{\totvi}\label{tot6} \\
   \scalebox{0.5}{\totvii} \label{tot7}\\
   \scalebox{0.5}{\totviii} \label{tot8}\\
   {\totix} \label{tot9}\\
   {\totx} \label{tot10}\\   
   {\totxi} \rule{0pt}{3em} \label{tot11}  \\
   {\totxii} \rule{0pt}{3em} \label{tot12} 
\end{gather}
  \caption{Rewrite sequence 2(b)}
  \label{fig:rwiib}
\end{figure}
\end{document}